\documentclass[12pt,a4paper]{article}

\usepackage{mathtools}
\usepackage{amssymb,bm,bbold}
\usepackage{enumerate}
\usepackage{comment}
\usepackage{esvect}
\usepackage[hidelinks]{hyperref}
\usepackage{a4wide}
\usepackage{cite}
\usepackage{import}

\usepackage{shuffle}

\usepackage{tikz}
	 \usetikzlibrary{arrows,decorations.markings}
	 \usetikzlibrary{calc}
	 \usetikzlibrary{3d}

\usepackage{pgfplots}
\pgfplotsset{compat=newest}
\usepgfplotslibrary{fillbetween}

\newcommand{\dd}{\text{d}}
\newcommand{\br}[1]{\langle #1 \rangle }
\DeclareMathOperator{\sign}{sign}

\def\cA{{\cal A}}
\def\cN{{\cal N}}
\def\cR{{\mathcal R}}

\def\centerarc[#1](#2)(#3:#4:#5){ \draw[#1] ($(#2)+({#5*cos(#3)},{#5*sin(#3)})$) arc (#3:#4:#5); }

\begin{document}

\null\vskip-12pt \hfill  \\
\null\vskip-12pt \hfill   \\

\vskip2.2truecm
\begin{center}
	\vskip 0.2truecm {\Large\bf
		{\Large Internal boundaries of the loop amplituhedron}
	}\\
	\vskip 1truecm
	{\bf Gabriele Dian, Paul Heslop and Alastair Stewart
	}
	
	\vskip 0.4truecm
	
	{\it
		Mathematics Department, Durham University, \\
		Science Laboratories, South Rd, Durham DH1 3LE, \vskip .2truecm                        }
\end{center}

\vskip 1truecm 

\begin{abstract}

    The strict definition of positive geometry implies that all maximal residues of its canonical form are  $\pm 1$. 
We observe, however, that the loop integrand of the amplitude in planar $\cN=4$ super Yang-Mills
has maximal residues not equal to $\pm 1$.
We find the reason for this is that deep in the boundary structure of the loop amplituhedron there are geometries which contain
 internal boundaries: codimension  one defects  separating two regions of opposite orientation. This phenomenon requires a generalisation of the concept of positive geometry and canonical form to include  such  internal boundaries and also suggests the utility of a further generalisation to `weighted positive geometries'.  
We re-examine the deepest cut of $\cN=4$ amplitudes in light of this and obtain new all order residues. 

\end{abstract}

\medskip

\noindent

\newpage
\tableofcontents
\newpage

\section{Introduction}

The amplituhedron is a geometrical object introduced in~\cite{Arkani-Hamed:2013jha,Arkani-Hamed:2013kca} and was discovered to yield a beautiful intrinsic  definition  of the perturbative expansion of planar amplitudes in $\mathcal{N}=4$ super Yang-Mills (SYM), allowing for entirely novel expressions for amplitudes to be found (see~\cite{Herrmann:2022nkh} for a review). In this framework, amplitude integrands are obtained as a differential form, called the canonical form, of the amplituhedron. 
The boundary structure of the amplituhedron then encodes  the full singularity structure of the integrand.
More precisely, residues of the  amplitude are the canonical forms of the corresponding amplituhedron boundaries,  which recursively give further residues as boundaries of the boundaries etc.
Thus the amplituhedron provides a fascinating way to analyse the singularity structure of amplitude integrands,  which in turn  is intimately connected to the branch cut structure of the integrated amplitude~\cite{Dennen:2016mdk, Prlina:2017azl}.
In this paper, we investigate  multiple residues of amplitudes and the corresponding amplituhedron boundary structure.  In particular, we point out new features which have not been appreciated previously.

First, we note that a direct consequence of amplitudes  arising from positive geometries is that the  amplitude should have unit maximal residues. This arises geometrically simply from the fact that maximal residues correspond to dimension $0$ geometries, ie points, which can only differ by their orientation. 
Tree-level amplitudes indeed appear to have unit maximal residues.
The  tree level superamplitudes can be computed by summing a certain set of on-shell diagrams~\cite{Arkani-Hamed:2016byb} arising from the BCFW recursion relation~\cite{Britto:2005fq}. The on-shell diagrams manifestly have only logarithmic singularities and non-vanishing maximal residues equal to $\pm1$ and  have a natural geometric interpretation in amplituhedron space~\cite{Arkani-Hamed:2013jha}. It was recently proven that they provide a tessellation of the amplituhedron~\cite{Even-Zohar:2021sec} and it would be interesting to see if the details of this proof can also be used to prove  that the non-vanishing maximal residues only equal $\pm1$.%
\footnote{It does not automatically follow since there are simple examples of geometries which can be tessellated with positive geometries but which themselves are not positive geometries, as we will see.}

What we will observe however is that, unlike at  tree level, the maximal residues of the {\em loop} amplitude integrand take many different values in $\mathbb{Z}$.
Examining the corresponding geometry, the loop amplituhedron,  we find starting from 2 loops that it contains a novel feature, namely  {\em internal boundaries}, deep within its boundary structure and find these are the geometric source of the non-unit residues. 
By an internal boundary we mean a codimension 1 surface separating two regions of opposite orientation, as below:
\begin{align*}
    \begin{tikzpicture}
    \node at (-6,0) {internal boundary:};
    \draw[-latex,thin] (-3,0) -- (1.1,0);
    \draw[thin]  (1,0) -- (5,0);
    \draw[-latex,thin] (-3,0.05) -- (1.1,0.05);
    \draw[thin]  (1,0.05) -- (5,0.05);
    \draw[-latex,semithick,red] (1,1) arc[radius=.2, start angle=60, end angle=300]-- ++(20:1pt);
    \draw[-latex,semithick,red] (1,-1) arc[radius=.2, start angle=300, end angle=60]-- ++(-20:1pt);
    \end{tikzpicture}
\end{align*}
We emphasise that such internal boundaries do not appear in the amplituhedron itself, but deep within its boundary structure. That is, we claim that a certain boundary component of a boundary component of a...of the amplituhedron will contain an internal boundary.

We find that  the internal boundaries appear to be closely associated with `composite singularities' (see~\cite{Arkani-Hamed:2009ljj}). A simple example of a composite singularity is given by $2/(z(z+xy))$. This only has two factors, but when taking the residue at $z=0$ the irreducible factor $z+x y$ factorises to  $x y$. Indeed the above is the canonical form of the ordinary looking geometry  given by points in 3d satisfying $z>0, z+xy>0$, which pictured from below looks like the following:  
\begin{center}
\includegraphics[width=.3\textwidth]{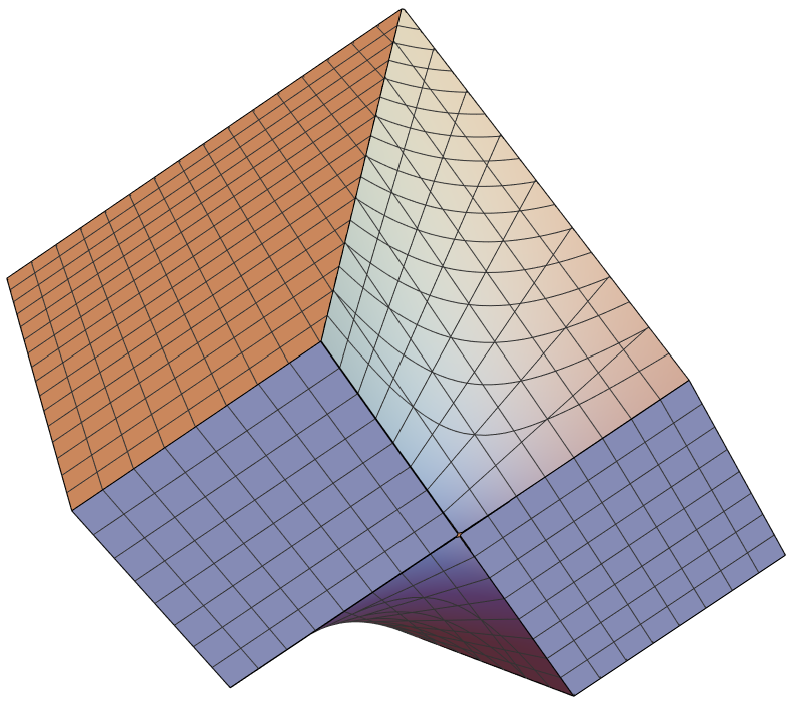}
\end{center}
This geometry contains a boundary (nearest to the viewer in the above picture) at $z=0$, which is given by $xy>0$; two opposite quadrants of a plane. This 2d boundary in turn has a boundary at $y=0$, consisting of two $1$-dimensional regions $x>0,x<0$ with opposite orientation. Finally, this 1d region contains a $0$-dimensional internal boundary at $x=0$.
This type of structure appears in the boundaries of the loop amplituhedron, implying
 that the loop amplituhedron is not a `positive geometry' in the sense  usually understood~\cite{Arkani-Hamed:2017tmz}.\footnote{Note that maximal residues differing from $\pm1$ have been observed previously in  the square of the super amplitude, where they appear already at tree level~\cite{Dian:2021idl}. 
In fact, there they appear in a less subtle way geometrically:
the tree squared amplituhedron interior is disconnected and internal boundaries correspond to boundaries shared between the different connected components, whereas the loop amplituhedron interior is connected.} 

 There has been tremendous progress in understanding the geometry of the tree level amplituhedron and its canonical form~\cite{Galashin:2018fri, Mohammadi:2020plf, Kojima:2020tjf, Even-Zohar:2021sec, Blot:2022geq}.
The loop amplituhedron and its tilings on the other hand are much less well understood.  A first exploration of the boundaries of the MHV loop amplituhedron was started in~\cite{Arkani-Hamed:2013kca}, where the 2-loop MHV amplitude was computed by triangulating the amplituhedron and several cuts were discussed. Then a systematic investigation of the boundaries of the MHV loop amplituhedron was carried out up to four loops in~\cite{Franco:2014csa} and extended to negative mutual positivity conditions in~\cite{Galloni:2016iuj}.
Internal boundaries, however, appear to have been missed in the construction of the stratification of the loop amplituhedron in previous works. One possible reason for this is that these boundaries cannot be labelled by the Pl\"ucker coordinates that are naturally used to describe the amplituhedron: $\br{Yijkl},\br{ABij}$ and $\br{A_iB_iA_jB_j}$. For example, internal boundaries arise when computing the all-in-one-point cut of~\cite{Arkani-Hamed:2018rsk} via consecutive single residues. By carefully looking at the boundary corresponding to three loop lines $(A_1B_1),(A_2B_2),(A_3B_3)$ all intersecting the point $A$, one finds that  $\br{AB_1B_2B_3}=0$ represents an internal boundary.

 We conclude that we need to generalise the definition of `positive geometry' to allow for such internal boundaries and to incorporate the loop amplituhedron. Thus, we define {\em generalised positive geometries} (GPGs) which include internal boundaries. 
Then we introduce a corresponding   extension of the recursive definition of the canonical form, by adding an additional term for internal boundaries which should appear with a factor of 2.  In doing so,  the loop level amplitude can still  be  obtained as the canonical form of the amplituhedron geometry. 

In practise,  the most convenient way to compute canonical forms of positive geometries is via tessellation (eg via cylindrical decomposition) rather than explicit use of the recursive definition. It is important to note that 
tessellations work for these generalised positive geometries (GPGs) just as for positive geometries. The canonical form of a GPG can be computed simply by summing over the canonical forms of the tiles in a tessellation. Indeed, the space of GPGs is closed under the disjoint-union: any geometry that can be tiled in GPGs is itself a GPG.  This differs from positive geometries, which are incomplete under tessellation.  We will see examples of geometries that can be tessellated in positive geometries, but which are not themselves a positive geometry.

While the above generalisation of positive geometry to include internal boundaries is perfectly  adequate to deal with the loop amplituhedron, it immediately suggests an even more general type of geometry which may be of wider use, namely `weighted positive geometry'. 
Any oriented geometry is defined by specifying a region (the geometry) together with its orientation form. However, in order to compute the canonical form of a geometry containing internal boundaries, such as  the loop amplituhedron, one needs some extra information encoding which points belong to an external boundary and which belong to an internal boundary.  It then seems very natural to define a new object called the weighted geometry (WG). A WG is given by a pair $(w,O)$, where $w$ is an integer valued function we call the weight function and $O$ is the orientation form. The value of the weight function on a point intuitively represents the number of coinciding  oriented geometries at that point.  For example, an ordinary oriented geometry $X_{\geq0}$ can be described as a weighed geometry with weight function $w(x)=1$ for all $x\in X_{\geq0}$ and zero elsewhere,  while an internal boundary will have $w(x)=2$ instead   (an internal boundary can be viewed as two external boundaries coinciding).

The space of weighted geometries is naturally equipped with two key operators: a sum and a projection onto boundary. The sum generalizes the union of disconnected oriented geometries by allowing for overlaps (the weights on the overlap simply sum). The projection onto boundary operator instead allows one to  define the induced weight function and orientation on the boundary of a WG. As a consequence, boundaries of WGs are WGs. This construction allows for a recursive definition of the canonical form and weighted positive geometries (WPG), that treats internal and external boundaries on the same footing and makes the tiling properties of the canonical form trivial. In fact, the canonical form turns out to be a linear operator with respect to the sum of WPGs. Then, because a  tiling of a geometry $X_{\geq 0}$ in this new language is nothing but a sum of WPGs, it follows trivially that the canonical form of a sum (union) of WPGs is equal to the sum of the canonical forms.

Another feature of the geometry of the amplituhedron which 
seems to have not been emphasised in the literature previously is the geometrical equivalent of the fact that multiple residues are not in general uniquely defined. One  way to define multiple residues is via the residue form~\cite{mathresidues} (see also~\cite{Arkani-Hamed:2009ljj}) which essentially defines it via a sequence of single residues. However, taking these in different orders can give completely   different results.
There is a direct analogue of this fact in terms of taking boundaries of the corresponding geometry. Rather than talking about codimension 2 boundary components, instead it is the precise boundary component of the boundary component  which will give the  multiple residue defined by taking the corresponding simple poles in sequence. Taking these boundary components in different orders can give different results.

Consider the solid 3d geometry below, which will serve as a very simple example to illustrate the dependence on the order of taking boundaries.
\begin{center}
\label{fig:solid}
\includegraphics[width=.3\textwidth]{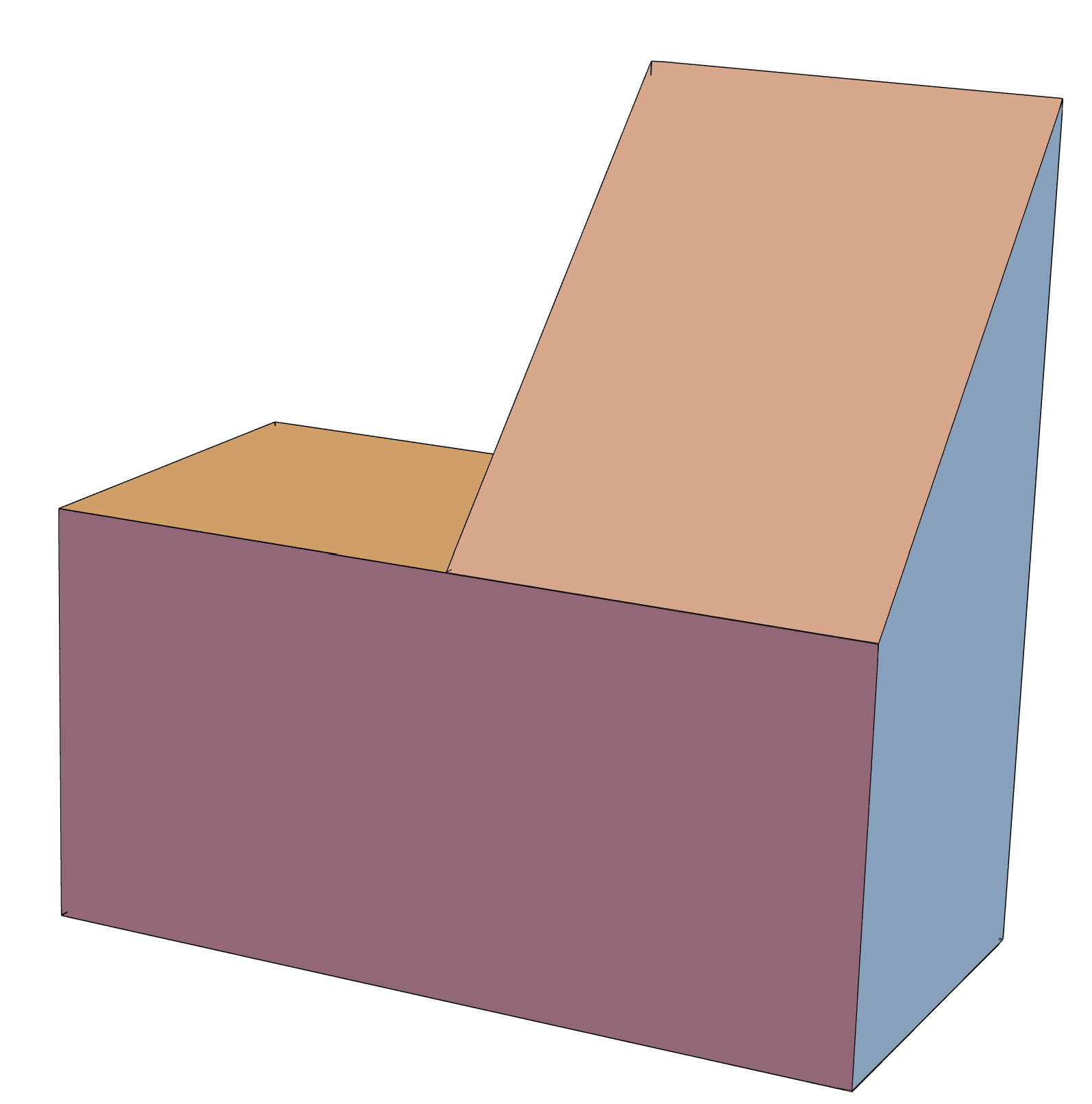}
\end{center}
Here we have  a 3d shape which looks like a house with a flat roof on the left and an  angled roof on the right.
The planes in which the two roofs lie intersect along the top of the front wall which we will call the `front eaves'.
While one might wish to talk about the  codimension two boundary component corresponding to the entire front eaves, taking appropriate boundaries of boundaries can give results which differ from this. 
One could  first  take the flat roof boundary component and then take the front eave boundary of that. This results in only the left half of the front eaves. If on the other hand we first  take the slanted roof boundary component and then the font eaves boundary of that, we obtain the right  half of the front eaves.
 Finally we could instead first take the boundary component on the front wall, and then the boundary component of that at the top, resulting in the entire length of the front eaves.
We thus arrive at three different results from taking a sequence of  boundary components.

This is completely consistent with what we get from multiple residues. 
Indeed, this example can be made completely precise and the corresponding canonical form and residues taken. We choose coordinates such that the flat roof lies on the plane $z=0$, the slanted roof on $y=z$ and the front wall $y=0$ (we also put the two side walls at $x=-1,x=1$, the back wall at $y=1$ and the floor at $z=-1$).  The corresponding canonical form, which can easily be obtained by summing the canonical form of the living space and the roof, is 
 \begin{align}
     \Omega=\frac{1}{(x-1) x (y-1) z (y-z)}-\frac{2}{(x-1) (x+1) (y-1) y z (z+1)}\ .
 \end{align}
 The residue corresponding to the front boundary of the flat roof is Res${}_{y=0}$Res${}_{z=0} \Omega = -dx/(x(x+1))=\Omega[-1,0]$, which is the canonical form of the one dimensional interval $-1 \leq x \leq 0$. On the other hand, the residue corresponding to the front boundary of the slanted roof is
Res${}_{y=0}$Res${}_{z=y} \Omega = -dx/(x(x-1))=\Omega[0,1]$, the canonical form of the one dimensional interval $0 \leq x \leq 1$. Finally, the residue corresponding to the top boundary of the front wall is
Res${}_{z=0}$Res${}_{y=0} \Omega = 2dx/((x-1)(x+1))=\Omega[1,-1]$.
All cases correspond in the end to the codimension two line $z=0,y=0$ but the precise way we get there gives different results.

Another example of this phenomenon is the case of all $l$ loop lines intersecting in a single  point, which is a configuration closely related to the deepest cut of~\cite{Arkani-Hamed:2018rsk}.  At 4 points,  from 4 loops onward,
different orderings of single residues give algebraically different results, as we will show in section \ref{HigherLoop}.
With this in mind, we do a detailed analysis of the all-in-one-point cut and find quite a complicated structure in general, although it seems one can always keep taking further loop loop residues to  reduce to the 3 loop all-in-one-point-and-plane cut. 

This paper is structured as follows. In section 2 we present a simple example of a maximal residue at two loops equal to $\pm 2$ rather than $\pm 1$, and give  its geometrical interpretation as an internal boundary. In section 3 we formally define generalized positive geometries (GPGs), which allow internal boundaries, and their canonical forms. We discuss tilings of GPGs and the key property that {\em any} geometry that can be triangulated by GPGs is a  GPG,  a property not shared by positive geometries. We describe the algebraic cylindrical decomposition algorithm, originally presented in~\cite{Eden:2017fow}, to compute the canonical form, and use it to identify a class of GPGs. Then, we introduce a generalisation of GPGs which we call weighted positive geometries. In section 4 we turn our attention to a specific boundary of the loop amplituhedron, that is the all-in-one-point cut, and we compute its geometry and discuss its internal boundary.  Finally, in section 5 we take further cuts on the result of the all-in-one-point cut and obtain a new all loop formula.

\section{Two loop maximal cuts and internal boundaries}
\label{sec:2}

The purpose of this section is to show that the loop level amplitude can have maximal residues which are not $\pm 1,0$ and give the  geometrical interpretation of this fact. Consider the four point two-loop MHV amplitude integrand  written as a volume form in momentum twistor space
\begin{align}\label{2loopMHV}
\text{MHV}(2)=\frac{\br{A_1B_1\dd^2A_1}\br{A_1B_1\dd^2B_1}\br{A_2B_2\dd^2A_2}\br{A_2B_2\dd^2B_2}\br{1234}^3}{\br{A_1B_1A_2B_2}\br{A_1B_114}\br{A_1B_112}\br{A_2B_223}\br{A_2B_234}}\times \nonumber\\\times \left[ \frac{1}{\br{A_1B_134}\br{A_2B_212}}+ \frac{1}{\br{A_1B_123}\br{A_2B_214}} \right] \quad +\quad A_1B_1\ \leftrightarrow \  A_2B_2 \; .
\end{align}
Here  we have external momentum twistors $Z_1,..,Z_4 \in \mathbb{C}^4$ and loop integration variables $A_i B_i\in \mathbb{C}^4$ which  define  a plane through the origin $a A_i + b B_i \in \mathbb{C}^4$ i.e. a line in projective twistor space. The bracket notation denotes the determinant of the $4\times4$ matrix formed by taking the four twistors inside as columns. We also surpress the $Z$s, so eg   $\br{A_1B_112}:=\det (A_1,B_1,Z_1,Z_2)$.

Now we  compute the multi-residue corresponding to taking a sequence of residues on \begin{align}\label{mres}
    \br{A_1B_112}=0, \; \br{A_1B_134}=0, \; \br{A_2B_212}=0 , \; \br{A_2B_234}=0, \;\br{A_1B_1A_2B_2}=0\; ,
\end{align}
followed by a residue on a hidden pole which appears at $\br{12B_1B_2}=0$. 
To do this we first parametrise the 4$\times$4 $Z=(Z_1Z_2Z_3Z_4)$ matrix as the identity and the loops as
\begin{align}
    \begin{pmatrix}A_i \\B_i\end{pmatrix}= \begin{pmatrix} 1& a_i& 0& -b_i\\ 0&c_i&1&d_i
    \end{pmatrix}\; .
    \label{parametrization}
\end{align}
For this choice, the brackets read
\begin{align}
  \br{A_iB_i12}=b_i, \quad \br{A_iB_i23}= d_i,\quad \br{A_iB_i34}=c_i\; , \quad \br{A_iB_i14}=a_i\; ,  \nonumber\\
  \br{A_1B_1A_2B_2}= -(b_1 - b_2) (c_1 - c_2) - (a_1 - a_2) (d_1 - d_2)\; , \quad \br{1234}=1\; , \nonumber \\
  \br{A_iB_i\dd^2 A_i}\br{A_iB_i\dd^2 B_i}=\dd a_i \dd b_i  \dd c_i  \dd d_i \;.  
\end{align}
Omitting the differentials, the amplitude~\eqref{2loopMHV} in these coordinates reads
\begin{align}
 \text{MHV}(2)=-\frac{a_2 d_1+a_1 d_2+b_2 c_1+b_1 c_2}{a_1 a_2 b_1 b_2 c_1 c_2 d_1 d_2 \left(\left(a_1-a_2\right) \left(d_1-d_2\right)+\left(b_1-b_2\right) \left(c_1-c_2\right)\right)} \; .
\end{align}
Now we take the first four residues in~\eqref{mres}, namely $b_1=0,c_1=0,b_2=0,c_2=0$. We see that the complicated factor in the denominator factorises thus revealing a new pole,%
\footnote{Such poles have been observed in the amplitude previously, and have been dubbed composite residues~\cite{Arkani-Hamed:2009ljj}. See also the example in the introduction.}
giving 
\begin{align}
     -\frac{a_2 d_1+a_1 d_2}{a_1 a_2  d_1 d_2  \left(a_1-a_2\right) \left(d_1-d_2\right) } \; .
     \label{laddercut}
\end{align}
Then we take the residue in $a_1$ at $a_1=a_2$, obtaining 
\begin{align}
     -\frac{ (d_1+ d_2)}{ a_2  d_1 d_2 \left(d_1-d_2\right)}\; .
\end{align}
We continue by taking the residue in $d_1$ at $d_1=d_2$, obtaining 
\begin{align}
     -\frac{ 2}{ a_2  d_2 }\; ,
     \label{finalres}
\end{align}
up to an overall sign due to the ordering of the differential $\dd^2 a_i\dd^2 b_i\dd^2 c_i\dd^2 d_i$. From \eqref{finalres}, it is clear that this form has maximal residues equal to $\pm 2$, contradicting the consequence of this being the canonical form of a positive geometry.

\subsection*{Geometrical interpretation}

Let's  try to understand how this factor of two appears  geometrically from the amplituhedron.
First we take boundaries corresponding to the four residues at
$\{\br{A_1B_112}=0,\br{A_1B_134}=0,\br{A_2B_212}=0,\br{A_2B_234}=0\}$.
The order in which these are performed is not important and the resulting geometry has each loop line $(A_iB_i)$ described by a point $A_i$ in the segment $\overline{12}$ and a point $B_i$ in $\overline{34}$, together with a further mutual positivity condition $\br{A_1B_1A_2B_2}>0$.
It is natural then to parametrise $A_i,B_i$ as $A_i=Z_1+a_iZ_2$ and $B_i=Z_3+d_iZ_4$, so that the geometry is described by the inequalities
\begin{align}
    a_i>0, \quad d_i>0, \quad - (a_1 - a_2) (d_1 - d_2)>0\; .
\end{align}
Notice that the mutual positivity inequality factorizes in to the product of two terms $a_1>a_2,d_1<d_2$ or $a_1<a_2,d_1>d_2$. This is just the geometrical version of composite residues mentioned in the introduction and  above~\eqref{laddercut}.
The factorisation results in a corresponding geometry given by two regions
\begin{equation}
    \begin{split}
    \mathcal{R}_1:=\{a_1,a_2,d_1,d_2 \ | \ a_1>a_2>0  \land d_2>d_1>0 \}\; ,  \\
    \mathcal{R}_2:=\{a_1,a_2,d_1,d_2 \ | \ a_2>a_1>0  \land d_1>d_2>0 \}\; .
    \end{split}
\end{equation}

 This geometry is illustrated in the following picture, 
\begin{align}\label{pic}
   \begin{tikzpicture}
     \fill[fill=gray!20] (0,0)--(2,0)--(2,2)--(0,2)--(0,0);
    \fill[fill=gray!20] (0,0)--(-2,0)--(-2,-2)--(0,-2)--(0,0);
    \draw[-latex] (0,2) -- (0,.9);
    \draw[-latex] (0,1) -- (0,0) -- (1.1,0);
    \draw  (1,0) -- (2,0);
 \draw[-latex] (0,-2) -- (0,-.9);
    \draw[-latex] (0,-1) -- (0,0) -- (-1.1,0);
    \draw  (-1,0) -- (-2,0);
    \draw[-latex,semithick,red] (1,1) arc[radius=.2, start angle=60, end angle=300]-- ++(20:1pt);
    \node at (1.4,1.4) {$\mathcal{R}_1$};
    \draw[-latex,semithick,red] (-1,-1) arc[radius=.2, start angle=-120, end angle=120]-- ++(200:1pt);
    \node at (-1.4,-1.4) {$\mathcal{R}_2$};
    \node at (0,0) {$\bullet$};
    \end{tikzpicture}
\end{align}
where the $x$ axis corresponds to increasing $a_1-a_2$ and the $y$ axis increasing $d_2-d_1$. 

The two regions share only a codimension 2 boundary that is contained on the surface $(a_1-a_2)=0,(d_1 - d_2)=0$. Both regions $\cR_1,\cR_2$ come equipped with an orientation induced by the bulk geometry,  which in this case is the same for both regions. Each of the two regions is clearly a positive geometry, with canonical forms
\begin{equation}
    \begin{split}
    \Omega(\mathcal{R}_1)=- \frac{1}{a_1d_2(a_1-a_2)(d_1-d_2)}\; ,  \\
    \Omega(\mathcal{R}_2)=- \frac{1}{a_2d_1(a_2-a_1)(d_2-d_1)}\; .
    \end{split}
\end{equation} 
The sum of these correctly reproduces the corresponding residue of the amplitude~\eqref{laddercut}. Since  the two regions share a lower codimension boundary we have to see what happens on this boundary to decide if the union is or isn't a positive geometry.  We can consider for example the $(d_2-d_1)=0$ boundary by sending $d_2 \to d_1$. This corresponds to projecting onto the $x$ axis of~\eqref{pic} and thus looks as

\begin{equation}
    \begin{tikzpicture}
        \node at (0,0) {$\bullet$};
         \draw[-latex] (0,0) -- (4,0);                
         \draw[-latex] (0,0) -- (-4,0);
         \node at (2.5,-.5) {$ \cR'_1$};
         \node at (-2.5,-.5) {$ \cR'_2$};
    \end{tikzpicture}
\end{equation}
We again get two regions
\begin{align}
    \mathcal{R'}_1:=\{a_1,a_2,d_1 \ | \ d_1 >0 \land a_1>a_2>0 \} \; ,\nonumber \\
    \mathcal{R'}_2:=\{a_1,a_2,d_1 \ | \ d_1 >0 \land a_2>a_1>0 \}\; ,
\end{align}
but since we approach the boundary from two different directions, the two induced orientations are opposite (see appendix \ref{OrientationAppendix} for a detailed explanation). The region $\mathcal{R'}_1$ and $\mathcal{R'}_2$ share a codimension 1 boundary $(a_1-a_2)=0$ where the orientation changes sign. We call this an {\em internal boundary}. 

 We see that this boundary  is in fact not oriented (rather it flips orientation on the internal boundary $a_1=a_2$). 
 Part of the definition of positive geometry in~\cite{Arkani-Hamed:2017tmz} is that it is oriented and, by the recursive nature of the definition, so are all boundaries etc. We conclude that a generalisation  of the concept of positive geometry is needed to accommodate the loop amplituhedron.

Since both  $\mathcal{R'}_1$ and  $\mathcal{R'}_2$ by themselves  {\em are} positive geometries  on the other hand, their respective residues at  $(a_2-a_1)=0$ are equal to the canonical form of this. Therefore the residue on the internal boundary of $\Omega(\mathcal{R'}_1)+\Omega(\mathcal{R'}_2)$ will be equal to twice the canonical form of a positive geometry. Note that if $\cR_1' $ and $\cR_2'$ instead had the same orientation as each other 
then $(a_2-a_1)=0$ would be a spurious boundary and the resulting residue would vanish.

We have thus seen that the loop amplitude has non-unit maximal residues and the geometrical origin of this is that the amplituhedron contains internal boundaries.
In the next section we will see how to formalize what we have observed in this simple example and generalize the definition of the canonical form to geometries with internal boundaries, which will then accommodate the loop amplituhedron.

\section{Generalized Positive Geometries and Weighted Positive Geometries}

\subsection{Positive geometry and its Canonical Form}

First we recall the definition of  a positive geometry, $X_{\geq 0}$, and its canonical form, $\Omega$, as defined  recursively in~\cite{Arkani-Hamed:2017tmz}, before we generalise this to accommodate  internal boundaries. 
A positive geometry $X_{\geq 0}$ is an oriented region in some space (an algebraic variety) whose boundary consists of a set of  positive geometries of one dimension less, with their orientation inherited from that of the neighbouring bulk geometry. Each component in this set lies inside some region of the form  $f(x_i)=0$ for some non-factorisable polynomial $f$.
Then the associated canonical form, $\Omega$,  of $X_{\geq 0}$ is related to the  associated canonical form,  $\omega$, of this boundary component via the residue form on $f(x_i)=0$
\begin{equation}\label{cfrecurs}
{df} \wedge \text{Res}_{f=0}\Omega=\lim_{f\rightarrow 0}  f \Omega = {df} \wedge \omega\; .
\end{equation}
Since the dimension of the boundary component $f=0$ is one dimension less than the dimension of $X$, this gives  a recursive definition which stops when we reach dimension 0 positive geometries, which are just single points. These points are then defined to have canonical form $\pm1$ according to their orientation inherited from the original geometry. In the appendix \ref{OrientationAppendix} we review how the induced orientation is defined and the convention we use for the signs.

Apart from the 0-dimensional oriented points,  the simplest examples of positive geometries are 1-dimensional. The most general 1d positive geometry is  a disjoint union of intervals of arbitrary orientation. The canonical form of the  interval from $a$ to $b$ is
\begin{align}
  [a,b] = 
  \raisebox{-0.5\height}{\begin{tikzpicture}
          \draw[very thick,-latex](0,0)node[label={[label distance=-.1cm]270:$a$}]{$\scriptscriptstyle \bullet$}--(1.1,0);
          \draw[very thick](1,0)--(2,0)node[label={[label distance=-.2cm]270:$b$}]{$\scriptscriptstyle \bullet$};
  \end{tikzpicture}}\qquad \qquad
  \Omega([a,b])=\frac{dx}{x-b}-\frac{dx}{x-a}\; .
\end{align}
The boundaries of this segment are given by $f(x)=x-b=0$ and $f(x)=x-a=0$, and we see from~\eqref{cfrecurs} that $b$ has canonical form 1 and $a$ has canonical form -1.
The canonical form of the disjoint union of such intervals is simply the sum of the canonical forms of the intervals. 

Notice that two intervals of the same orientation which share a common boundary point are equivalent to the larger interval
\begin{align}
  [a,b] \cup [b,c]= 
  \raisebox{-0.5\height}{\begin{tikzpicture}
          \draw[very thick,-latex](0,0)node[label={[label distance=-.1cm]270:$a$}]{$\scriptscriptstyle \bullet$}--(1.1,0);
          \draw[very thick](1,0)--(2,0)node[label={[label distance=-.2cm]270:$b$}]{$\scriptscriptstyle \bullet$};
          \draw[very thick,-latex](2,0)--(3.1,0);
          \draw[very thick](3,0)--(4,0)node[label={[label distance=-.2cm]270:$c$}]{$\scriptscriptstyle \bullet$};
  \end{tikzpicture}}\qquad = \qquad
\raisebox{-0.5\height}{\begin{tikzpicture}
          \draw[very thick,-latex](0,0)node[label={[label distance=-.1cm]270:$a$}]{$\scriptscriptstyle \bullet$}--(2.1,0);
          \draw[very thick](2,0)--(4,0)node[label={[label distance=-.2cm]270:$c$}]{$\scriptscriptstyle \bullet$};
  \end{tikzpicture}}
\end{align}
with the shared boundary point absent. This  is reflected in the addition of the corresponding canonical forms
\begin{align}
    \Omega([a,b])+\Omega([b,c])=\Omega([a,c])\; .
\end{align}
In this case, the point $b$ is sometimes called a spurious boundary.

However, two intervals of different  orientations sharing  a common boundary point 
\begin{align}\label{ex1}
  [a,b] \cup [c,b]= 
  \raisebox{-0.5\height}{\begin{tikzpicture}
          \draw[very thick,-latex](0,0)node[label={[label distance=-.1cm]270:$a$}]{$\scriptscriptstyle \bullet$}--(1.1,0);
          \draw[very thick](1,0)--(2,0)node[label={[label distance=-.2cm]270:$b$}]{$\scriptscriptstyle \bullet$};
          \draw[very thick,-latex](4,0)node[label={[label distance=-.2cm]270:$c$}]{$\scriptscriptstyle \bullet$}--(2.9,0);
          \draw[very thick](3,0)--(2,0);
  \end{tikzpicture}}
\end{align}
does not constitute a positive geometry (for example, it is not oriented).
Nevertheless, it is natural to associate to this geometry the corresponding  canonical form 
\begin{align}\label{ex1form}
    \Omega([a,b])+\Omega([c,b])\ =\  2\frac{dx}{x-b}-\frac{dx}{x-a}-\frac{dx}{x-c}\; .
\end{align}
The point $b$ is then special, separating two regions of opposite orientation, and we refer to this as   an `internal boundary'. It  has residue twice that of each of the two individual boundaries there. This is exactly what  we observed occurring for the two loop amplitude in the previous section.

\subsection{Generalised Positive Geometry and its Canonical Form}

The discussion above motivates a generalisation of the concept of positive geometry to incorporate internal boundaries.
Internal boundaries separate two regions of opposite orientation. So, we define a generalised positive geometry as one whose internal and external boundaries are both generalised positive geometries.  Both external and internal boundaries must  lie inside a space defined by $f(x_i)=0$ for some non-factorisable polynomial $f$. A particular subspace $f(x_i)=0$ could contain both internal and external boundaries, each of which must be a (generalised) positive geometry with canonical form $\omega_{\text{int}}$ and $\omega_{\text{ext}}$ respectively.  Then we define the canonical form recursively as  
\begin{equation}\label{cfrecurs2}
\text{Res}_{f=0}\Omega=\omega_{\text{ext}}+2\omega_{\text{int}} \; ,
\end{equation}
or equivalently,
\begin{equation}
 \lim_{f\rightarrow 0}  f \Omega \ =\  {df} \wedge (\omega_{\text{ext}}+2\omega_{\text{int}})\; .
\end{equation}
We see an extra term compared to the original canonical form for positive geometries~\eqref{cfrecurs}, giving the factor of 2 associated with internal boundaries.
The starting point for the recursion is the same as before; the 0-dimensional geometries, oriented points,  with canonical form $\pm 1$ according to their orientation. Note that the orientation of the interior boundary is unambiguously inherited from that of the bulk just as for the exterior boundaries.

So consider the 1d example in~\eqref{ex1}. This has external boundaries (with negative  orientation) at $f_a(x)=x-a=0$ and $f_c(x)=x-c=0$ and an internal boundary with positive orientation at $f_b(x)=x-b=0$. One can see that the canonical form~\eqref{ex1form} satisfies the recursive relation~\eqref{cfrecurs2} at all three points.

Now consider a 2d example. 
\begin{align}
    \begin{tikzpicture}
    \node at (-3,0) {$R_1\quad=$};
    \fill[fill=gray!20] (0,2) -- (0,0.05)--(2,0.05)--(0,2);
    \fill[fill=gray!20] (-2,0) -- (2,0)--(0,-2)--(-2,0);
    \draw[-latex] (0,1) -- (0,0.05)--(1.1,0.05);
    \draw[-latex] (1,0.05) -- (2,0.05) -- (.9,1.1);
    \draw[-latex]  (1,1) -- (0,2)--(0,.9);
    \draw[-latex] (-1,0) -- (2,0)--(.9,-1.1);
    \draw[-latex] (1,-1) -- (0,-2) -- (-1.1,-.9);
    \draw[-latex]  (-1,-1) -- (-2,0)--(-.9,0);
    \draw[-latex,semithick,red] (.7,.7) arc[radius=.2, start angle=60, end angle=300]-- ++(20:1pt);
    \draw[-latex,semithick,red] (0,-1) arc[radius=.2, start angle=300, end angle=60]-- ++(-20:1pt);
    \end{tikzpicture}
\end{align}
Here the $x$ axis contains both an external boundary and an internal boundary. The canonical form can be obtained straightforwardly by simply adding together the canonical forms of the two triangles,%
\footnote{This key tessellation feature  of positive geometries and the canonical forms   is inherited by (and is indeed more powerful for) the generalised positive geometries as we will see.}
giving 
\begin{align}
\label{eq:2DExample}
    \Omega(R_1)=\frac{dx\,dy}{xy(x+y-1)}+\frac{2dx\,dy}{y(x+y+1)(x-y-1)}\; .
\end{align}
Now we can see how this satisfies the recursive definition~\eqref{cfrecurs2} along the $x$ axis. Taking the residue of \eqref{eq:2DExample} at $y=0$ gives
\begin{align}
    \text{Res}_{y=0}\, \Omega = dx \left( \frac{1}{x}{-}\frac{1}{x{+}1}\right){+}2dx \left( \frac{1}{x{-}1}{-}\frac{1}{x} \right)=\Omega([-1,0])+2\Omega([0,1])\; ,
    \end{align}
which is exactly as predicted by~\eqref{cfrecurs2} with $f = y$ since the external boundary on the $x$ axis is the interval $[-1,0]$ and the internal boundary is $[0,1]$.

Note that internal boundaries give a contribution to the canonical form of twice that of a standard external boundary. One might think therefore that a leading singularity of any such a generalised positive geometry must  be $0,\pm 1$ (as for a positive geometry)  or $\pm 2$ if there is an internal boundary present. However, there can be internal boundaries inside internal boundaries, leading to higher maximal residues.
A very simple example of this is the region consisting of the entire plane, but with the four quadrants having alternating  orientations
 \begin{align}
    \begin{tikzpicture}
    \label{maxRes4}
    \node at (-3,0) {$R_2\quad =$};
     \fill[fill=gray!20] (0.05,0.05)--(2,.05)--(2,2)--(.05,2)--(0.05,0.05);
    \fill[fill=gray!20] (0,0.05)--(0,2)--(-2,2)--(-2,.05)--(0,0.05);
    \fill[fill=gray!20] (0.05,0)--(2,0)--(2,-2)--(.05,-2)--(0.05,0);
    \fill[fill=gray!20] (0,0)--(-2,0)--(-2,-2)--(0,-2)--(0,0);
    \draw[-latex] (0.05,-2) -- (0.05,-.9);
    \draw[-latex] (0.05,-1) -- (0.05,0) -- (1.1,0);
    \draw  (1,0) -- (2,0);
    \draw[-latex] (0.05,2) -- (0.05,.9);
    \draw[-latex] (0.05,1) -- (0.05,0.05) -- (1.1,0.05);
    \draw  (1,0.05) -- (2,0.05);
    \draw[-latex] (0,2) -- (0,.9);
    \draw[-latex] (0,1) -- (0,0.05) -- (-1.1,0.05);
    \draw  (-1,0.05) -- (-2,0.05);
\draw[-latex] (0,-2) -- (0,-.9);
    \draw[-latex] (0,-1) -- (0,0) -- (-1.1,0);
    \draw  (-1,0) -- (-2,0);
    \draw[-latex,semithick,red] (1,1) arc[radius=.2, start angle=60, end angle=300]-- ++(20:1pt);
    \draw[-latex,semithick,red] (-1,1) arc[radius=.2, start angle=120, end angle=-120]-- ++(-200:1pt);
    \draw[-latex,semithick,red] (-1,-1) arc[radius=.2, start angle=-120, end angle=120]-- ++(200:1pt);
    \draw[-latex,semithick,red] (1,-1) arc[radius=.2, start angle=300, end angle=60]-- ++(-20:1pt);
    \end{tikzpicture}
\end{align}
Here, each quadrant has exactly the same canonical form $dx\,dy/(xy)$ and so the geometry has non-zero canonical form
\begin{align}
    \Omega(R_2)=\frac{4dx\,dy}{xy}\; .
\end{align}
Taking the residue on the (internal) boundary  $x=0$ gives $\lim_{x\rightarrow 0} (x \Omega) =2dx \wedge \left(\frac{2dy}{y}\right)=2dx \wedge \omega_{int}$ with $\omega_{int}=\frac{2dy}{y}$, satisfying~\eqref{cfrecurs2}. Here the internal  boundary in $x=0$ is $(-\infty,0)\cup (\infty,0)$ and this itself has an internal boundary at $y=0$. Thus it has canonical form  $\frac{2dy}{y}$ (as we  get directly  from~\eqref{ex1form} by taking $b\rightarrow 0$, $a,c \rightarrow \infty$ and $x\rightarrow y$). So the leading singularity at $x=y=0$ is 4.

It is also possible to get a leading singularity 3. Simply take three of the four quadrants from the previous example
 \begin{align}
    \begin{tikzpicture}
    \label{maxRes5}
    \node at (-3,0) {$R_3\quad =$};
    \fill[fill=gray!20] (0.05,0.05)--(2,.05)--(2,2)--(.05,2)--(0.05,0.05);
    \fill[fill=gray!20] (0,0.05)--(0,2)--(-2,2)--(-2,.05)--(0,0.05);
    \fill[fill=gray!20] (0.05,0)--(2,0)--(2,-2)--(.05,-2)--(0.05,0);
    \draw[-latex] (0.05,-2) -- (0.05,-.9);
    \draw[-latex] (0.05,-1) -- (0.05,0) -- (1.1,0);
    \draw  (1,0) -- (2,0);
    \draw[-latex] (0.05,2) -- (0.05,.9);
    \draw[-latex] (0.05,1) -- (0.05,0.05) -- (1.1,0.05);
    \draw  (1,0.05) -- (2,0.05);
    \draw[-latex] (0,2) -- (0,.9);
    \draw[-latex] (0,1) -- (0,0.05) -- (-1.1,0.05);
    \draw  (-1,0.05) -- (-2,0.05);
    \draw[-latex,semithick,red] (1,1) arc[radius=.2, start angle=60, end angle=300]-- ++(20:1pt);
    \draw[-latex,semithick,red] (-1,1) arc[radius=.2, start angle=120, end angle=-120]-- ++(-200:1pt);
    \draw[-latex,semithick,red] (1,-1) arc[radius=.2, start angle=300, end angle=60]-- ++(-20:1pt);
    \end{tikzpicture}
\end{align}
This geometry has  canonical form
\begin{align}\label{cf4Quadrants}
    \Omega(R_2)=\frac{3dx\,dy}{xy}\; .
\end{align}
Here taking the residue on the  boundary  $x=0$ gives $\lim_{x\rightarrow 0} (x \Omega) =dx \wedge \left(\frac{dy}{y}\right)+ 2dx \wedge \left(\frac{dy}{y}\right)$
in agreement with~\eqref{cfrecurs2}. This time $x=0$ contains the  external  boundary  $(-\infty,0)$ as well as the internal boundary $(0,\infty)$ both of which have canonical forms $\frac{dy}{y}$.  Therefore, the leading singularity at $x=0$ then $y=0$ is 3.

Note that in~\cite{Dian:2021idl}, in the context of the squared amplituhedron, another   generalisation of positive geometry was considered, and the associated canonical form called the globally  oriented canonical form was defined.
These geometries are in fact also examples of GPGs and we discuss this relation in more detail in  appendix~\ref{oriented}.

\subsection{Weighted Positive Geometry and its Canonical Form}

Although the definition of the canonical form~\eqref{cfrecurs2} is extremely compact, it has the downside of  treating external and internal boundaries on a different footing. On the right hand side we have a weighted sum of canonical forms, so it's tempting to rewrite this as the canonical form of a weighted sum of geometries. In this section we make this intuition precise by generalising what we mean by a geometrical region slightly, to include a weight taking  arbitrary integer values.
This  articulates the idea of having multiple coinciding geometries which we naturally have when two regions meet on an internal boundary.
This concept will allow us to give an explicit formula for maximal residues.

Firstly, recall that orientation on a space can be described by a top form where we are only really interested in the sign of the top form. So orientation is the equivalence class of real top forms modulo positive rescaling,
$O \in \Omega^d/\sim$ where $O \sim \lambda O$ for any $\lambda>0$.
Now we extend this to  define the weighted orientation as a pair  $(w,O):X \rightarrow \left(\mathbb{Z},\Omega^d(X)\right)/\sim$  where here the equivalence relation involves positive or negative rescaling with the negative case also flipping the weight $w$
\begin{align}\label{equiv}
    (w,O) \sim (\sign(\lambda)w,\lambda O')\ \qquad \lambda\neq0 \; .
\end{align}
Thus changing the orientation is equivalent to flipping the sign of $w$. 
In practise we can of course always choose coset representatives of~\eqref{equiv} such that $w>0$ and we will mostly assume this from now on.

Weighted geometries have a natural additive  structure. 
At any point $x \in X$ we define the  sum of two weighted geometries $(w_1,O_1)\oplus (w_2,O_2)$ as
\begin{align}\label{WGsum}
    (w_1,O_1) \oplus (w_2,O_2) =  (w_1+\text{sign}(\lambda) w_2, O_1)  \; ,
\end{align}
where $\lambda$ is such that $O_1=\lambda O_2$ \footnote{  Note that the $x$ dependence is suppressed in the equation and that the function $\text{sign}(\lambda(x))$ is negative if the two orientations $O_1(x)$ and $O_2(x)$ are opposite and positive if  they match.}.
Notice, that because of~\eqref{equiv} this sum is symmetric and the identity element $(0,O)$ is unique.

So now we define  a weighted geometry entirely by specifying its  weight function and its  orientation, rather than directly defining a region $X_\geq$. The  region $X_\geq$ can be reconstructed   by simply defining it as the set of points where $w\neq0$. 
Boundaries are then places where the weighted orientation is discontinuous and they divide regions inside which the weighted orientation is continuous (and therefore $w$ is constant).
 We will shortly  define a canonical form for weighted geometries, the existence and uniqueness of which will   define weighted positive geometries (WPG). For multivariate residues to be well defined, we insist that these boundaries  must be subsets of algebraic varieties -- so these regions  are semi-algebraic sets.

A key aspect of this construction is that both weights and orientation on the boundaries are uniquely induced from those in the bulk. This happens as follows. 
Any boundary component, $\mathcal{C}$ can be defined through a polynomial $p(x)=0$. Then one side of the boundary is $p(x)>0$, with weight and orientation  $(w^+,O^+)$, whereas the other side is   $p(x)<0$ with  $(w^-,O^-)$. 
Now the region $p(x)>0$ naturally induces the orientation $O^+|_\mathcal{C}$ on the boundary of the region $p>0$ in the standard way (see appendix  \ref{OrientationAppendix}), so  $O^+= dp \wedge O^+|_\mathcal{C}$.  The region $p(x)<0$ on the other hand naturally induces the orientation $ O^-|_\mathcal{C}$ on the boundary,  where $O^-= -dp \wedge O^-|_\mathcal{C}$\footnote{The minus sign arises from the fact that the normal vector pointing inward the region $p<0$ is $-\partial_p$. }.

Given a codimension-1 variety $\mathcal{C}\in X$, then we define a projection operator $\Pi_\mathcal{C}$ that maps weighted orientations $(w,O)$ on $X$ to weighted orientations on $\mathcal{C}$ as
\begin{align}\label{projdef}
\Pi_\mathcal{C}(w,O)=(w^+|_\mathcal{C},O^+|_{\mathcal{C}}) \oplus(w^-|_\mathcal{C},O^-|_{\mathcal{C}}) \; .
\end{align}
Choosing representatives such that $w>0$, we can give the following two dimensional  illustration of the induced weights  and orientations (denoted with arrows):
\begin{align}
    \begin{tikzpicture}
    \draw[-latex,thin] (-3,0) -- (1.1-2,0);
    \draw[thin]  (1-2,0) -- (5-4,0);
    \draw[-latex,thin] (-3,0.05) -- (1.1-2,0.05);
    \draw[thin]  (1-2,0.05) -- (5-4,0.05);
    \draw[-latex,thin] (-3+5,0) -- (1.1+3,0);
    \draw[thin]  (1+3,0) -- (5+1,0);
    \draw[-latex,thin] (-3+5,0.05) -- (1.1+3,0.05);
    \draw[thin]  (1+3,0.05) -- (5+1,0.05);
    \draw[-latex,thin] (5+6,0) -- (.9+8,0);
    \draw[thin]  (1+8,0) -- (-3+10,0);
    \draw[-latex,thin] (5+6,0.05) -- (.9+8,0.05);
    \draw[thin]  (1+8,0.05) -- (-3+10,0.05);
    \draw[-latex,semithick,red] (-1,1) arc[radius=.2, start angle=60, end angle=300]-- ++(20:1pt);
    \node at (0,1) {$w$};
    \draw[-latex,semithick,red] (-1,-1) arc[radius=.2, start angle=300, end angle=60]-- ++(-20:1pt);
    \node at (0,-1) {$w'$};
    \draw[-latex,semithick,red] (4,1) arc[radius=.2, start angle=60, end angle=300]-- ++(20:1pt);
    \node at (5,1) {$w$};
    \draw[-latex,semithick,red] (4,-1) arc[radius=.2, start angle=240, end angle=490]-- ++(200:1pt);
    \node at (5,-1) {$w'$};
    \draw[-latex,semithick,red] (9,1) arc[radius=.2, start angle=60, end angle=300]-- ++(20:1pt);
    \node at (10,1) {$w$};
    \draw[-latex,semithick,red] (9,-1) arc[radius=.2, start angle=240, end angle=490]-- ++(200:1pt);
    \node at (10,-1) {$w'$};
    \node at (-1,0.25) {$\scriptscriptstyle w+w'$};
      \node at (4,0.25) {$\scriptscriptstyle w-w'$};
      \node at (9,0.25) {$\scriptscriptstyle w'-w$};
      \node at (4,-1.5) {$(w>w')$};
      \node at (9,-1.5) {$(w<w')$};
      \end{tikzpicture}
\end{align}
Note that $w$ or $w'$ could have been zero in which case we have a conventional external boundary. This definition implies that, for $w(x)=1$ in $X_{>0}$ and 0 otherwise,  $w|_\mathcal{C}$ will be equal to $2$ on internal boundaries, to $1$ on external boundaries and $0$ otherwise.  In this formulation internal and external boundaries are not distinguished.
Furthermore note that if  $w^+= w^-$ and $\lambda<0$ then there are equivalent weighted orientations on both sides (meaning the induced orientations are opposite) and thus there is no genuine boundary there (it is a spurious boundary). 

An important observation now is that the projection $\Pi$ is a linear operator
\begin{align}\label{projSum}
    \Pi_\mathcal{C}\Big( (w_1,O_1)\oplus  (w_2,O_2)\Big)=  \Pi_\mathcal{C}(w_1,O_1)\oplus  \Pi_\mathcal{C}(w_2,O_2) \; ,
\end{align}
which can be easily proven from the definitions.

Now we can define a weighted positive geometry as a weighted geometry possessing a canonical form. The definition of the canonical form of a weighted geometry is defined recursively such that the residue of the canonical form on $\mathcal{C}$ is the canonical form of the geometry projected on $\mathcal{C}$:
\begin{equation}\label{cfrecursNew}
\text{Res}_{\mathcal{C}} \Omega(w,O)  =  \Omega(\Pi_{\mathcal{C}}(w,O))\; .
\end{equation}
The recursion starts by defining the canonical form of a zero dimensional weighted geometry (for which $O$ is a 0 form, simply a scalar) as the product of  $w$ with the sign of $O$
\begin{align}\label{cf0}
    \Omega(w,O)= w \sign( O)  \qquad \qquad \text{(In zero dimensions)}\ .
\end{align}

In zero dimensions therefore the canonical form is a linear operator
\begin{align}
\Omega\Big((w_1,O_1)\oplus(w_2,O_2)\Big)& =\Omega\Big(w_1+ \text{sign}(\lambda) w_2 ,O_1\Big) =(w_1+ \text{sign}(\lambda) w_2) \times \text{sign}(O_1) \notag\\
&= w_1 \text{sign}(O_1) +w_2 \text{sign}(O_2)=
\Omega(w_1,O_1)+\Omega(w_2,O_2)
\end{align}
where recall $O_1=\lambda O_2$. It follows by induction  from the recursive definition~\eqref{cfrecursNew} and linearity of the projection operator $\Pi$~\eqref{projSum} that this linearity property of $\Omega$ then holds for spaces of arbitrary dimension:
\begin{align}\label{lin}
    \Omega
    \Big((w_1,O_1)\oplus (w_2,O_2)\Big)=\Omega
    (w_1,O_1)+ \Omega
    (w_2,O_2)\ .
\end{align}
Remarkably  we have the feature that we can freely sum arbitrary (even overlapping) WPGs!
It also follows directly from this that 
\begin{align}
    \Omega
    \Big((\lambda w,O)\Big)=\lambda \Omega
      (w,O)\ .
\end{align}

Now given a sequence of boundaries $\{\mathcal{C}_1,\cdots,\mathcal{C}_n\}$, we can follow $n$ steps of the recursion~\eqref{cfrecursNew} and write the multi-residue of a canonical form as the canonical form of the multiply induced boundary
\begin{equation}\label{cfrecurs4}
\text{Res}_{\mathcal{C}_1,\cdots,\mathcal{C}_n} \Omega(w,O)  =  \Omega(\Pi_{\mathcal{C}_1,\cdots,\mathcal{C}_n}(w,O))\; ,
\end{equation}
Then taking $n=d$, the dimension of $X$, we obtain an expression for the maximal residues in terms of the canonical form at a point~\eqref{cf0}
\begin{equation}\label{maximalresidueNew}
\text{Res}_{\mathcal{C}_1,\cdots,\mathcal{C}_d} \Omega(w,O)  =  w_{\mathcal{C}_1,\cdots,\mathcal{C}_d} \times  \sign\left( O_{\mathcal{C}_1,\cdots,\mathcal{C}_d}\right) \; ,
\end{equation}
where $( w_{\mathcal{C}_1,\cdots,\mathcal{C}_d},   O_{\mathcal{C}_1,\cdots,\mathcal{C}_d})=\Pi_{\mathcal{C}_1,\cdots,\mathcal{C}_d} (w,O)$.  This last equation can also be used as a direct, non-recursive, definition of the canonical form by giving all its maximal residues (the canonical form is completely determined by its maximal residues).

Note that generalised positive geometries, defined in previous subsections, should simply be WPGs for which 
the weight function (in the bulk) is $w=\pm1,0$ everywhere. Similarly positive geometries are WPGs for which the weight function $w=\pm1,0$ everywhere (so they are also GPGs) but also the induced weight function on all nested boundary components is also always $\pm 1,0$.

To check this we need to show that the canonical form for the GPGs defined by~\eqref{cfrecurs2} and that for the WPGs~\eqref{cfrecursNew} are equivalent. By equivalent we mean that  given any GPG $X_{\geq0}$ with orientation $O$ we associate a weighted geometry with orientation $O$ and weight $w$ such that $w(x)=1$ for all $x\in X_{\geq0}$ and zero otherwise,  and then 
\begin{align}
    \Omega(X_{\geq0})=\Omega(w,O) \; .
\end{align}
Now  notice that the projection of $(w,O)$ onto $\mathcal{C}$ (described in~\eqref{projdef} and above)  will have induced weight 1 or 2, depending on whether it is an external or internal boundary.  So we can write $\Pi_\mathcal{C}(w,O)=(w_\text{ext},O_\text{ext})\oplus(w_\text{int},O_\text{int}) $ where $w_\text{ext}=1$  on external boundaries and zero elsewhere whereas $w_\text{int}=2$ on internal boundaries and zero elsewhere.   Then, it follows that if we apply~\eqref{cfrecursNew} we get
\begin{align}
\text{Res}_{\mathcal{C}} \Omega(w,O) & =  \Omega\Big(\Pi_\mathcal{C}(w,O)\Big)= \Omega((w_\text{ext},O_\text{ext})\oplus(w_\text{int},O_\text{int}))=\notag \\ &=\Omega(w_\text{ext},O_\text{ext})+2\Omega(\frac{w_\text{int}}{2},O_\text{int})\; .
\end{align}
Now since $w_\text{ext}$ and $\frac{1}{2}w_\text{int}$ are both functions respectively equal to $1$ on internal and external boundaries and equal to 0 otherwise, they represent with their orientations the  external and internal boundaries as GPGs. This is then precisely the original defining equation of the canonical form~\eqref{cfrecurs2}. Since we showed that the recursion~\eqref{cfrecursNew} and~\eqref{cfrecurs2} have the same form it follows that the two definitions of the canonical form give the same result.

Finally, let us illustrate with a slightly more involved example, returning to the case considered in~\eqref{maxRes5} from this new perspective 
 \begin{align}
    \begin{tikzpicture}
    \label{maxRes522}
    \node at (-3,0) {$R_3\quad =$};
    \fill[fill=gray!20] (0.05,0.05)--(2,.05)--(2,2)--(.05,2)--(0.05,0.05);
    \fill[fill=gray!20] (0,0.05)--(0,2)--(-2,2)--(-2,.05)--(0,0.05);
    \fill[fill=gray!20] (0.05,0)--(2,0)--(2,-2)--(.05,-2)--(0.05,0);
    \draw[-latex] (0.05,-2) -- (0.05,-.9);
    \draw[-latex] (0.05,-1) -- (0.05,0) -- (1.1,0);
    \draw  (1,0) -- (2,0);
    \draw[-latex] (0.05,2) -- (0.05,.9);
    \draw[-latex] (0.05,1) -- (0.05,0.05) -- (1.1,0.05);
    \draw  (1,0.05) -- (2,0.05);
    \draw[-latex] (0,2) -- (0,.9);
    \draw[-latex] (0,1) -- (0,0.05) -- (-1.1,0.05);
    \draw  (-1,0.05) -- (-2,0.05);
    \draw[-latex,semithick,red] (1,1) arc[radius=.2, start angle=60, end angle=300]-- ++(20:1pt);
        \node at (1.3,.8) {1};
    \draw[-latex,semithick,red] (-1,1) arc[radius=.2, start angle=120, end angle=-120]-- ++(-200:1pt);
            \node at (-1.3,.8) {1};
    \draw[-latex,semithick,red] (1,-1) arc[radius=.2, start angle=300, end angle=60]-- ++(-20:1pt);
    \node at (1.3,-.8) {1};
     \node at (-1.3,-.8) {0};
     \node at (.2,1) {$\scriptscriptstyle 2$};
    \node at (.2,-1) {$\scriptscriptstyle 1$};
     \node at (1,.2) {$\scriptscriptstyle 2$};
    \node at (-1,.2) {$\scriptscriptstyle 1$};
    \end{tikzpicture}
\end{align}
Here we see the induced weights 2,1 on the codimension 1 boundaries $x=0,y=0$. Considering these boundaries themselves they then induce the weight $2+1=3$ at the origin with positive orientation on the $y$ axis, negative on the $x$ axis. This is in line with~\eqref{maximalresidueNew} and the corresponding maximal residue $\text{Res}_{y=0,x=0}\Omega_{R_3}= - \text{Res}_{x=0,y=0}\Omega_{R_3}=3$.

Note that in \cite{Arkani-Hamed:2017tmz} a generalisation of positive geometries was defined, the Grothendieck group of pseudo-positive geometries, consisting of the formal sum of positive geometries modded out by geometries with vanishing canonical form. The Grothendieck group of pseudo-positive geometries and the WPGs   are closely related but different.
The Grothendieck group of pseudo-positive geometries is presumably equivalent to the space of WPGs after modding out by elements with zero canonical form. This equivalence relation implies that
\begin{equation}
  \Omega((w_1,O_1))=\Omega((w_2,O_2)) \quad \Rightarrow \quad (w_1,O_1)\sim (w_2,O_2)  \;.
\end{equation}

\subsection{Uniqueness of the Canonical Form}

The uniqueness of the canonical form for GPG/WPGs is equivalent to the statement that the algebraic variety on which the positive geometry lives has geometric genus zero ie has no non-zero holomorphic volume forms, just as for PGs~\cite{Arkani-Hamed:2017tmz}.
 Since holomorphic forms have no poles, they could be added to any canonical form to obtain a new canonical form satisfying all the requirements and thus we would not have a unique canonical form.
 Conversely, under the assumption that  GPGs/WPGs have  no holomorphic forms,  we can proceed by induction assuming that the canonical form in $d-1$ dimensions is always unique. 
 Consider two canonical forms $\Omega_1,\Omega_2$ for the same GPG or WPG.
 By definition, both forms have poles only on the boundary components. The residue on a boundary component is the canonical form of the boundary component, by the recursive definition of the canonical form. But since by induction we assumed that the canonical form in $d-1$ dimensions is unique then we conclude that for any residue $\text{Res}(\Omega_1-\Omega_2)=0$, and so $\Omega_1-\Omega_2$ has no poles and is thus a holomorphic form and so must vanish. We conclude that $\Omega_1=\Omega_2$ and so the canonical form is unique.

\subsection{Tilings}
\label{Triangulations}

A fundamental property of the canonical form is that given a positive geometry $X_{\geq 0}$ and a set of positive geometries $X^{(i)}_{\geq 0}$ tiling $X_{\geq 0}$,\footnote{By a tiling we mean the $X^{(i)}_{\geq 0}$ cover $X_{\geq 0}$ with non-overlapping regions. We will often  also  call such a tiling a triangulation.} then the canonical form of $X_{\geq 0}$ is the sum of that of the tiles
\begin{align}
\label{triangulationCF}
\Omega(X_{\geq 0})=\sum_i\Omega(X^{(i)}_{\geq 0})\; .
\end{align}
However it can happen that  a non positive geometry can be triangulated by positive geometries - so the space of positive geometries is not closed under the union.  This is because even if the orientation of the $X^{(i)}_{\geq 0}$ tiling $X_{\geq 0}$ matches on codimension 1 boundaries this does not imply that they will necessarily match on the boundaries of boundaries etc. This can then  give rise to internal boundaries. As examples consider the following two geometries:
	 \begin{align}
	 \label{ex2}
	 \begin{tikzpicture}[x=1.5cm,y=1.5cm,decoration={markings, 
	mark= at position 0.5 with {\pgftransformscale{1.5}\arrow{latex}}}
] 
	 \filldraw[fill=gray!20] (0,0) node[above left]{\footnotesize 3}--(-1,0) node[above left]{\footnotesize 2}--(0,-1)node[below left]{\footnotesize 1}--(0,0);
	 \draw[postaction={decorate}] (0,0)--(-1,0);
	 \draw[postaction={decorate}] (-1,0)--(0,-1);
	 \draw[postaction={decorate}] (0,-1)--(0,0);
	  \filldraw[fill=gray!20] (0,0)--(1,0)node[below right]{\footnotesize 5}--(0,1)node[above right]{\footnotesize 4}--(0,0);
	 \draw[postaction={decorate}] (0,0)--(1,0);
	 \draw[postaction={decorate}] (1,0)--(0,1);
	 \draw[postaction={decorate}] (0,1)--(0,0);
	 \draw[->,>=latex,semithick,red] (.5,.3) arc[radius=.2, start angle=0, end angle=300]-- ++(20:1pt);
	 	 \draw[->,>=latex,semithick,red] (-.1,-.3) arc[radius=.2, start angle=0, end angle=300]-- ++(20:1pt);
	 \end{tikzpicture}
	 \qquad \qquad &
	 	 \begin{tikzpicture}[x=1.5cm,y=1.5cm,decoration={markings, 
	mark= at position 0.5 with {\pgftransformscale{1.5}\arrow{latex}}}
] 
	 \filldraw[fill=gray!20] (1,.5) node[below]{\footnotesize 1}--(2.5,.5) node[below]{\footnotesize 2}--(2.5,2.5) node[above]{\footnotesize 3}--(0.5,2.5) node[above]{\footnotesize 4}--(0.5,1) node[left]{\footnotesize 5}--(1,1) node[ below left]{\footnotesize 6}--(1,1.5) node[above]{\footnotesize 7}--(1.5,1.5) node[above]{\footnotesize 8}--(1.5,1) node[right]{\footnotesize 9}--(1,1)--(1,0.5);
	 	 \draw[postaction={decorate}] (1,0.5)--(2.5,0.5);
	 	 \draw[postaction={decorate}] (2.5,0.5)--(2.5,2.5);
	 	 \draw[postaction={decorate}] (2.5,2.5)--(0.5,2.5);
	 	 \draw[postaction={decorate}] (0.5,2.5)--(0.5,1);
	 	 \draw[postaction={decorate}] (0.5,1)--(1,1);
	 	 \draw[postaction={decorate}] (1,1)--(1,1.5);
	 	 \draw[postaction={decorate}] (1,1.5)--(1.5,1.5);
	 	 \draw[postaction={decorate}] (1.5,1.5)--(1.5,1);
	 	 \draw[postaction={decorate}] (1.5,1)--(1,1);
	 	 \draw[postaction={decorate}] (1,1)--(1,0.5);
	 	 \draw[->,>=latex,semithick,red] (2,2) arc[radius=.2, start angle=0, end angle=300]-- ++(20:1pt);
	 \end{tikzpicture}
	 \end{align}
	 Both can be triangulated by positive geometries.
	 The first can be triangulated as the oriented union of two triangles while the second as the oriented union of 4 rectangles with matching orientation.
	 Both examples are not positive geometries themselves however. This can be seen graphically observing the orientation of the boundary, the edge $14$ in the first example and $59$ in the second look like~\eqref{ex1} and have an internal boundary.

    Both of these examples {\em are} generalised positive geometries however.
And we claim more generally 
that if $X_{\geq 0}$ is triangulated by a set of generalized positive geometries $X^{(i)}_{\geq 0}$ tiling $X_{\geq 0}$  then $X_{\geq 0}$ is a generalized positive geometry and its canonical form is given by~\eqref{triangulationCF}.  Thus the space of GPGs is closed under the disjoint-union. This is essentially trivial from linearity of the WPG canonical form and the definition of a GPG as a WPG with weight $0,\pm1$ everywhere. 

A beautiful consequence of the WPG formalism is that it also yields a simple proof of the tiling property of positive geometries~\eqref{triangulationCF}. Indeed this follows trivially from the fact that the canonical form is a linear operator for weighted positive geometries~\eqref{lin}.  Translating~\eqref{triangulationCF} into WPG language, on the right hand side we have the canonical form of a region which is equivalently  a weight function $(w,O)$ with $w=1$ in the region and 0 outside. This region has a tesellation with tiles $(w_i,O_i)$ with 
\begin{align}\label{eq:hyp}
(w,O)=\bigoplus_i (w_i,O_i) \; . 
\end{align}
Now linearity of WPGs give 
\begin{align}
\label{triangulationWG}
\Omega(w,O)=\sum_i\Omega(w_i,O_i)\; ,
\end{align}
which proves~\eqref{triangulationCF}.

\subsection{Speculations on an explicit characterisation of GPGs / WPGs}

In the previous subsection we gave an implicit (recursive)  definition of generalised positive geometry and weighted positive geometries. Here we consider whether it is possible to give more explicit characterisations,
so we can know in advance if a particular region is a GPG/WPG or not.
The fact that GPGs/WPGs are closed under union (or sum for WPGs) as discussed in the previous subsection already suggests they ought to be more amenable to a direct characterisation than PGs. For example any characterisation of PGs would have to  exclude the two examples in~\eqref{ex2}.

We  first note that  if we  restrict ourselves to a specific class of geometry which we call {\em multi-linear geometries} then the characterisation is very simple. Multi-linear geometries are geometries defined by {\em multi-linear inequalities} in some coordinates. Note that although this may be  a big restriction of the full space of positive geometries it nevertheless provides a very wide class of cases. Crucially 
it is straightforward to see that the amplituhedron is a multi-linear geometry. The defining inequalities of the amplituhedron are given in terms of either minors of a $C$ matrix, or alternatively determinants of the form $\langle Y L_i ... \rangle$. Thus by choosing components of either the $C$ matrix, and/or the $Y, L_i$ as coordinates, the resulting inequalities will be multi-linear in those coordinates (simply because the determinant is a multi-linear function of its components).

Note here that it is important not to confuse multi-linear geometries with linear geometries. Many of the toy examples one considers are linear geometries where defining inequalities can be given which are linear in all variables. These then have straight edges, flat planes etc. Multi-linear geometries can however be curvey.
For example in 2d, boundaries of multi-linear geometries have the form $a x y +b x+cy +d=0$ (in some coordinates) which correspond to hyperbolas as well as straight lines. On the other hand circles or ellipses would include the non multi-linear terms $ x^2,y^2$ and are not multi-linear. They can however be boundaries of positive geometries~\cite{Arkani-Hamed:2017tmz}.
We will shortly return to this point. 

We first claim that  {\em any} multi-linear geometry is a (generalised) positive geometry. We can show this by explicitly and uniquely computing the canonical form for multi-linear geometries. Given a multi-linear geometry, first use cylindrical decomposition (see also~\cite{Eden:2017fow}) which recasts any region as a disjoint union of regions $\mathcal{R}_i$ of the form 
\begin{align}
  \mathcal{R}_{i}:= \{x_1,\cdots,x_{d}\} \quad \text{st} \qquad \begin{cases} \hfil  a_1<x_1<b_1  \\  \hfil  a_2(x_1)<x_2<b_2(x_1)\\
  \hfil   \cdots \\
  a_{d}(x_1,\cdots,x_{d-1})<x_{d}<b_{d}(x_1,\cdots,x_{d-1})
  \end{cases}
  \label{easyregion}
\end{align}
for some functions $a_j,b_j$.
Now  changing  variables to:
 \begin{align}
 x_j'=-\frac{x_j-a_j}{x_j-b_j}\,,
 \label{map}
 \end{align}
 then $\mathcal{R}_i$ becomes 
 \begin{align}
  \mathcal{R}_{i}:= \{x'_1,\cdots,x'_{d}\} \quad \text{st} 
 \qquad x'_j>0 \qquad \text{for all} \ j\;.    
 \end{align}
In the new coordinates $\mathcal{R}_i$ is thus a simplex-like  positive geometry with canonical form 
 \begin{align}\label{simpleCF}
     \Omega(\mathcal{R}_i)= \prod_{j=1}^{d}\frac{d x_j'}{x_j'}\; .
 \end{align}
 But as discussed in~\cite{Arkani-Hamed:2017tmz} under a {\em rational} map 
the canonical forms map to each other. So as long as the change of variables~\eqref{map} is rational then we have that  in the original coordinates  
 \begin{align}\label{replace}
       \Omega(\mathcal{R}_i)=\prod_{j=1}^{d}
       \left(\frac{1}{x_j{-}a_j}{-} \frac{1}{x_j{-}b_j}\right)
    d x_j
 \end{align}
 and the canonical form of the full region can then be obtained by summing the contributions from all the $\mathcal{R}_i$. 
 We see how multi-linearity is crucial here. The inequalities in~\eqref{easyregion} must arise from the defining inequalities of our region which are multi-linear. This ensures that the resulting functions $a_i$ and $b_i$ will  be rational functions and thus the change of variables~\eqref{map} is rational. 

\begin{figure}[h!]
\begin{center}
\begin{tikzpicture}
\begin{axis}[height=5cm,set layers,axis equal,
    xlabel = {$R_1$},
    ylabel = {},
    xmin=0, xmax=8,
    ymin=0, ymax=8,minor tick num=1,minor tick style={draw=none},
    grid=both,restrict y to domain=0:8]
 
\addplot [name path = A,
    domain = 0:8,
    samples = 1000] {7/x} 
    node [very near end, right] {$ $};
 
\addplot [name path = B,
    domain = 0:8] {-x+8} 
    node [pos=1, right ] {$ $};
 
\addplot [gray!50] fill between [of = A and B, soft clip={domain= 1:7}];
\end{axis}
\end{tikzpicture}
\qquad
 \begin{tikzpicture}
\begin{axis}[height=5cm,set layers,axis equal,
    xlabel = {$R_2$},
    ylabel = {},
    xmin=-.9, xmax=.9,
    ymin=-1, ymax=1,minor tick num=4,minor tick style={draw=none},
    grid=both]
 
\addplot [name path = A,
    domain = -1:1,
    samples = 1000] {sqrt(1-x^2)} 
    node [very near end, right] {$ $};

\draw (axis cs:0,0) circle [black, radius=1];
 
\addplot [name path = C,
    domain = -4:4] {1/10} 
    node [pos=1, right ] {$ $};
 
\addplot [gray!50] fill between [of = A and C, soft clip={domain= -0.9949:0.9949}];
\end{axis}
\end{tikzpicture}
\end{center}
\caption{Two examples of positive geometries obtained by sandwiching a conic and a straight line. The first, involving  a hyperbola, is a multi-linear geometry whereas the second is not multi-linear, but is still a positive geometry.}
\label{plots}
\end{figure}
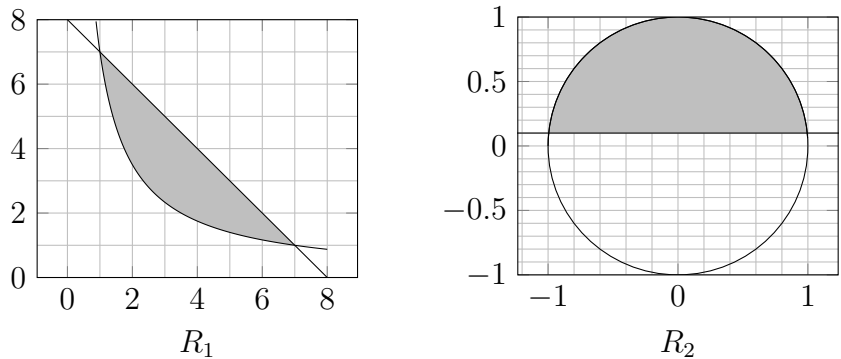

Let us illustrate some of these points now with a couple of examples shown in figure~\ref{plots}. 

Firstly we have a region $R_1$ sandwiched between a hyperbola and a line. It is defined by the inequalities $x>0,xy>7,x+y<8$. Cylindrical decomposition rewrites this as a single region written in the form of~\eqref{easyregion} as $1<x<7 \quad \frac7x<y<8-x$. Then the simple replacement rule of ~\eqref{replace} yields the canonical form
\begin{align}
\Omega(R_1) = \left(\frac{dx}{x-1}-\frac{dx}{x-7}\right)
   \left(\frac{dy}{y-\frac{7}{x}}-\frac{dy}{x+y-8}\right)=-\frac{6\,dx\,dy}{(x+y-8) (x y-7)}\; .
\end{align}

The second example $R_2$, found in~\cite{Arkani-Hamed:2017tmz},  is not a multi-linear geometry. This is the region between a circle  and a line and is defined by the inequalities $x^2+y^2<1, y>1/10$. Let us see what happens if we  attempt the same procedure to obtain its canonical form. 
Here cylindrical decomposition rewrites the region as
\begin{align}\label{nonrat}
    -\frac{3 \sqrt{11}}{10}<x<\frac{3 \sqrt{11}}{10},\quad
   \frac{1}{10}<y<\sqrt{1-x^2}\; .
\end{align}
But now we encounter a problem. We see that, due to the square root $\sqrt{1-x^2}$, the change of variables needed in~\eqref{map} will no longer be rational and the above procedure no longer works.

So we have seen that cylindrical decomposition gives the unique canonical form as long as all the resulting functions $a_i,b_i$ in~\eqref{easyregion} are  rational. This is clearly the case for multi-linear geometries,
but could also be the case for more general geometries. 
Further it may  be possible  to change coordinates so that only in the new coordinates the cylindrical decomposition map~\eqref{map} is rational. 
So in general we can characterise generalised positive geometries to be those for which there exist coordinates and an ordering of these coordinates such that cylindrical decomposition yields a map~\eqref{map} which is rational.

For example let us 
return to the region $R_2$ in figure~\ref{plots} for which the cylindrical decomposition method of obtaining the canonical form didn't work as it produces an irrational map~\eqref{map}.
Now the circle is the classic  example of a rational variety.  This is a variety that has a parametrisation $t_i$ in terms of which its embedding coordinates $x_i(t_j)$ are rational functions and there is a rational inverse map $t_i(x_j)$.
In this case it has a rational parametrisation given by
\begin{align}
    x(t)=\frac{2t}{1+t^2} \qquad y(t)=\frac{1-t^2}{1+t^2}\; .
\end{align}
with an inverse map from $\mathbb{R}$ onto the circle embedded in $\mathbb{R}^2$ which is also rational
\begin{align}
    t(x,y) = \frac x{1+y}\; .
\end{align}
Here the parameter $t$ has the  geometrical interpretation of  the projection of a point on the circle,   from the point $(0,-1)$ to the $x$ axis.
But this projection can clearly be extended to any point in $\mathbb{R}^2$ not just those points on the circle. So  consider the change of variables from $x$ to $t$, 
$(t,y) \rightarrow (x,y)=( t(1+y),y)$. In the new variables the region $R_2$ then has cylindrical decomposition
\begin{align}
    -\frac{3}{\sqrt{11}}<t<\frac{3}{\sqrt{11}},\ \ \  \frac{1}{10}<y<\frac{1-t^2}{t^2+1}\; ,
\end{align}
which is now rational. The replacement rule~\eqref{replace}, then gives the canonical form
\begin{align}\label{cfCircle}
 \Omega(R_2) = -\frac{6 \sqrt{11} dt\, dy}{(10 y-1) \left(t^2 (y+1)+y-1\right)}= -\frac{6 \sqrt{11} dx dy}{(10 y-1) \left(x^2+y^2-1\right)}\ ,
\end{align}
which is in precise  agreement with the canonical form for this geometry found in~\cite{Arkani-Hamed:2017tmz} (see figure 1) using the recursive definition of the canonical form.

In general, if a codimension 1 boundary of a region is a rational variety, then changing coordinates from $x_i$ to $t_i,x_n$ will rationalise the final step in the cylindrical decomposition involving that boundary. In other words cylindrical decomposition in those variables will give  the boundary in the form $x_n<x_n(t_i)$ which is a  rational function as we saw in the above example which gave  $y<y(t)=(1{-}t^2)/(1{+}t^2)$). 
This all  suggests there should be  a more intrinsic definition of a GPG/WPG in terms of rational varieties.

\section{All-in-one-point cut}

We now look at a particular boundary of the loop amplituhedron related to a set of cuts on the integrand of MHV amplitudes explored in~\cite{Arkani-Hamed:2018rsk,Langer:2019iuo}, referred to as the {\em deepest cut}.  This provides another example of an internal boundary as well as illustrating the other important point mentioned in the introduction, namely that the order of taking residues (or going to boundaries) can yield completely different results.

In  general the deepest cut places all internal propagators on-shell
\begin{equation}
    \label{eq:OnShellInternal}
    \left< (AB)_{\alpha}(AB)_{\beta}\right> = 0 \;\;\;\;\;\; \forall \;\; \alpha, \beta = 1, ..., l \; ,
\end{equation}
while leaving all external propagators $\left< (AB)_{\alpha}i i+1\right>$ generic.  Geometrically there are two possible final configurations which solve~\eqref{eq:OnShellInternal}: first, all loop lines passing through a single point $A$, or second, all loop lines lying on the same plane. In~\cite{Arkani-Hamed:2018rsk} the canonical form corresponding to these two  solutions was found at any loop order. We find that this form can not be reproduced from 
any  sequence of single residues (or any linear combination of such) acting on the amplitude and so some more complicated operation is presumably needed to reproduce it\footnote{We thank Nima Arkani-Hamed and Jaroslav Trnka for valuable discussions on this point}.
Furthermore there are many inequivalent ways of approaching this final all-in-one-point configuration via different sequences of single residues, as becomes especially apparent starting at four loops.  In this section we  systematically investigate all cuts ending in the all-in-one-point configuration.

We will begin by discussing the three loop all-in-one-point cut, computing its geometry and discussing the internal boundary that arises, before considering higher loops.
Although we will limit the discussion to the 4-point MHV amplituhedron geometry, the derivation of the geometry is completely independent of the tree level inequalities $\br{Yijkl}$ and $\br{ABij}$. The results obtained in section~\ref{HigherLoop} for the loop-loop inequalities of the all-in-one-point cut hold for any multiplicity and any NMHV degree by simply promoting the brackets $\br{AB_iB_jB_k}$ to $\br{YAB_iB_jB_k}$.

\subsection{Three-loop all-in-one-point cut}\label{ss:3PointDC}

The first case of an all-in-one-point cut is at two loops. Although we saw above that this contains a previously undetected internal boundary, since the  all-in-one-point cut {\em is} this boundary it doesn't affect anything and the corresponding residue is simply the canonical form of two lines satisfying 1 loop inequalities as predicted in~\cite{Arkani-Hamed:2018rsk}. We will return to this in section~\ref{ll2}. 

We thus turn to three loops.
The integrand of the three-loop MHV amplitude is given by \small
\begin{equation}
    \begin{split}
    \label{eq:3LoopMHV}
    \text{MHV(3)} &= \frac{\prod^3_{i=1} \br{A_i B_i \dd^2A_i} \br{A_i B_i \dd^2 B_i} \br{1234}^3}{\br{A_1B_114}\br{A_1B_112}\br{A_1B_134}\br{A_2B_212}\br{A_2B_223}\br{A_3B_334}\br{A_1B_1A_3B_3}\br{A_2B_2A_3B_3}} \times \\ 
    & \times \;\; \left[\frac{1}{2} \frac{\br{1234}}{\br{A_3B_312}\br{A_2B_234}} + \frac{\br{A_1B_123}}{\br{A_3B_323}\br{A_1B_1A_2B_2}}\right] \;\;\; + \;\;\;  \text{symmetry} \; .
    \end{split} 
\end{equation} \normalsize
Here the `+ symmetry' is a sum of 23 more terms:  the  $3!$ terms  generated by permutation symmetry over the loop variables (simultaneous permutation of  $A_i$ and $B_i$) together with the four terms from  cyclic symmetry of the external twistors, giving 24 terms in total in the sum. The  first term in square brackets is  the three-loop  ladder integrand, and after summing it generates 12 unique terms (all with coefficient 1), while the second term is the so-called 3-loop `tennis court' diagram generating 24 unique terms (again all will have coefficient 1). 

In order to achieve an all-in-one-point final configuration -- with three loop lines passing through a single point $A$ -- we must take three residues at $\br{A_iB_iA_jB_j}=0$~\eqref{eq:OnShellInternal}.  By inspection, one can see that the first term in square brackets in~\eqref{eq:3LoopMHV} (the ladder integral) does not contain all three poles $\br{A_iB_iA_jB_j}$ and therefore vanishes after taking these residues. From this point forward then we will only concern ourselves with the second term.

\begin{figure}
\centering
\begin{tikzpicture}
\draw[thick,-latex] (2,2) -- (-3,-3);
 \node[label={[label distance=-.2cm]0:$A_1$}] at (1.5,1.5) {$\bullet$};
 \node[label={[label distance=-.2cm]0:$B_1$}] at (-2,-2) {$\bullet$};
\node[label={[label distance=-.1cm]0:$A_3{=}A_1{+} a_3 B_1$}] at (0.5,0.5) {$\bullet$};
\draw[thick,-latex] (-1,1) -- (5,-1);
 \node[label={[label distance=-.2cm]90:$B_3$}] at (4,-.666) {$\bullet$};
 \draw[thick,-latex] (-3,-1) -- (3,-3);
 \node[label={[label distance=-.2cm]90:$B_2$}] at (2,-2.666) {$\bullet$};
 \node[label={[label distance=-.1cm]0:$A_2{=}A_1{+}a_2 B_1$}] at (-1.5,-1.5) {$\bullet$};
 \draw[thin,dashed,-latex] (0-0.5,0) -- (-1-0.5,-1);
 \end{tikzpicture}
 \caption{A geometrical depiction of the third residue taken when calculating the three loop all-in-one-point cut.  Note the lines that pass through the points $B_2$ and $B_3$ do not in general lie on the same plane, but they do not (yet) intersect.}
 \label{fig:3}
\end{figure}
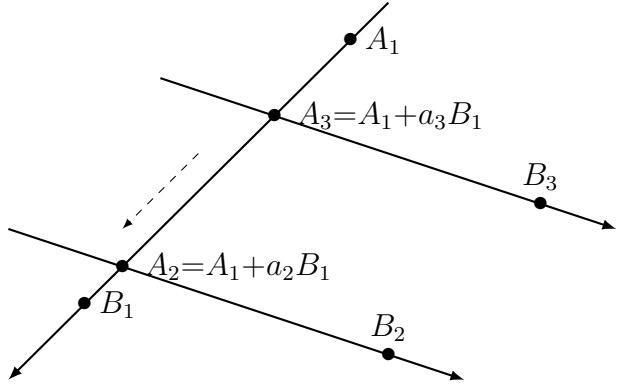

The key factor which all surviving terms contain is
\begin{equation}
    \label{eq:TopForm3Loop}
  F=  \frac{\prod^3_{i=1}\br{A_iB_i \dd^2A_i}\br{A_iB_i\dd^2B_i}}{\br{A_1B_1A_2B_2}\br{A_2B_2A_3B_3}\br{A_1B_1A_3B_3}} \; .
\end{equation}
We will then first consider the residue at $\left<A_1B_1A_2B_2\right>=0$ followed by $\left<A_1B_1A_3B_3\right>=0$. This corresponds geometrically to first intersecting the line $A_2B_2$ with $A_1B_1$ and then $A_3B_3$ with $A_1B_1$ (see~\ref{fig:3}).  To do this, we parametrise $A_2$ and $A_3$ as
\begin{equation}
    \begin{split}
    \label{eq:A2Loop3}
    A_2 &= A_1 + a_2 B_1 + b_2 Z_* \; , \\
    A_3 &= A_1 + a_3 B_1 + b_3 Z_* \; ,
    \end{split}
\end{equation}
where $Z_*$ is an arbitrary twistor.  In this parametrisation, the limits $b_2 \to 0$ and $b_3 \to 0$ correspond to the points $A_2$ and $A_3$ moving to lie on the line $A_1B_1$ respectively. Using this parametrization, we have for example that $\br{A_2B_2\dd^2A_2}=\dd b_2 \dd a_2\br{A_1B_2 Z_* B_1}$ etc. and the factor~\eqref{eq:TopForm3Loop} produces $\dd b_2 \dd b_3/(b_2 b_3)$. Taking the residue then gives
\begin{align}
   \underset{\substack{\left<A_1B_1A_2B_2\right>=0\\\left<A_1B_1A_3B_3\right>=0}}{\text{Res}} F 
    &=\frac{\dd a_2\dd a_3 \br{A_1B_1\dd^2A_1}\prod_{i=1}^3\br{A_iB_i\dd^2B_i}}{\br{A_2B_2A_3B_3}} \; .
    \label{eq:Intersect1Loop3}
\end{align}

With the  parametrisations~\eqref{eq:A2Loop3}, with $b_2=b_3=0$ we see that the remaining singularity of $F$ factorises into two terms
$\langle A_2 B_2 A_3 B_3 \rangle = (a_2{-}a_3)\langle A_1 B_1 B_2 B_3 \rangle$. This is another example of composite residues discussed in the introduction and section~\ref{sec:2}.  The first factor $a_2{-}a_3=0$ corresponds to the three loop lines intersecting in one point, while $\br{A_1B_1B_2B_3}=0$ corresponds to the three lines lying in the same plane (and thus intersecting pairwise).
We focus on the all-in-one-point case $(a_2 {-} a_3) \to 0$ (intersecting $A_3B_3$ with $A_2B_2$ by sliding the intersection point $A_3$ along the line $A_1B_1$ to meet the intersection point $A_2$ see figure~\ref{fig:3}).  
We change variables from $(a_2,a_3)$ to $(a_2,\xi)$ where $\xi = (a_2 {-} a_3)$, so  $\dd a_2 \dd a_3= \dd\xi \dd a_2$ and the residue  at $\xi \rightarrow 0$ of  ~\eqref{eq:Intersect1Loop3} is
\begin{equation}
    \label{eq:Intersect2Loop3_3}
    \underset{\xi = 0}{\text{Res}}\; \left(\frac{\dd\xi\dd a_2}{\xi} \frac{\br{A_1 B_1 \dd^2 A_1}\prod_{i=1}^3\br{A_iB_i\dd^2B_i}}{\br{A_1B_1B_2B_3}}\right) = \frac{\br{A\dd^3A}\prod_{i=1}^3\br{AB_i\dd^2B_i}}{\br{AB_1B_2B_3}} \; .
\end{equation}
On the right-hand side we have written the expression manifestly as a function of the common intersection point  of all three lines $A  = A_1 + a_2 B_1$, thus $\dd a_2\br{A_1 B_1 \dd^2 A_1}= \br{A\dd^3A}$. 

Substituting~\eqref{eq:Intersect2Loop3_3} into~\eqref{eq:3LoopMHV} gives the all-in-one-point cut of the three loop amplitude
:
\begin{align}
\label{eq:DC3Loop}
    &\frac{\br{A\dd^3A}\prod_{i=1}^3\br{AB_i\dd^2B_i}}{\br{AB_1B_2B_3}} \times \notag \\\times &\left(  \frac{\br{1234}^3\br{AB_123}}{\br{AB_114}\br{AB_112}\br{AB_134}\br{AB_212}\br{AB_223}\br{AB_323}\br{AB_334}} \;\;\; + \;\;\;  \text{symmetry}\right)
\end{align} 
where the sum occurs by applying cyclic symmetry of the external momenta and permutation symmetry of the $B_i$s. 
We see that after taking the three consecutive residues required to reach this final configuration, a new pole $\br{AB_1B_2B_3}$ appears explicitly in the denominator of the integrand. 
This is not the same as the result given for the all-in-one-point cut in~\cite{Arkani-Hamed:2018rsk} which  is instead the canonical form of the intersection of the hyperplane $\br{A_iB_i A_jB_j}=0$ with the amplituhedron. \footnote{Note that if one instead takes an antisymmetric sum over the $B_i$ permutations in~\eqref{eq:DC3Loop}, the result produces a zero in $\br{AB_1B_2B_3}$, cancelling the pole and reproducing the deepest cut given in~\cite{Arkani-Hamed:2018rsk}. We can not obtain this via an operation acting on the amplitude however but rather one would have to take a different operation on each contributing diagram.}
Taking further residues of this all-in-one-point cut, starting with the  pole $\br{AB_1B_2B_3}$ one ends up with a maximal residue of 2 (see appendix~\ref{sec:ibls} for this computation) which, as discussed in section~\ref{sec:2} suggests the existence an internal boundary.

In the next subsection we therefore look at the geometrical region corresponding to taking the above all-in-one-point cut.   We will find that the pole at $\br{AB_1B_2B_3} =0$ indeed corresponds geometrically to  an internal boundary of the codimension 3 boundary of the four-point three-loop amplituhedron corresponding to the all-in-one-point cut.

\subsection{Geometric all-in-one-point cut} \label{Geometric Deepest Cut}

We now look to derive the geometry of the all-in-one-point cut.  Following precisely the residues taken in section \ref{ss:3PointDC}, we first intersect line $L_1$ with $L_2$, then intersect $L_3$ with  $L_1$, and finally intersect $L_2$ and $L_3$ by sliding $A_3$ along $L_1$.  

The four-point loop level amplituhedron is defined as the set of loop lines $L_i = (A_iB_i)$ with $i=1,..,L$ satisfying
\begin{equation}
    \cA^{(L)}=\Big\{ A_iB_i:  \br{A_iB_i\bar{k}\bar{l}}>0, \ \br{A_iB_iA_jB_j}>0,   \quad 1\leq i,j\leq L,\ 1\leq k<l\leq 4     \Big\}
\end{equation}
Here for each loop we have the inequalities of the one loop amplituhedron
\begin{equation}
    \label{eq:1LoopIneq}
    \begin{split}
        &\br{A_iB_{i}12} > 0, \;\;\;\;\; \br{A_iB_{i}13} < 0, \;\;\;\;\; \br{A_iB_{i}14} > 0, \\
        &\br{A_iB_{i}23} > 0, \;\;\;\;\; \br{A_iB_{i}24} < 0, \;\;\;\;\; \br{A_iB_{i}34} > 0,
    \end{split}
\end{equation}
which can all be conveniently rewritten in terms of the conjugate planes as~\cite{Arkani-Hamed:2018rsk}
\begin{align}  \label{eq:1LoopIneq2}
    \br{A_iB_i\bar{j}\bar{k}}>0  \qquad 1\leq j<k\leq 4
\end{align}
where
 $\bar{j} \equiv (j-1jj+1)$ and $\br{A_iB_i\bar{j}\bar{k}} \equiv \br{A_iB_i (j-1jj+1) \cap (k-1kk+1)}$.
  We then also have 
the loop-loop inequalities, $\br{A_iB_iA_jB_j}>0$.
 The all-in-one point configuration
occurs when all loop lines $L_i$ pass through a single point $A$, so we simply set $A_i=A$. Then the loop-loop inequalities trivialise and this all-in-one-point cut geometry is
\begin{equation}\label{aldcdef}
    \cA^{(L)}_{dc} =\cA^{(L)}|_{A_i=A}=\Big\{ A,B_i:  \br{AB_i\bar{k}\bar{l}}>0, \ \ 1\leq k<l\leq 4     \Big\}\ .
\end{equation}
This is the codimension 3 configuration of the amplituhedron  corresponding to all loop lines intersecting in one point. 

We now however wish to examine in detail what happens when we take a  sequence of codimension 1 boundaries in order to reach such a configuration. Using the same  parametrisation as~\eqref{eq:A2Loop3}, $
    A_2 = A_1 {+} a_2 B_1 {+} b_2 Z_*$, $A_3 = A_1 {+} a_3 B_1 {+} b_3 Z_*$ the loop-loop inequalities become
\begin{equation}
    \label{eq:mutualPosDC}
    \begin{split}
        \br{A_1B_1A_2B_2} &= -b_2\br{A_1B_1B_2Z^*} > 0\; , \\
        \br{A_1B_1A_3B_3} &= b_3\br{A_1B_1Z^*B_3} > 0\; , \\
        \br{A_2B_2A_3B_3} &= (a_2 {-} a_3)\br{A_1B_1B_2B_3}+(b_2 {-} b_3)\br{A_1Z_*B_2B_3} > 0 \; .
    \end{split}
\end{equation}
We then consider  the boundary at $a_2=a_3$ of the boundary at $b_3=0$ of the boundary at $b_2=0$, which  corresponds precisely  to taking the consecutive residues of~\eqref{eq:Intersect1Loop3} and below. 
Here  $Z_*$ is chosen arbitrarily and we can arrange it so that $\br{A_1B_1B_2Z_*}<0$ and $\br{A_1B_1Z_*B_3}>0$
and thus $b_2,b_3>0$.
Notice that the third inequality factorises when $b_2,b_3 \rightarrow 0$. This is the geometric version of the factorisation discussed  below~\eqref{eq:Intersect1Loop3}, related to composite residues and reducible varieties. 
Thus the boundary at $b_2,b_3 \rightarrow 0$ is the union of two disconnected regions $\mathcal{R}_1 \cup \mathcal{R}_2$
\begin{equation}
    \label{eq:ineqDC3Loop}
    \begin{split} 
     \mathcal{R}_1&: \;\; \;\;\; a_2 > a_3, \;\;\; \br{AB_1B_2B_3} > 0\; ,\\
     \mathcal{R}_2&: \;\; \;\;\; a_2 < a_3, \;\;\; \br{AB_1B_2B_3} < 0\; ,
    \end{split}
\end{equation}
where $A = A_1 + a_2B_1$.
 The inequalities,~\eqref{eq:ineqDC3Loop}, carve out a region consisting of two almost disconnected pieces of the same orientation.
 This geometry is illustrated in the picture (the same as for the two-loop internal boundary case~\eqref{pic})
\begin{align}
   \begin{tikzpicture}
     \fill[fill=gray!20] (0,0)--(2,0)--(2,2)--(0,2)--(0,0);
    \fill[fill=gray!20] (0,0)--(-2,0)--(-2,-2)--(0,-2)--(0,0);
    \draw[-latex] (0,2) -- (0,.9);
    \draw[-latex] (0,1) -- (0,0) -- (1.1,0);
    \draw  (1,0) -- (2,0);
 \draw[-latex] (0,-2) -- (0,-.9);
    \draw[-latex] (0,-1) -- (0,0) -- (-1.1,0);
    \draw  (-1,0) -- (-2,0);
    \draw[-latex,semithick,red] (1,1) arc[radius=.2, start angle=60, end angle=300]-- ++(20:1pt);
    \node at (1.4,1.4) {$\mathcal{R}_1$};
    \draw[-latex,semithick,red] (-1,-1) arc[radius=.2, start angle=-120, end angle=120]-- ++(200:1pt);
    \node at (-1.4,-1.4) {$\mathcal{R}_2$};
    \node at (0,0) {$\bullet$};
    \end{tikzpicture}
\end{align}
where the $x$ axis corresponds to the region $a_2=a_3$ and the $y$ axis corresponds to $\br{AB_1B_2B_3}=0$. The all-in-one-point cut corresponds to the boundary $a_2 {=} a_3$ (so the $x$ axis). We can then clearly see that the all-in-one-point cut  consists of two regions, $\br{AB_1B_2B_3} \lessgtr 0$, with opposite orientation
separated by an (internal)  boundary at $\br{AB_1B_2B_3} = 0$:
\begin{equation}
    \begin{tikzpicture}
        \node at (0,0) {$\bullet$};
         \draw[-latex] (0,0) -- (4,0);                
         \draw[-latex] (0,0) -- (-4,0);
         \node at (2.5,-.5) {$ \br{AB_1B_2B_3} > 0$};
         \node at (-2.5,-.5) {$ \br{AB_1B_2B_3} < 0$};
    \end{tikzpicture}
\end{equation}

Geometrically the two regions arise from the intersection point  $A_3$ approaching $A_2$ from two different directions along the line $A_1B_1$ (see figure~\ref{fig:3}).  Importantly, after approaching the all-in-one-point cut,  the mutual positivity conditions between the loops of~\eqref{eq:mutualPosDC}  do not trivialise, but  instead new inequalities emerge dictated by the sign of $\br{AB_1B_2B_3}$.

 So altogether then, incorporating the inequalities resulting from~\eqref{eq:1LoopIneq} (rewritten as in~\eqref{eq:1LoopIneq2}) we see that the 
 the full geometry of the three-loop all-in-one-point cut is given by the two regions with opposite orientation

  \begin{equation}\label{eq:3loopsDeepest}
  \begin{split}
       \mathcal{R}^{\text{dc}}&= \mathcal{R}^{\text{dc}}_{1}\cup \mathcal{R}^{\text{dc}}_{2} \\
       \mathcal{R}^{\text{dc}}_1&= \cA_{dc}^{(3)} \,\cap\, \left\{ \br{AB_1B_2B_3}>0\right\}  \;  \qquad \text{positive orientation} \\
     \mathcal{R}^{\text{dc}}_2&= \cA_{dc}^{(3)} \, \cap \, \Big\{  \br{AB_1B_2B_3}<0 \Big\} \; \qquad    \text{negative orientation}\; .
     \end{split}
 \end{equation}
 Note that the deepest cut geometry $\cA_{dc}^{(3)}$
is the union of these two regions with the {\em same} orientation, but the actual result of taking boundaries of boundaries of boundaries requires the regions  to have opposite orientation separated by an internal boundary.
Also note that we made a choice of which loop lines to intersect first and which to slide etc. and one might expect different choices to give different results. This is indeed the case at higher loops. At three loops however the resulting geometry~\eqref{eq:3loopsDeepest} is the unique geometry one obtains from approaching the all-in-one-point cut.

So to summarise we find that the all-in-one-point cut as computed as  a residue corresponds to two regions of opposite orientation separated by an internal boundary.
At higher loops it turns out that the all-in-one-point cut is no longer even unique but depends on the precise sequence of codimension 1 boundaries taken to reach it.

\subsection{Higher loop all-in-one-point cut} 
\label{HigherLoop}

We commented that at three loops the multiple residue leading to the all-in-one-point  cut is unique and the corresponding geometry given by~\eqref{eq:3loopsDeepest}.  For higher loops, however, there are a number of inequivalent resulting geometries depending on the sequence of single residues taken.  Here, we generalise the discussion of the previous section to give the inequalities associated to any all-in-one-point cut for any loop.  We show that while the final configuration is always the same -- that is $L$ lines intersecting in a point -- distinct paths to reach this configuration can carve out different oriented regions.

\begin{figure}
\centering
\begin{tikzpicture}
\draw[thick] (-3.85,0.15) -- (2-4,2);
\draw[thick,-latex] (-4.15,-0.15) -- (-3-4,-3);
\node[label={[label distance=-.2cm]0:$L_i$}] at (-2.3,1.5){};

\draw[thick,-latex] (-6,2/3) -- (-1,-1);
\node[label={[label distance=-.2cm]0:$L_j$}] at (-1.4,-5/6+0.25){};

\node[label={[label distance=-.1cm]0:}] at (-4.7,-0.7) {$\bullet$};
\node[label={[label distance=-.1cm]0:}] at (-4.7,7/30) {$\bullet$};
\draw[thin,red,-latex] (-4.7,1/30) -- (-4.7,-0.5);

\draw[thick,-latex] (2+4,2) -- (-3+4,-3);
\node[label={[label distance=-.2cm]0:\small$L_k$\normalsize}] at (-2.3+7.4,2){};

\draw[rotate around={-15:(0.5+4,0.5)},thick,-latex] (-1+4,1) -- (-1+8,-1/3);
\draw[rotate around={10:(0.5+4,0.5)},thick,-latex] (-1+4,1) -- (-1+8,-1/3);
\draw[rotate around={25:(0.5+4,0.5)},thick,-latex] (-1+4,1) -- (-1+8,-1/3);
\centerarc[thick,dashed](0.5+4,0.5)(-30:-12:2.6);
\centerarc[thick,dashed](0.5+4,0.5)(166:150:1.5);

\node[label={[label distance=-.2cm]0:\small $L_{i_1}$ \normalsize}] at (-2.3+9.2,1.5-0.45){};
\node[label={[label distance=-.2cm]0:\small $L_{i_2}$ \normalsize}] at (-2.3+9.25,1.5-1.15){};
\node[label={[label distance=-.2cm]0:\small $L_{i_n}$ \normalsize}] at (-2.3+8.3,1.5-2.6){};

\draw[rotate around={-15:(-1.5+4,-1.5)},thick,-latex] (-3+4,-1) -- (-3+8,-7/3);
\draw[rotate around={10:(-1.5+4,-1.5)},thick,-latex] (-3+4,-1) -- (-3+8,-7/3);
\draw[rotate around={25:(-1.5+4,-1.5)},thick,-latex] (-3+4,-1) -- (-3+8,-7/3);
\centerarc[thick,dashed](-1.5+4,-1.5)(-30:-12:2.6);
\centerarc[thick,dashed](-1.5+4,-1.5)(166:150:1.5);

\node[label={[label distance=-.2cm]0:\small $L_{j_1}$ \normalsize}] at (-2.3+9.2-2,1.5-0.45-2){};
\node[label={[label distance=-.2cm]0:\small $L_{j_2}$ \normalsize}] at (-2.3+9.25-2,1.5-1.15-2){};
\node[label={[label distance=-.2cm]0:\small $L_{j_{m}}$ \normalsize}] at (-2.3+8.3-2,1.5-2.6-2){};

\node[label={[label distance=-.1cm]0:}] at (-1.5+4,-1.5) {$\bullet$};
\node[label={[label distance=-.1cm]0:}] at (0.5+4,0.5) {$\bullet$};

\draw[thin,red,-latex] (4,0.3) -- (3,-0.7);
 
 \end{tikzpicture}
 \caption{A graphical representation of the two types of residues discussed here.  On the left is an intersection between lines $L_i$ and $L_j$, which we label as $(i,j)$.  On the right is a sliding between the sets of lines $L_{i_1}, L_{i_2},...,L_{i_n}$ and $L_{j_1},L_{j_2},...,L_{j_m}$, which we label as $(I,J) \equiv (i_1i_2...i_nk , \;j_1j_2...j_mk)$.}
 \label{fig:DCResidues}
\end{figure}
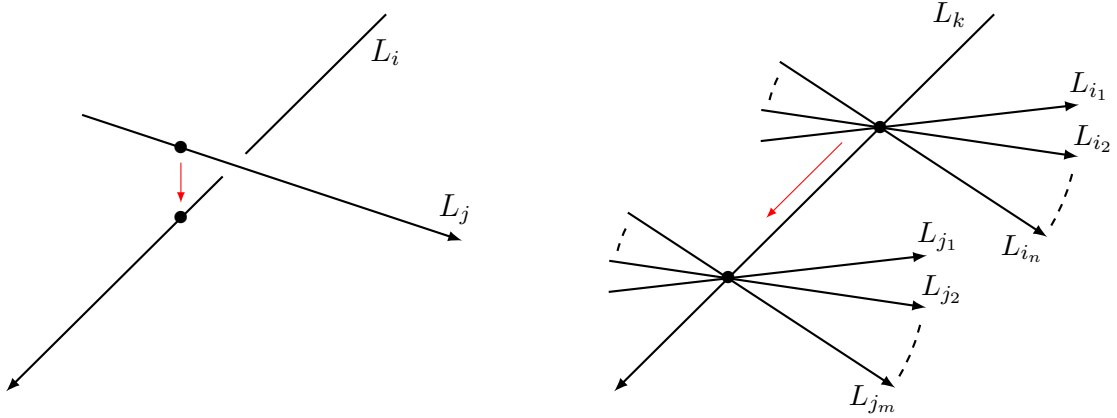

Enforcing that a line in 3d (which has four degrees of freedom)  intersects a specified point kills two degrees of freedom. Thus making all $L$ lines go through a  specified point would reduce by $2L$ degrees of freedom. However the intersection  point itself $A$ is not fixed and has 3 degrees of freedom, thus only $2L-3$ degrees of freedom are lost, corresponding to taking $2L-3$ single residues.
We distinguish between two types of residue, each of which has a different geometrical interpretation.  The first is the intersection of two loop lines  which are currently not connected by any set of intersecting lines (see figure~\ref{fig:DCResidues} on the left). Taking the maximal possible number of such intersections results in a maximal tree configuration.\footnote{More precisely the graph obtained by replacing each loop line with a vertex joined by edges if and only if the respective loop lines intersect should be a maximal tree on $L$ vertices.} These  we will refer to simply as {\em intersections}. The second type occurs when we merge two separate intersection points along a line (see figure~\ref{fig:DCResidues} on the right) which  we shall call a {\em sliding}.  An all-in-one-point cut then consists of $(L{-}1)$ intersections and $(L{-}2)$ slidings to make a total of $2L-3$.

To perform an intersection, for example $(A_iB_i) \cap (A_jB_j)$ depicted on the left in figure \ref{fig:DCResidues}, we parameterise the point $A_j$ as $A_j = A_i + aB_i + bZ_*$ and take the residue $b = 0$.  Similarly to the discussion at three loops leading to~\eqref{eq:Intersect1Loop3}, any such intersection saturates  one positivity condition, $\br{A_iB_iA_jB_j}=0$, and does not generate any new inequalities. The order in which these are performed is also not important. The all-in-one-point cut consists of $L-1$ intersections, therefore $(L-1)$ of the $(2L-3)$ mutual positivity conditions are trivialised and no new inequalities arise.

The remaining $(L-2)$ mutual positivity conditions are handled by slidings.  However, unlike the residues corresponding to the intersections, here new inequalities {\em are} generated.  Let us begin by determining what happens when a residue is taken corresponding to a single sliding, for example the one depicted on the right in figure \ref{fig:DCResidues}.  We start off with two sets of lines intersecting at two different points, with one common loop that the two intersection points lie on.  

Let $I$ and $J$ be the sets of labels of the two groups of intersecting lines,  $k=I\cap J$ labels the line in common, and $A, A^{'}$ label the intersection points of the groups of lines $I$ and $J$ respectively, so
\begin{equation}
    \label{eq:DCNotation}
    \begin{split}
    &A = \bigcap_{i \in I}\,L_i = A_k + c_1B_k \; , \\
    &A^{'} = \bigcap_{j \in J}\,L_j = A_k + c_2B_k \; .
    \end{split}
\end{equation}
In this parametrization the mutual positivity relation between loops in $I$ and $J$ reads 
\begin{equation}
    \label{eq:MutualPositivitySliding}
    \begin{split}
    \br{A_iB_iA_jB_j} = (c_2{-}c_1)\br{A_kB_iB_kB_j} > 0 \; , \qquad \text{for all } \ i\in I \; ,\;  j\in J \; .
    \end{split}
\end{equation}
As in the three loop case, the brackets factorize, giving rise to two almost disconnected regions (see~\eqref{eq:3loopsDeepest}.
The geometric sliding residue is then calculated by taking the limits $(c_2 - c_1) \to 0^{\pm}$, leaving two regions with opposite orientation.  A ``positive" region for which $\br{A_kB_iB_kB_j} > 0 $ for all $i \in I,\ j\in J$ and a ``negative" region for which $\br{A_kB_iB_kB_j} < 0$ for all $i \in I,\ j\in J$.

To compute an all-in-one-point cut we must take $L-2$ sliding residues, each of which splits the geometry of the boundary in two parts. If we label slidings by the index $a = 1, ..., L{-}2$ then  we can identify a sub region with fixed orientation through the string  $\vec{s}=\{s_1,\cdots,s_{L-2}\}$,  where $s_a=\pm 1$ and keeps track of the signs of positively and negatively oriented regions. The resulting geometry is the union of these regions
\begin{align}
    \label{eq:DCFinal1Reg}
    \mathcal{R}^{\text{dc}}&=\bigcup_{\vec{s}} \mathcal{R}^{\text{dc}}_{\vec{s}}\notag\\
    \mathcal{R}^{\text{dc}}_{\vec{s}} &=
    \cA^{(L)}_{dc} \cap \; \left\{ A,B_i: s_a\br{AB_iB_{k_a}B_j} > 0, \quad a=1,..,L{-}2,\  i\in I_a,\ j\in J_a     \right\}\;\\
    &\qquad \qquad \qquad \text{orientation of } \mathcal{R}^{\text{dc}}_{\vec{s}}= \prod_a s_a \notag
    \end{align}
where we recall that $\cA^{(L)}_{dc} $ is the deepest cut  geometry, obtained by trivialising the loop-loop inequalities of the amplituhedron and sending $A_i\rightarrow A$~\eqref{aldcdef}.
In particular this region depends explicitly on the sequence of boundaries we took to approach the geometry through the sets $I_a,J_a$.   

Note that~\eqref{eq:DCFinal1Reg} generalises directly to describe the all-in-one-point cut geometry for amplituhedrons at any number of points. One just needs to add a $Y \in Gr(k,k+4)$ into each bracket and modify $\cA^{(L)}_{dc}$   appropriately.

\subsubsection*{Example: Four Loop all-in-one-point Cuts}

\begin{figure}
\centering
\begin{tikzpicture}
\draw[thick,-latex] (2-4,2) -- (-3-4,-3);
\node[label={[label distance=-.2cm]0:$L_1$}] at (-2.3,1.5){};

\draw[thick,-latex] (-1-4,1) -- (-1,-1/3);
\node[label={[label distance=-.2cm]0:$L_4$}] at (-0.9,-1/3+0.1){};
\draw[thick,-latex] (-3-4,-1) -- (-3,-7/3);
\node[label={[label distance=-.2cm]0:$L_2$}] at (-2.9,-7/3+0.1){};
\draw[thick,-latex] (-6,0) -- (-2,-4/3);
\node[label={[label distance=-.2cm]0:$L_3$}] at (-1.9,-4/3+0.1){};

\node[label={[label distance=-.1cm]0:}] at (-1.5-4,-1.5) {$\bullet$};
\node[label={[label distance=-.1cm]0:}] at (0.5-4,0.5) {$\bullet$};
\node[label={[label distance=-.1cm]0:}] at (-4.5,-0.5) {$\bullet$};

\draw[thin,red,-latex] (0.5-4.5,0.5) -- (-4.5,0);
\node[label={[label distance=0cm]0:\small$c_2$\normalsize}] at (-4.9-0.1,0.4){};
\draw[thin,red,-latex] (-5,-0.5) -- (-5.5,-1);
\node[label={[label distance=0cm]0:\small$c_1$\normalsize}] at (-5.9-0.05,-0.6){};
 
\draw[thick,-latex] (2+4,2) -- (-3+4,-3);
\node[label={[label distance=-.2cm]0:$L_1$}] at (-2.3+8,1.5){};
\draw[thick,-latex] (-1+4,1) -- (-1+8,-1/3);
\node[label={[label distance=-.2cm]0:$L_3$}] at (-0.9+8,-1/3+0.1){};
\draw[thick,-latex] (-3+4,-1) -- (-3+8,-7/3);
\node[label={[label distance=-.2cm]0:$L_2$}] at (-2.9+8,-7/3+0.1){};
\node[label={[label distance=-.1cm]0:}] at (-1.5+4,-1.5) {$\bullet$};
\node[label={[label distance=-.1cm]0:}] at (0.5+4,0.5) {$\bullet$};

\draw[thin,red,-latex] (4,0.5) -- (3,-0.5);
\node[label={[label distance=0cm]0:\small$c_1$\normalsize}] at (2.8,0.2){};

\draw[thick,-latex] (2.5,-3) -- (6.5,-1);
\node[label={[label distance=0cm]0:$L_4$}] at (-1+7.15,-1/3-1.1){};
\node[label={[label distance=0cm]0:}] at (43/10,-21/10) {$\bullet$};

\draw[thin,red,-latex] (43/10-0.5,-21/10-0.05) -- (3.4-0.5,-1.8-0.05);
\node[label={[label distance=0cm]0:\small$c_2$\normalsize}] at (3.0-0.2,-1.8-0.4){};
 \node[label={[label distance=-.2cm]0:$\mathcal{D}_1$}] at (-4.5,-3.5){};
 \node[label={[label distance=-.2cm]0:$\mathcal{D}_2$}] at (3.8,-3.5){};
 
 \end{tikzpicture}
 \caption{Graphical representation of the four loop all-in-one-point cut labelled in~\eqref{eq:DCs4Loops}.  The all-in-one-point cut corresponds to drawing a tree configuration, and collapsing the graph so that only one intersection point remains.  The \textit{intersections} are given by pairs of intersecting lines $L_i$, $L_j$.  The \textit{slidings} are labelled in the order they should be done, $c_1, ..., c_{L-2}$.  Each slide corresponds to moving one intersection point along a line in the direction dictated by the red arrow until it meets another intersection point.}
 \label{fig:4LoopExample}
\end{figure}
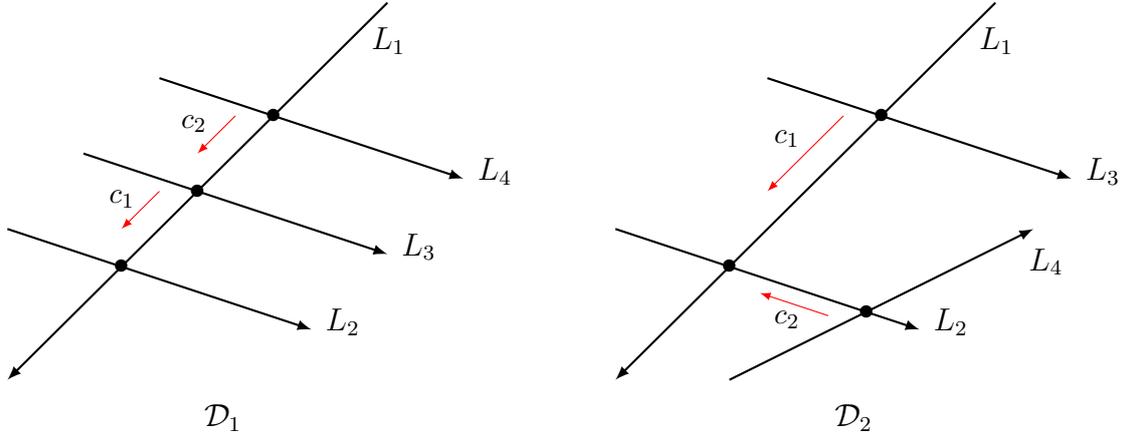

 Let us illustrate~\eqref{eq:DCFinal1Reg} by giving an explicit example at four loops.
  Each 4 loop all-in-one-point cut is given by 3 intersections and 2 sidings. Denoting the intersection between lines $L_i$ and $L_j$ by $(i,j)$ and a sliding between the sets of lines $I$ and $J$ as $(I,J)$, we explore  the cut 
\begin{equation}
    \label{eq:DCs4Loops}
  \{ (1,2), (1,3), (1,4) \; ; \; (12,13), (123,14) \} \; ,
\end{equation}
represented in Figure~\ref{fig:4LoopExample}a.
From~\eqref{eq:DCFinal1Reg}, the resulting geometry is given by a union of four regions:
\begin{equation}
    \label{eq:DC4LoopGeom1}
    \begin{split}
    \mathcal{R}_1(\mathcal{D}_1) &\ =\  \cA^{(L)}_{dc} \; \land \; \br{AB_2B_1B_3}>0 \; \land \;  \br{AB_2B_1B_4}>0 \land \br{AB_3B_1B_4}>0   \; ,\ +\\
    \mathcal{R}_2(\mathcal{D}_1) &\ =\  \cA^{(L)}_{dc} \; \land \; \br{AB_2B_1B_3}<0 \; \land \;  \br{AB_2B_1B_4}>0 \land \br{AB_3B_1B_4}>0   \; ,\  -\\
    \mathcal{R}_3(\mathcal{D}_1) &\ =\  \cA^{(L)}_{dc} \; \land \; \br{AB_2B_1B_3}>0 \; \land \;  \br{AB_2B_1B_4}<0 \land \br{AB_3B_1B_4}<0   \; ,\  -\\
    \mathcal{R}_4(\mathcal{D}_1) &\ =\  \cA^{(L)}_{dc}\; \land \; \br{AB_2B_1B_3}<0 \; \land \; \br{AB_2B_1B_4}<0 \land \br{AB_3B_1B_4}<0   \; ,\ +
    \end{split}
\end{equation}
where $A$ denotes the final point that all loops intersect, and the orientation of the regions is indicated on the right by a $`+'$ or $`-'$.  In particular, $\mathcal{R}_1$, $\mathcal{R}_4$ have the same orientation and $\mathcal{R}_2$, $\mathcal{R}_3$ have the same orientation but opposite to $\mathcal{R}_1$, $\mathcal{R}_4$. 

We see that this four loop all-in-one-point cut geometry has an internal boundary at $\br{AB_2B_1B_3}=0$ and external boundaries at $\br{AB_2B_1B_4}=0,\br{AB_3B_1B_4}=0$. The corresponding multiple residue has poles in these positions. 

Recall that at 3 loops all possible ways of reaching the all-in-one-point cut configuration result in the same geometry~\eqref{eq:3loopsDeepest}. At 4 loops on the other hand there are twelve different possible geometries. They are   all equivalent to each other up to permutations. That is they are all  given by~\eqref{eq:DC4LoopGeom1} after permuting the $B_i$ (permuting $B_2,B_3$ in~\eqref{eq:DC4LoopGeom1} gives back the same geometry up to swapping the overall orientation and so there are only 12 inequivalent permutations rather than 24).\footnote{The choice of all-in-one-point cut $\mathcal{D}_{2} = \{ (1,2), (1,3), (2,4) \; ; \; (12,13),  (123, 24) \}$ illustrated in figure~\ref{fig:4LoopExample}b
 looks like a different case at first sight but in fact results in the same geometry as~\eqref{eq:DC4LoopGeom1} after permuting $B_1$ and $B_2$.}
 Thus the corresponding action of taking residues on a permutation invariant object such as  the loop integrand yields the same result for  all twelve four-loop all-in-one-point cuts.
 From 5 loops however there are  genuinely different all-in-one-point cuts giving different results when the corresponding residues are taken on  a permutation invariant object.

\section{ All-in-one-point-and-plane cuts}

One of the attractive features of the deepest cut defined in~\cite{Arkani-Hamed:2018rsk} was that its canonical form was defined by a simple formula at all loops. This was because the all-in-one-point   configuration~\eqref{aldcdef} consists of $L$ independent one loop inequalities and no loop-loop inequalities and the resulting geometry thus factorises.  We have seen however  that any action of taking consecutive boundary components to reach the deepest cut configuration gives non-unique geometries which are more involved than~\eqref{aldcdef} and in particular new loop-loop inequalities of the form $\br{AB_iB_jB_k}>0$ are generated, spoiling this factorisation. The presence of these brackets makes the computation of the canonical form much more challenging and also dependent on the particular sequence off boundaries taken to reach all-in-one-point configuration. 

In this section we will show that despite this complication it can  still be possible to find fairly simple all loop geometries by taking further residues after reaching the all-in-one-point cut configuration  that trivialize all the new $\br{AB_iB_jB_k}$ inequalities. The further cuts constrain the loop lines to all lie in the same plane as well as going through the same point. They are thus simultaneously  all-in-one-point and all-in-one-plane cut configurations. We will thus refer to them as all-in-one-point-and-plane cuts or point-and-plane cuts for short.
They are defined in terms of the point $A$ which all loop lines go through together with the plane $(AP_1P_2)$ which all loop lines lie on. It is useful also to project through the point  $A$ and thus reduce the geometry to 2d, in which case we refer to the plane $P$ instead as a line.

If we project through the common intersection point $A$, the geometry of the cut correspond to $L$ points $B_i$ on an oriented line $P$ in $\mathbb{P}^2$. Starting at 4 loops, the $\br{AB_iB_jB_k}$ inequalities force some ordering between the points on $P$. As a practical consequence, this implies that an all-in-one-point-and-plane cut can itself also have further loop-loop type  boundaries at $B_i=B_j$. Thus taking the residues / boundaries on these  effectively reduces the number of free loop variables further. We call a cut for which we have exhausted all loop-loop  type residues a {\em maximal loop-loop} cut. 
All the maximal loop-loop cuts that we have considered correspond  -- up to a permutation of the $B$s and an integer factor arising from the number of internal boundaries taken in reaching there -- to the three loop maximal loop-loop cut (which is also the unique all-in-one-point-and-plane cut)
\begin{equation}\label{eq:looploopCut}
  \cA^{L=3}_{\text{mll}} = \left\{A,P,B_i=P_1+b_iP_2:  \br{AB_i \bar j \bar k}>0:\quad i=1,2,3,\ 1\leq j<k\leq 4  \right\}\ .
\end{equation}
We conjecture this to hold in general, that is the maximal loop loop cut always reduces to the three loop one, $\cA^{L}_{\text{mll}}= \cA^{L=3}_{\text{mll}}$.

In this section we will show how to compute the geometry and the canonical form of the all-in-one-point-and-plane cut from the amplituhedron. We will start with the 2 and 3 loop cases, which contain all the main features of the problem. Then we will look at the geometry of the only $2$ independent (up to permutations of the loop lines)  all-in-one-point-and-plane cuts at 4 loops,  and finally we will define a particular cut at  arbitrary loops  and compute its canonical form.

\subsection{All-in-one-point-and-plane canonical form  at 2 loops}
\label{ll2}

At higher loops one can take further boundaries of the all-in-one-point configuration so that the lines all lie in a single plane.
But at two loops we only have two lines intersecting in a point so they automatically lie in the same plane. Thus the all-in-one-point and the all-in-one-point-and-plane cases are identical. 
Nevertheless it is useful to rewrite the two loop all-in-one-point case in the same variables we will use at higher loops, namely in terms of a single line $P$ in $\mathbb{P}^2$ (after projection through $A$) on which the $B_i$s lie (each now with 1 degree of freedom).

At two loops the all-in-one-point cut is obtained simply by taking the  residue in $\br{A_1B_1A_2B_2}$ of~\eqref{eq:factorChange} and is thus given by
\begin{align}\label{eq:2loopdc}
\cA^{(2)}_{dc} =&\frac{\br{AB_1d^2B_1}\br{AB_2d^2B_2}\br{1234}^3\br{A\dd^3A}}{\br{AB_114}\br{AB_112}\br{AB_223}\br{AB_234}}\times \nonumber\\&\times \left[ \frac{1}{\br{AB_134}\br{AB_212}}+ \frac{1}{\br{AB_123}\br{AB_214}} \right] \quad +\quad B_1\ \leftrightarrow \  B_2 \; .
\end{align}
The deepest cut formula of~\cite{Arkani-Hamed:2018rsk} is however a completely different-looking yet identical  formula for $\cA^{(2)}_{dc}$ obtained by computing the canonical form of its corresponding geometry~\eqref{aldcdef} (we recall that  at two loops this correctly reproduces the corresponding residue but not beyond). 

Following~\cite{Arkani-Hamed:2018rsk}, the first step  in computing the canonical form is to triangulate the $A$ geometry into regions where the brackets $\br{Aijk}$ have a well defined sign.
Let's derive such a tiling for the intersection point  $A=A_1B_1\cap A_2B_2$. Since the intersection point $A$ can occur at any point along a loop line, the allowed space for $A$ can be computed as the linear combination $A= c_1 A_1+ c_2 B_1$, where $A_1B_1$ lives in the amplituhedron. Notice, that the intersection point $A$ is defined up to a sign, so we can fix for example $c_1>0$. Solving the inequalities one finds that the allowed regions for $A$ correspond to $4$ twisted cyclic permutations\footnote{$Z_i\rightarrow Z_{i+1}$ for $i=1,2,3$ and $Z_4\rightarrow -Z_1$.} of the solution
\begin{equation}\label{eq:regionPicture}
    \br{A123}>0, \quad   \br{A124}>0,\quad  \br{A134}<0,\quad  \br{A234}>0 \; . 
\end{equation}
 All these cyclically related $A$ regions are tetrahedra, and their canonical forms $\omega_i(A)$
 (where we assign the label $i=1$ to region~\eqref{eq:regionPicture} and the other values to its cyclic twisted permutations)
 can be written as $\omega_i(A)=(-1)^i\omega(A)$, with
\begin{equation}
    \omega(A)= \frac{\br{A\dd^3A}\br{1234}^3}{\br{A123}\br{A234}\br{A134}\br{A124}} \; .
\end{equation}

We can now project through $A$ onto a plane not containing $A$ and the remaining geometry is two dimensional. The configuration of $Z_i$ arising from~\eqref{eq:regionPicture} 
 is such that $1,2,3$ form an anti-clockwise oriented triangle containing $4$:
\begin{equation}
\label{fig:projectionA}
\begin{tikzpicture}[scale=1.8]
\filldraw[fill=gray!20] (0,1) --(1,1)--(0.5,0.5)--(0,1); 
\filldraw[black] (0,0)
circle (0.4pt)
node[anchor=south east]{$1$};
	 \filldraw[black] (0,1) circle (0.4pt) node[anchor=south east]{$3$};
	  \filldraw[black] (1,1) circle (0.4pt) node[anchor=south]{$23\cap14$};
	  \filldraw[black] (0.5,0.5) circle (0.4pt) node[anchor= west]{$4$};
	  \filldraw[black] (2,1) circle (0.4pt) node[anchor=south east]{$2$};
	  \draw (0,1)--(2,1);
	 \draw (0,0)--(1,1);
	 \draw (0,0)--(2,1);
	 \draw (0,0)--(0,1);
	 \draw (0,1)--(2/3,1/3);
	 \end{tikzpicture}
	 \end{equation}
Now we can analyze the $B$ inequalities
\begin{align}
   \br{AB14}>0\, , \quad \br{AB23}>0\,, \quad \br{AB34}>0\,, \quad \br{AB12}>0\; , 
\end{align}
which one can see puts $B$ inside the shaded triangle in~\eqref{fig:projectionA},
with  vertices $\{4,3,(23)\cap(14)\}$.  
For general $i$ after cycling we have that the $B_i$ are in the triangle with edges $(i{+}2\,i{+}3),\,(i\,i{+}4),\,(i{+}1\, i{+}2)$ 
and vertices 
\begin{equation}
    W_{i1}= i{+}2\; , \\
    W_{i2}= i{+}3 \; ,\\
    W_{i3}= (i{+}1\,i{+}2)\cap (i\,i{+}3)\; .
\end{equation}
For fixed $i$, our problem now simply reduces to computing the canonical form of two points $B_1$ and $B_2$ living independently inside the triangle $(W_{i1}W_{i2}W_{i3})$.
Each point $B_i$ thus has the canonical form of a triangle and  we obtain the two loop deepest cut form as
\begin{align}
\cA^{(2)}_{dc}=\sum_{i=1}^4  \frac{(-1)^i  \br{A\dd^3A}\br{1234}^3}{\br{A123}\br{A234}\br{A134}\br{A124}}  \prod_{L=1}^2  \frac{\br{A B_L  d^2B_L}\br{AW_{i1}W_{i2}W_{i3}}^2}{\br{AB_LW_{i1}W_{i2}}\br{AB_LW_{i2}W_{i3}}\br{AB_LW_{i3}W_{i1}}}\ .
\end{align}
Remarkably, this is indeed equal to~\eqref{eq:2loopdc}.

But we now wish to rewrite this further in  a way appropriate for the higher loop all-in-one-point-and-plane cut.
So instead of considering the $B_i$ living in 2d,  we consider first fixing a line $P$ and then two points $B'_1,B'_2$ living on the 1d line $P$.

 The Jacobian of the transformation from $B_1,B_2$ to $P,B_1',B_2'$ is given by 
\begin{equation}\label{eq:factorChange}
    \br{AB_1d^2B_1}\br{AB_2d^2B_2}= \frac{\br{AZ_*B'_1B'_2}\br{AZ_*B'_1\dd B'_1}\br{AZ_*B'_2\dd B'_2}\br{AP\dd P_1}\br{AP\dd P_2}}{\br{AZ_*P}^3} \; ,
\end{equation}
where $Z_*$ is a fixed element of $\mathbb{P}^3$ such that $\br{AB_1B_2Z_*}\neq0$. Notice that the 2-loop deepest cut is symmetric in $B_1$ and $B_2$, but becomes anti-symmetric in $B'_1,B'_2$. This corresponds geometrically to the fact that  switching  $B_1,B_2$ flips the orientation of the configuration on  the right. We will shortly see how this symmetry is reflected in the amplituhedron geometry in the new variables. From now on we will drop the primes on the $B_i$s.

The task is now to translate the geometry of two points in a triangle to that of a line through a triangle and two points on that line.
We start by observing that the geometry of a line through a triangle can be triangulated into 3 regions. These correspond to the combination of the 3 ways in which the line $P$ can intersect the edges of the triangle. However it will turn out  that we also need to consider which side $B_1$ is of $B_2$ (due to the orientation switch mentioned above) and so we in fact need to split into 6 regions. 
\begin{equation}
    \label{fig:onetriangle}
\begin{tikzpicture}[x=1.5cm,y=1.5cm,decoration={markings, 
	mark= at position 0.5 with {\pgftransformscale{1.5}\arrow{latex}}}
] 
	 \filldraw[fill=gray!20] (-1,0) node[below left]{\footnotesize $W_{i,3}$}--(0,1) node[above]{\footnotesize $W_{i,2}$}--(1,0)node[below right]{\footnotesize $W_{i,1}$}--(-1,0);
	 \draw (-1,0)--(0,1);
	 \draw (0,1)--(1,0);
	 \draw (1,0)--(-1,0);
	  \filldraw[fill=gray!20] (-1,0)--(0,1)--(1,0)--(-1,0);
	 \filldraw[black] (-0.3,0.5) circle (1pt) node[below]{$B_1$};
	  \filldraw[black] (0.3,0.5) circle (1pt) node[below]{$B_2$};
	  \draw[postaction={decorate}] (-1.5,0.5)--(1.5,0.5);
	  	  \node at (-1,.6) {$P$};
	 \end{tikzpicture}
	 \qquad
	 \begin{tikzpicture}[x=1.5cm,y=1.5cm,decoration={markings, 
	mark= at position 0.5 with {\pgftransformscale{1.5}\arrow{latex}}}
] 
	 \filldraw[fill=gray!20] (-1,0) node[below left]{\footnotesize $W_{i,3}$}--(0,1) node[above]{\footnotesize $W_{i,2}$}--(1,0)node[below right]{\footnotesize $W_{i,1}$}--(-1,0);
	 \draw (-1,0)--(0,1);
	 \draw (0,1)--(1,0);
	 \draw (1,0)--(-1,0);
	  \filldraw[fill=gray!20] (-1,0)--(0,1)--(1,0)--(-1,0);
	 \filldraw[black] (-0.3,0.5) circle (1pt) node[below]{$B_2$};
	  \filldraw[black] (0.3,0.5) circle (1pt) node[below]{$B_1$};
	  \draw[postaction={decorate}] (-1.5,0.5)--(1.5,0.5);
	    \node at (-1,.6) {$P$};
	 \end{tikzpicture}
\end{equation}
Let's consider one of these 6 regions, the one on the left in~\eqref{fig:onetriangle}. It is  described by the inequalities
\begin{equation}\label{region2triangles}
    \begin{split}
    &\br{PW_{i,1}}>0\; , \quad \br{PW_{i,2}}<0\; , \quad \br{PW_{i,3}}>0\; , \\
   & \br{B_l W_{i,j}W_{i,j+1}}>0\, , \qquad \text{with} \quad j=1,2,3 \text{ and } l=1,2 \; ,
   \\ 
   &\br{W_{i,2}B_1B_2}<0 \; ,
    \end{split}
\end{equation}
with the last inequality ensuring that $B_1$ and $B_2$ are ordered.

Then all $6$ configurations can be generated by cyclic permutations of~\eqref{region2triangles} together with $B_1\leftrightarrow B_2$. We will use $p=1,2,3$ to label the cyclic permutations of~\eqref{region2triangles}, with $p=2$ corresponding to the case~\eqref{region2triangles}.

The canonical form $\lambda_{i,p}(P)$ corresponding to a line through the triangle is the same for all $p=1,2,3$ and equal to
\begin{align}
    \lambda_{i,p}(P)=\lambda_{i}(P)= -\frac{\br{P \dd P_1}\br{P \dd P_2}\br{W_{i,1}W_{i,2}W_{i,3}}}{\br{PW_{i,1}}\br{PW_{i,2}}\br{PW_{i,3}}} \; , \qquad \text{with } p=1,2,3  \; .
\end{align}
For fixed $i$ and $p$, the geometry of $B_1$ and $B_2$ corresponds to two points living on the segment with vertices $I_{i,p}=(W_{i,p-1}W_{i,p})\cap P$ and $J_{i,p}=(W_{i,p}W_{i,p+1})\cap P$.

The canonical form of a point $B$ on a segment $(I J)$ in $\mathbb{P}^1$ can be written in general as
\begin{align}
 [I;B;J]&:= \frac{\br{J\dd B}}{\br{JB}}-\frac{\br{I \dd B}}{\br{IB}}= \frac{\br{B \dd B}\br{IJ}}{\br{BI}\br{JB}} \; .
\end{align}

For  our application the 1d segment  lives in $\mathbb{P}^2$ (well really in $\mathbb{P}^3$ but we already projected through $A$ onto $P^2$). We can choose any point to project onto the segment, call this $Z_*$,%
\footnote{Note that a natural point to choose for $Z_*$ is the intersection point of the two edges that $P$ is passing through, $W_{i,p}$. Then the intersection points  $I_{ip}\sim W_{i,{p-1}}$ and $J_{ip} \sim W_{i,p+1}$ and thus the formulae dramatically simplify.}
then all the two brackets in the above formula can be viewed as 3-brackets with an additional $Z_*$ and in turn eventually as 4-brackets with an additional $A$ (so eg $\br{B \dd B}=\br{Z_* B \dd B}=\br{AZ_* B \dd B}$ etc.)

Then, the  canonical form for $B_1,B_2$ ordered on the segment $(IJ)$, with  $I<B_1<B_2<J$ can be written as
\begin{equation}
[I;B_1,B_2;J]:= [I;B_1;J][B_1;B_2;J] = [I;B_1;B_2][I;B_2;J]\  ,
\end{equation}
and similarly for arbitrary numbers of ordered $B_i$ on $(IJ)$ we define inductively
\begin{equation}
[I;B_1,..,B_L;J]:= [I;B_1,..,B_{L-1};J][B_{L-1};B_L;J]\  .
\end{equation}

Now we claim that the canonical form for two free  points $B_1,B_2$ in a triangle translates as follows
\begin{equation}
  \prod_{l=1}^2  \frac{\br{A B_l  d^2B_l}\br{AW_{i1}W_{i2}W_{i3}}^2}{\br{AB_lW_{i1}W_{i2}}\br{AB_lW_{i2}W_{i3}}\br{AB_lW_{i3}W_{i1}}} = \lambda_{i}(P)\sum_{p=1}^3\Big( [I_{ip};B_1,B_2;J_{ip}]-[I_{ip};B_2,B_1;J_{ip}]\Big)\ .
\end{equation}
Note in particular the minus sign between the canonical forms for the two orderings of $B_1,B_2$. This is because  the orientation flips when $B_1$ passes through $B_2$  as discussed below~\eqref{eq:factorChange}.

We can finally put all the pieces together and write the canonical form $\cA^{(L)}_{dc}$ in terms of these variables as
\begin{align}\label{2loopsdcPlane}
    \cA^{(2)}_{dc} =   \omega(A)\sum_{i=1}^4(-1)^i \lambda_{i}(P)\sum_{p=1}^3\Big( [I_{ip};B_1,B_2;J_{ip}]-[I_{ip};B_2,B_1;J_{ip}]\Big) 
    \; .
\end{align}
An interesting aspect of this formula is that each term in the sum has a pole at $B_1=B_2$. For $i,p$ fixed this represent an internal boundary of the geometry. The sum of the residues over $p$ though, as expected from~\eqref{eq:2loopdc}, is equal to zero, which means that this pole is  actually a spurious one. Geometrically, we have that when the only two points on $P$ coincide the latter can rotate unconstrained on the pivotal point $B_1=B_2$ and therefore its canonical form will be zero.

\subsection{3-loops all-in-one-point-and-plane canonical form}

We can now generalize the two loop result to higher loops. To compute the canonical form of a point-plane cut, we triangulate the $A$ and $P$ geometry in the same way we did for the two loop case and then we consider the position of $B$s on the line $P$. The general structure of the canonical form of a specific  point-plane cut will depend on the details of how the cut is taken, but it will always have the general form
\begin{align}\label{eq:general-loop-plane}
    \cA^{(L)}_{\text{point-plane}} = 2^{n_I}  \omega(A)\sum_{i=1}^4(-1)^i \lambda_{i}(P)\sum_{p=1}^3 \sum_{\sigma \in S_L} c_\sigma \times \,[I_{ip};B_{\sigma_1},..,B_{\sigma_L};J_{ip}]\; ,
\end{align}
where $c_\sigma= \pm 1,0$ reflecting the orientation (or absence) of a certain ordering of the $B_i$s and 
where $n_I$ is the number of internal boundaries approached   to reach the  configuration.
So for example  the two loop case~\eqref{2loopsdcPlane} takes this form with  $c_{\text{id}}=1$ and $c_{(12)}=-1$.
Turning to the three loop case then, we find, by direct computation of the residues, that the point-plane cut is given by~\eqref{eq:general-loop-plane} with  
$c_\sigma=1$ for all 6 permutations $\sigma \in S_3$. Thus unlike the two loop case, for this case, the order of the $B_l$s on the line $P$ is not relevant and the canonical form simplifies to that of the product of three $B_l$s
\begin{align}\label{3looppp}
    \cA^{(3)}_{\text{point-plane}} =  2 \omega(A)\sum_{i=1}^4(-1)^i \lambda_{i}(P)\sum_{p=1}^3 \prod_{l=1}^3 [I_{i,p};B_l;J_{i,p}] \; .
\end{align}

Let us then see how this arises from the  geometry.
As we saw in section \ref{Geometric Deepest Cut}, the three loop all-in-one-point cut geometry is given by~\eqref{eq:3loopsDeepest}. In particular, we have a positively oriented region for $\br{AB_1B_2B_3}>0$ and a negatively oriented region for  $\br{AB_1B_2B_3}<0$. Now consider fixing the line $P$ (with  $B_1$ and $B_2$ lying on $P$) and fixing $B_3$.  Now consider  passing $B_1$ through $B_2$ on the line $P$. As we saw in the two loop case the orientation for the geometry involving $B_1,B_2$ will swap, but simultaneously $\br{B_1B_2B_3} \rightarrow -\br{B_1B_2B_3}$ and so the overall orientation will also swap (see ~\eqref{eq:3loopsDeepest}). The result is no orientation change at all. 
We are now interested in the geometry of the internal boundary $\br{AB_1B_2B_3}=0$ so moving $B_3$ also onto the line $P$. 
The point  $B_3$ is free to go anywhere on the line $P$ (inside the triangle). The resulting geometry is indeed just that of three free points on the line $P$ with the canonical form~\eqref{3looppp}, including the factor of 2 from taking an internal boundary.

Notice that in this case there are no remaining singularities of the form $B_i\to B_j$ so $ \cA^{(3)}_{\text{point-plane}}$ also represents what we call a maximal loop-loop cut and $ 2\cA^{L=3}_{\text{mll}}:=\cA^{(3)}_{\text{point-plane}} $.

\subsection{All 4-loop point-plane and maximal loop-loop cuts}

We have seen that at 3-loops the all-in-one-point-and-plane cut is unique~\eqref{3looppp}. At 4-loops this is not true anymore and we can have two types of geometry (modulo permutations) resulting from approaching the point-plane configuration in different ways. Each will be characterized algebraically by different coefficients $c_{\sigma}$ in~\eqref{eq:general-loop-plane}  and geometrically by different ordering constraints of the $B$s on the line $P$.

We start with the all-in-one-point cut which is unique up to permutations of the loop lines and the resulting geometry given by~\eqref{eq:DC4LoopGeom1}. We now consider taking further  boundaries of this geometry so the loops also lie in a plane. There are 3 possible loop-loop boundaries, $\br{AB_2B_1B_3}=0, \br{AB_2B_1B_4}=0$ and $\br{AB_3B_1B_4}=0$.  We start by looking at the geometry of the boundary when $B_3$ lies on the line $P$ (on which $B_1,B_2$ lie) followed by the boundary of that geometry found when  $B_4$ also approaches  $P$ (shortly we will switch the order of in which we take these boundaries). 

The resulting geometry is  of the four points $B_i$ on the line $P$ inside the triangle, with the following restrictions: $B_1$ is not allowed to be between $B_2$ and $B_3$ and the orientation depends on the relative position of $B_1$ and the pair $B_2,B_3$ (with the position of $B_4$ unconstrained). This geometry is derived in appendix~\ref{4loopgeom}.

The cut resulting from this geometry is then given 
by the general form~\eqref{eq:general-loop-plane} with 
\begin{equation}\label{4lppp}
n_I=2 \qquad \qquad     c_\sigma= \begin{cases} 
      \,1&  \sigma=(2,3,1)\shuffle(4)\ \text{or}\ (3,2,1)\shuffle(4)\, ,\\
    \text{-}1 & \sigma=(1,2,3)\shuffle(4)\ \text{or}\ (1,3,2)\shuffle(4)\, ,\\
    \,0 & \sigma=(2,1,3)\shuffle(4)\ \text{or}\ (3,1,2)\shuffle(4)\, ,
    \end{cases} 
\end{equation}
where $\shuffle$ is the shuffle operation (thus $4$ can appear in any position).
We have checked this is indeed correct by explicitly taking the corresponding residues of the 4 loop amplitude and finding perfect agreement.

Now note that even after taking the all-in-one-point {\em and} the all-in-one-plane configuration there are still  uncancelled loop-loop poles at $B_1=B_2$ and $B_1=B_3$, corresponding geometrically to external boundaries.

Using the very simple residue structure of the ordered points on an interval canonical form, namely
\begin{align}
    \text{Res}_{B_{k}\rightarrow B_{j}}[I;..,B_i,B_{j},B_k,B_l..;J]=-\text{Res}_{B_{k}\rightarrow B_{j}}[I;..,B_i,B_{k},B_j,B_l..;J]=[I;..,B_i,B_{j},B_l..;J]\ ,
\end{align}
one can quickly check that the residue of the four-loop point-plane cut~\eqref{4lppp} when $B_1=B_2$ or $B_1=B_3$ precisely reproduces the three-loop point-plane cut~\eqref{3looppp} with appropriate variables
and with weight 4 instead of 2 (since this time we approached two internal boundaries,  $\br{AB_2B_1B_3}=0$ and $\br{AB_2B_3B_4}=0$). This implies that the residue corresponding to this 4-loop maximal loop-loop boundary is equal to 2 times the all-in-one-point-and-plane 3-loop residue~\eqref{eq:resGeomDC2}.

Returning to the  all-in-one-point cut, we now consider the only other independent way of reaching the point-plane geometry (modulo permutation of the loop variables) by taking $\br{AB_2B_1B_4}=0$ by sending $B_4$ to the line $P$ followed by sending $B_3$ to the line $P$ (the other way around to what we did above). 
In appendix~\ref{4loopgeom} we again examine this carefully geometrically. The end result this time is the geometry of four points $B_i$ unconstrained on the line $P$ but with the overall  orientation dependent on the ordering of $B_1,B_2$. The resulting canonical form is thus 
 given by~\eqref{eq:general-loop-plane} with
\begin{equation}
n_I=1 \qquad \qquad     c_\sigma= \begin{cases}
  \,  1& \sigma=(2,1)\shuffle(3,4)\, ,\\
   \text{-}1 & \sigma=(1,2)\shuffle(3,4)\, ,\\

    \end{cases} 
\end{equation}
as we have confirmed by taking the residues explicitly and comparing.

Note that this time there is  a remaining loop-loop residue apparent at $B_1\to B_2$ (corresponding to an internal boundary). Taking the residue as above this again leads to the three loop point-plane result~\eqref{3looppp} after which no more loop loop residues are present. This final configuration corresponds to the 3-loop maximal cut $\cA_{\text{mll}}$ with weight 2.

\subsection{A cut at arbitrary loop order}
\label{alllppp}
We have already seen from the four loop examples of the previous section  that the point-plane cut depends on how you approach the configuration.
However one can give specific ways of approaching the point-plane geometry at any loop order and find the resulting cut.
So we conclude this section by giving precisely such an example of a cut that can be computed at arbitrary loop order. This means that we are now specifying an ordered set of residues and giving a closed formula for the result. The particular case is a generalisation of the second 4-loop case considered in the previous subsection.

We start defining what we call the simplest all-in-one-point cut. In this  all loop lines first intersect the line $A_1B_1$ and then they all slide to the same intersection point in the same order as their labeling.  The geometry of this boundary at $L$ loops can be obtained from~\eqref{eq:DCFinal1Reg2} and is given explicitly in appendix~\ref{alllpppap}.

After taking the above all-in-one-point cut we then constrain all loops to lie on the same line $P$, first $B_1,B_2$ then $B_L,B_{L-1},..,B_3$ thus  taking the ordered series of boundaries  $\{\br{AB_2B_1B_L}=0,\br{AB_2B_1B_{L-1}}=0,\cdots,\br{AB_2B_1B_3}=0\}$.

Carefully examining the resulting geometry as is done explicitly in  appendix~\ref{alllpppap}, we arrive at the final geometry corresponding to this point-plane cut. It is given by  $L$ points $B_i$ lying on the line $P$, with $B_3,B_L$ unconstrained and with $B_1$ always lying between $B_2$ and all of the points  $B_4,.., B_{L-1}$. The orientation of the geometry depends on the relative order of $B_1,B_2$.
The resulting canonical form at arbitrary loop order is thus
\begin{equation}\label{eq:predictionSimplest}
\begin{split}
    \cA^{(L)}_{\text{point-plane}} = &2\sum_{i=1}^4 \omega(A) \sum_{p=1}^3 \lambda_{i,p}(P)[I_{i,p};B_3;J_{i,p}][I_{i,p};B_L;J_{i,p}]\,\times\\
    &\times \left(  [I_{i,p};B_2,B_1;J_{i,p}]\prod_{l=4}^{L-1} [B_1;B_l;J_{i,p}]+(-1)^{L-1}[I_{i,p};B_1,B_2;J_{i,p}]\prod_{l=4}^{L-1}[I_{i,p};B_l;B_1]\right)\; .
    \end{split}
\end{equation}
We tested~\eqref{eq:predictionSimplest} by computing this simplest maximal loop-loop residue up to 7 loops from the explicit from of the amplitude obtained in~\cite{Bourjaily:2016evz} and found complete agreement. We see  that this point-plane cut has further poles when $B_l \rightarrow B_1$. We have checked up to 5 loops that taking further  residues in these poles eventually leads to  the 3 loop point-plane cut~\eqref{3looppp}. Indeed our investigations so far indicate that after taking any all-in-one-point-and-plane geometry at any loop order, there are always $L-3$ boundaries of the form $B_i\to B_j$ remaining. After further taking these boundaries we are then always lead to the three loop point-plane geometry~\eqref{3looppp}.
It might be possible to prove this starting from the explicit all-in-one-point geometry~\eqref{eq:DCFinal1Reg}.

\section{Conclusions}

We summarise the main points of the paper and point out topics for further work. 

We began by observing that non-vanishing maximal residues of loop amplitudes are not always $\pm1$ as has generally been assumed, but can take arbitrary  values in $\mathbb{Z}$, apparently contradicting the fact that the loop amplituhedron is a positive geometry. We found the source of this apparent contradiction geometrically to be the existence of {\em internal boundaries}  in the geometry where two regions of opposite orientation touch. This minimally requires including an extra term in the recursive definition of the canonical form to take into account these { internal boundaries}~\eqref{cfrecurs2}. 
In all the examples we have found, the internal boundaries arise from  loop-loop propagators factorizing into the product of two factors. Algebraically these are examples of  composite residues discussed in this context in~\cite{Arkani-Hamed:2009ljj} and  it would be interesting to explore the relation between composite residues and internal boundaries in more detail.

As well as internal boundaries we have also stressed another under emphasised feature of the boundary structure of the amplituhedron and multiple residues, namely the simple fact that multiple-residues and corresponding multiple boundaries are non-unique. On the algebraic side a multiple residue is defined as an (ordered) sequence of simple residues. In the same way, on the geometrical side the relevant quantity one must use is `boundaries of boundaries of...' rather than `codimension $k$ boundaries'.
It would be interesting to revisit previous computations of the boundary structure of the loop amplituhedron, and in particular its genus, for example~\cite{Franco:2014csa}, taking into account both internal boundaries and the above non uniqueness of codimension $k$ boundaries.

These two features of the amplituhedron, internal boundaries and ambiguity of multiple boundaries,  come into sharp focus when investigating  the deepest cuts of~\cite{Arkani-Hamed:2018rsk}. The all loop formulae for these cuts described there is not obtainable directly by taking multiple single  residues of the amplituhedron due to the presence of internal boundaries. 
We have shown that nevertheless all-loop formulae can be obtained. They will inevitably depend though on the details of the cut taken (that is the precise  sequence of simple residues taken) and they will in general contain  further loop-loop poles after reaching the all-in-one-point cut configuration. We give an example of an all loop formula for a particular all in plane and point cut. It would be fascinating to obtain a general formula taking as input the details of the cut taken and as output the corresponding canonical form.
In general it seems that, taking loop-loop type residues  there are always available more loop-loop type poles (or boundaries) until one eventually reaches a configuration with just three loop lines (so all other loop lines coincide with one of these three)  intersecting in a  point and lying in a plane. This is arguably then the true `deepest cut' or `maximal loop-loop cut' one can take.  That is to say it is the maximal cut involving only loop-loop type poles. It has a universal form given by the  three loop all in plane and point cut,  but with an integer valued coefficient related to the number of internal boundaries one takes in arriving there. 

The four-point planar amplitude integrand in $\cN=4$ SYM (and its closely related half BPS correlator) is known to ten loops using various graphical rules together with correlator insights~\cite{Bourjaily:2016evz}.
There have also been investigations o it directly using amplituhedron insights~\cite{Mond1, Mond2, Mond3}.
A key question in this context then is whether the above all-loop cuts can be used practically to actually compute the 4-point amplitude/correlator at higher loops. Taking maximal residues (eg first the all-in-one-point cut, then further external cuts) yields a vast amount of information about the amplitude. 
It also has the tantalising chance of being a constructive approach: rather than using a huge basis of graphs and determining their coefficients, most of which are zero, it might be possible to use the cuts to  construct only the relevant graphs and non-zero  coefficients with which they appear.

Although we have focused on 4 points, the all-in-one-point cut and more general loop-loop  cuts are largely independent of the number of points, and also the MHV degree,  since they involve only the mutual geometry between loop lines rather than the details of the external geometry. 
There has  been some nice recent progress in computing the  amplitude for arbitrary multiplicity directly from the loop amplituhedron~\cite{Kojima:2018qzz,Kojima:2020gxs,Kojima:2020tjf,Herrmann:2020qlt}. 
Points worthy of note in the current context are that taking the maximal multiple  residues involving loops at higher points   
yields leading singularities of amplitudes -- rational coefficients -- which have been extensively analysed and are given by Yangian invariant Grassmann integrals~\cite{Arkani-Hamed:2009ljj,Mason:2009qx}. 
Furthermore in the works~\cite{Dennen:2016mdk, Prlina:2017azl} a method for extracting a list of the
physical amplitude's branch points  from the amplituhedron  is suggested. Here the boundaries are derived by intersecting the closure of the amplituhedron with the boundary components corresponding to vanishing brackets of the form $\br{ABij},\br{A_iB_iA_jB_j}$. 
It would be extremely interesting to revisit both the above points using the insights and technology developed here.

Unlike positive geometries, the space of generalised positive geometries is  closed under union: regions that are triangulated by GPGs will be GPGs. This then suggests they might be a better  arena to seek a direct characterisation -- that is a simple straightforward answer to the question `is the following region a GPG or not' -- than PGs. In particular it seems clear that all multi-linear geometries are  GPGs, and their canonical form can be computed algorithmically via cylindrical decomposition. Nicely  all Grassmannian-type geometries, the type seen in physics,  are multi-linear geometries and thus GPGs. But this is not the most general class and the most general classification needs more investigation.

The generalisation of positive geometry to include internal boundaries also suggests the utility of  a further generalisation to {\em weighted positive geometries} where instead the weights contain the information about internal boundaries, and the canonical form has a much more natural definition~\eqref{cfrecursNew}. 
This concept also makes certain proofs very direct since one can add two WPGs together without first making sure they are disjoint
(see eg the proof of triangulations of positive geometries  in section~\ref{Triangulations}). 

In \cite{Arkani-Hamed:2013kca,Arkani-Hamed:2021iya} the geometry of the log of the MHV amplitude is considered  and defined as a union of geometries with negative mutual positivity condition $\br{A_iB_iA_jB_j}<0$. One of these has negative mutual positivity condition for all $i,j$ and its canonical form is equal to the amplituhedron canonical form. The latter is the only term in the log of the amplitude surviving the all-in-one-point cut and therefore the all-in-one-point cut of the amplitude and the log of the amplitude are the same. The log of amplitude should most naturally be  described by a WPG.

We believe that the above insights will also  have utility in the increasing number of wider  applications of positive geometry  concepts in physics beyond the amplituhedron. One closely related case is the momentum amplituhedron.
For the tree-level momentum amplituhedron~\cite{Damgaard:2019ztj} a lot is known about its boundary stratification~\cite{Lukowski:2020bya,Ferro:2020lgp} and its Euler characteristic has been proven to be equal to one~\cite{Moerman:2021cjg},  a strong indication that the geometry is free from internal boundaries.  However, despite the very solid understanding achieved at tree level, finding the geometry of the loop momentum amplituhedron remains an open problem. In this case we expect internal boundaries to appear and the language of WPGs could give the right framework to define a loop momentum amplituhedron. Also  in the search for a non-planar   amplituhedron  interesting ideas involving the sum of geometries over different orderings have  been explored in~\cite{Arkani-Hamed:2014bca,Damgaard:2021qbi} and might benefit from being viewed as weighted positive geometries. 
Other wider applications of positive geometry which one could revisit include 
~\cite{He:2021llb, Huang:2021jlh, Ferro:2021dub, Arkani-Hamed:2019vag, Arkani-Hamed:2022cqe, John:2020jww, Jagadale:2022rbl, Lukowski:2022fwz}. 
Similarly weighted positive geometry  may provide the right mathematical framework to deal with 
 cosmological correlators, which contrarily to the wavefunction of the universe described by the cosmological polytopes~\cite{Arkani-Hamed:2017fdk,Benincasa:2022gtd}s, do not currently have a geometrical description. Their maximal residues are not +/- 1 and they naively appears as a weigheted sum of canonical forms of cosmological polytopes (P. Benincasa, private communication).

\section*{Acknowledgments}

We thank Nima Arkani-Hamed, Paolo Benincasa, Tomasz Lukowski and  Jaroslav Trnka for helpful discussions.  
GD and PH acknowledge support from the European Union's Horizon 2020 research
and innovation programme under the Marie Sk\l{}odowska-Curie grant
agreement No.~764850 {\it ``\href{https://sagex.org}{SAGEX}''}. GD would like also to thank Iñaki García Etxebarria for many valuable and stimulating discussions. PH
also acknowledge support from the Science and Technology Facilities
Council (STFC) Consolidated Grant ST/P000371/1.
AS acknowledges support from the Engineering and Physical Sciences Research Council (EPSRC) Doctoral Training Partnership 2021-22.
\\

\appendix

\section{Inducing orientations on boundary manifolds}
\label{OrientationAppendix}

Here we would like to review how given an oriented manifold one can derive the induced orientation of the boundary.
An oriented manifold $\mathcal{M}$ is oriented if it posses a continuous top differential form $\mathcal{O}$ that is always non-vanishing  in $\mathcal{M}$. The orientation is then equivalently defined by this volume form modulo positive scaling. Suppose then $\mathcal{M}$ has a boundary $\partial \mathcal{M}$ of codimension one. The orientation $\omega_\partial $ induced by $\mathcal{O}$ on $\partial \mathcal{M}$ will be the projection of $\mathcal{O}$ on $\partial \mathcal{M} $. We will now define what we mean by projection. 
Let the boundary be defined by $f=0$ with $f>0$ inside and $f<0$ outside the region (at least near by).
Then $\mathcal{O}_\partial $ is defined simply as
\begin{align}
\label{inducedOrientation}
     df \wedge \mathcal{O}_\partial = \mathcal{O}|_{\partial \mathcal{M}} \; \ .
\end{align}
   Note that the standard convention in math literature differ by a sign, that is $df$ represents an outward pointing differential. We make this choice so that the segment $x>0$ with positive orientation form $O(x)=\dd x$ has orientation $O|_{x=0}=1$. In this way the canonical form of the positively oriented segment  is $\Omega(x>0,\dd x)=\frac{\dd x}{x}$, in fact this form satisfies 
   \begin{align}
       \text{Res}_{x=0}(\frac{\dd x}{x}) = O|_{x=0}=1 \; .
   \end{align}

\section{Relation of generalised positive geometries to the globally oriented canonical form}
\label{oriented}

The globally  oriented canonical form was defined in~\cite{Dian:2021idl} to obtain the square of the super amplitude out of the geometry of the squared amplituhedron. The interior of the squared amplituhedron consists of several  disconnected components so the first problem is to define the relative orientation of these components. 
The second problem is that the square of the super-amplitude has non-normalizable maximal residues and therefore can not be possibly interpreted as a canonical form of a positive geometry. We solved both problems by noticing that the GCD algorithm (see~\cite{Eden:2017fow, Dian:2021idl}) correctly reproduces the canonical form of the squared amplituhedron. The GCD algorithm works in coordinates, that is on a $\mathbb{R}^{k m}$ patch of the oriented Grassmannian using the tiling property of canonical forms. The orientation of the subregions given by the GCD are fixed by the algorithm as the global orientation of the coordinate patch. Then it gives the result as the sum of the canonical forms of the positive geometries triangulating the region.  We then defined the oriented canonical form as the sum of the canonical form of the regions triangulating the squared amplituhedron.

With the new concept we introduced in this paper, we can simply say then that the squared amplituhedron is a GPG with the orientation fixed by the global orientation of the oriented Grassmannian. The canonical form of a generalized positive geometry $X_{\geq 0}$ in an orientable space $X$  with orientation coinciding with the orientation of $X$ is equal to its globally oriented canonical form.

\section{Three loop internal Boundary and its maximal residues}
\label{sec:ibls}
Here we consider taking further residues of the all-in-one-point cut~\eqref{eq:DC3Loop} to eventually arrive at the leading singularities. Then we will consider the same sequence geometrically. 
Indeed in~\cite{Langer:2019iuo} such maximal  residues were considered leading to a final configuration in which all loop lines intersect external twistors as well as intersecting each other at a single point $A$.

Specifically we consider the case where loop line $L_1$  intersects $Z_1$ and $L_2,L_3$ intersect $Z_2$.   We here show that the resulting residue depends on the path taken. Furthermore if the path taken involves taking a residue in $\br{AB_1B_2B_3}=0$ first,  then the resulting maximal residue has magnitude 2 suggesting that $\br{AB_1B_2B_3}=0$ corresponds to an internal boundary.

The two routes we consider to reach the above configuration are as follows. For route 2 we  first take further residues of the all-in-one-point cut~\eqref{eq:DC3Loop} in the following order
\begin{equation}
    \label{eq:res3LoopExt}
    \br{AB_112} = 0, \; \; \br{AB_114} = 0\; , \;\; \br{AB_212} = 0, \; \; \br{AB_223} = 0\; .
\end{equation}
This corresponds to intersecting line $L_1$ with the edge $Z_1Z_4$ and then with $Z_1$ followed by intersecting $L_2$ with the edge $Z_1Z_2$ and then to $Z_2$. In the process the pole $\br{A B_1 B_2 B_3}\rightarrow \br{A 12 B_3}$ and so the final step is to take residues in this pole $\br{A 12 B_3}=0$ followed by $\br{A 23 B_3}=0$ corresponding to $L_3$ intersecting $Z_1Z_2$ and then sliding to $Z_2$. We take the residues explicitly by parametrizing the $B_i$ as follows (with $Z_*$ an arbitrary twistor)
\begin{equation}
    \label{eq:ParamExternal1}
    \begin{split}
    B_1 &= c_1 Z_4 +  Z_1 + d_1 Z_*\; , \\
    B_2 &= c_2 Z_1 +  Z_2 + d_2 Z_*\; ,\\
    B_3 &= c_3 Z_1 + Z_2 + d_3 Z_*\; .
    \end{split}
\end{equation}
and considering the residues at zero in $d_1,c_1,d_2,c_2,d_3,c_3$ in that order. The residues are straightforward to compute and can be done covariantly, for example:
\begin{equation}
\label{eq:ExternalCut}
    \begin{split}
    \underset{d_1 = 0, \, c_1 = 0}{\text{Res}}\;\left(\frac{\br{AB_1\dd^2B_1}}{\br{AB_114}\br{AB_112}}\right) &= \frac{-1}{\br{A214}} \; .
    \end{split}
\end{equation}

Only the displayed term in~\eqref{eq:Intersect2Loop3_3} out of the 24 total terms survives this sequence of residues and it produces the final result
\begin{equation}
\label{eq:ExternalCut3Loop}
- \frac{\br{1234}^3\br{A\dd^3A}}{\br{A123}\br{A124}\br{A134}\br{A234}} = - \frac{\dd a \dd b \dd c}{abc} \; ,
\end{equation}
where on the right hand side we parametrise the point $A$ as 
\begin{equation}\label{aparam}
    A=Z_1 + aZ_2 + bZ_3 + cZ_4\ .
\end{equation} 
This is the canonical form of a tetrahedron with vertices $Z_i$ and is inline with the prediction of~\cite{Langer:2019iuo}. Indeed the mutual intersection point $A$ is now the only remaining freedom and restricting the amplituhedron geometry to this configuration results in the tetrahedron. However as we saw in the simple example in the introduction, simply restricting the geometry to a high codimension boundary will not always give the right answer and  the precise order in which one takes the residues can be important.

For route 2 therefore we change  the order in which we take the residues  on the all-in-one-point cut~\eqref{eq:DC3Loop}. We first take a residue in the pole $\br{A B_1 B_2 B_3}$ making $L_3$ coplanar with $L1$ and $L2$. Then proceed taking residues as previously in~\eqref{eq:res3LoopExt} moving $B_1\rightarrow Z_1$ and $B_2\rightarrow Z_2$.  Then finally we take a residue as $B_3\rightarrow Z_2$. 
Explicitly then, this time  we 
parametrize the $B_i$ as follows
\begin{equation}
    \label{eq:ParamExternal}
    \begin{split}
    B_3 &= c_3 B_1 + B_2 + d_3 Z_*\; \\
    B_1 &= c_1 Z_4 +  Z_1 + d_1 Z_*\; , \\
    B_2 &= c_2 Z_1 +  Z_2 + d_2 Z_*\; .
    \end{split}
\end{equation} and take the residues at zero in 
the order $d_3,d_1,c_1,d_2,c_2,c_3$.
The first residue is 
\begin{equation}
    \label{eq:MeasureInternalBoundary1}
    \underset{\br{AB_1B_2B_3} = 0}{\text{Res}}\;\left(\frac{\br{AB_3\dd^2 B_3}}{\br{AB_1B_2B_3}}\right) = -\dd c_3 \; ,
\end{equation}
and then use similar results to~\eqref{eq:ExternalCut} before finally taking the residue in the parameter $c_3$.
This time two terms in~\eqref{eq:Intersect2Loop3_3} survive, the displayed term together with the term
\begin{equation}
  \frac{\br{A\dd^3A}\prod_{i=1}^3\br{AB_i\dd^2B_i}\langle 1234\rangle ^3 \left\langle 23AB_1\right\rangle
   }{\left\langle 12AB_1\right\rangle  \left\langle 12AB_3\right\rangle 
   \left\langle 14AB_1\right\rangle  \left\langle 23AB_2\right\rangle 
   \left\langle 23AB_3\right\rangle  \left\langle 34AB_1\right\rangle 
   \left\langle 34AB_2\right\rangle \br{AB_1B_2B_3}  }\ .
\end{equation}
Note that this term only survives because a required pole $\br{12AB_2}$  appears from   the pole in $\br{12AB_3}$  after the residues in $d_3=0,d_1=0,c_1=0$ have been taken. 
Since there are two terms now surviving,  the final result turns out to be twice~\eqref{eq:ExternalCut3Loop}
\begin{equation}
\label{eq:ExternalCut3Loop2}
- \frac{2\br{1234}^3\br{A\dd^3A}}{\br{A123}\br{A124}\br{A134}\br{A234}} = -2 \frac{\dd a \dd b \dd c}{abc} \; .
\end{equation}

 Now the final configuration of these two routes is exactly the same in both cases:  $B_1 \rightarrow Z_1,B_2,B_3 \rightarrow Z_2$ and yet the results differ by a factor of 2.  We thus clearly see the importance of  path dependence when taking residues. We will shortly see that path dependence can give different results for  the all-in-one-point cut itself (rather than just when taking further residues).
Furthermore, taking three further residues as $a,b,c\rightarrow 0$ clearly gives us a maximal residue of magnitude 2 indicating that there is  an internal boundary present. Let us then consider this geometrically.

We can now redo the above computation geometrically by taking boundaries of the all-in-one-point cut geometry~\eqref{eq:3loopsDeepest} and using our formula of the canonical form of a GPG~\eqref{cfrecurs2}.
To do this, we parametrise $A$ and $B_{i}$ just as in~\eqref{aparam} and~\eqref{eq:ParamExternal}
\begin{equation}
    \label{eq:paramGeomDC}
    \begin{split}
           B_3 &= c_3 B_1 + B_2 + d_3 Z_*\; \\
    B_1 &= c_1 Z_4 +  Z_1 + d_1 Z_*\; , \\
    B_2 &= c_2 Z_1 +  Z_2 + d_2 Z_*\; ,  \\  
    A \, &=  Z_1 + a Z_2 + b Z_3 + c Z_4\; 
    \end{split}
\end{equation}
Then take boundaries in the same  order with which we took residues following~\eqref{eq:ExternalCut3Loop2}
\begin{equation}
    \label{eq:resGeomDC}
    d_3 =0, \;\;\; d_1 = 0, \;\;\; c_1 = 0, \;\;\; d_2 = 0, \;\;\; c_2 = 0, \;\;\; c_3 = 0\ .
\end{equation}
The first boundary corresponds to the above internal boundary $\br{AB_1B_2B_3} = 0$.  Therefore, from the recursive definition of the canonical form in the presence of internal boundaries~\eqref{cfrecurs2}, the canonical form of the all-in-one-point cut geometry satisfies
\begin{equation}
    \label{eq:resGeomDC2}
    \lim\limits_{d_3 \to 0} d_3\big(\Omega(\mathcal{R}^{\text{dc}}_1) + \Omega(\mathcal{R}^{\text{dc}}_2)\big) = 2\dd d_3 \wedge  \Omega(\mathcal{R}^{\text{dc}}\rvert_{\br{AB_1B_2B_3} = 0}).
\end{equation}
This simply means that the residue, $d_3 = 0$, on the canonical form determined by the inequalities describing the geometry of the all-in-one-point cut~\eqref{eq:ineqDC3Loop}, is equal to twice the canonical form of the interior boundary, i.e. the canonical form determined by the inequalities~\eqref{eq:1LoopIneq} with $B_3 = B_1 + a_3 B_2$.

Continuing with the remaining boundaries, the final geometry is described by the following set of inequalities,
\begin{equation}
    \label{eq:IntBoundary}
    \mathcal{R}^{\text{final}}: \;\;\;a<0, \;\;\; b>0, \;\;\; c>0 \; .
\end{equation}
The remaining residues were all on external boundaries, therefore the canonical form of the final region is $2\Omega(\mathcal{R}^{\text{final}})$, where the factor of two comes from the internal boundary residue~\eqref{eq:resGeomDC2}.
The final inequalities describe a tetrahedron with vertices $\{Z_1,-Z_2, Z_3, Z_4\}$ and therefore the final canonical form is in precise agreement with~\eqref{eq:ExternalCut3Loop}.

We see that the  internal boundary at $\br{AB_1B_2B_3} = 0$ is key to obtaining the correct leading singularity from the geometry.  
Just as we saw algebraically above, it is also possible to 
reach the same final loop configuration  by only going to consecutive {\em external} boundaries.
An example of this would be to follow the residues described in~\eqref{eq:ParamExternal1} geometrically.
Then the final canonical form would be, up to an overall sign, the canonical form of the tetrahedron without the factor of two,  as predicted  in~\cite{Langer:2019iuo}. We see that the precise sequence of codimension 1 boundaries taken  to approach higher codimension boundaries starting from the all-in-one-point cut configuration can give different results. At higher loops this is also true for the all-in-one-point cut itself (rather than just its maximal residues as here).

\section{Four loop point-plane boundary geometry}
\label{4loopgeom}
We here examine  the geometry of the loop-loop boundaries of the four loop all-in-one-point cut~\eqref{eq:DC4LoopGeom1}. We start by the boundary when $B_3$ lies on the line $P=B_1B_2$ followed by the boundary of that geometry when  $B_4$ also approaches  $P$. 
We observe that the regions $\mathcal{R}_1$ and $\mathcal{R}_2$ of~\eqref{eq:DC4LoopGeom1} touch on $\br{AB_2B_1B_3}=0$ as do $\mathcal{R}_3$ and $\mathcal{R}_4$. Since the orientations of $\mathcal{R}_1$ and $\mathcal{R}_2$ are opposite as are $\mathcal{R}_3$ and $\mathcal{R}_4$, this is an internal boundary.
Thus the geometry of the (internal) amplituhedron boundary $\br{AB_2B_1B_3}=0$, with $B_3$ living on $B_2B_1$, is given by $\cR_{12} \cup \cR_{34}$ where 
    \begin{equation}
    \label{eq:DC4LoopGeom1abbb2}
    \begin{split}
    \mathcal{R}_{12} &\ =\  \cA^{(4)}_{dc} \;  \; \land \;  \br{AB_2B_1B_4}>0 \land \br{AB_3B_1B_4}>0 |_{B_3\in B_2B_1},\ +  \; ,\\
    \mathcal{R}_{34}&\ =\  \cA^{(4)}_{dc} \; \land \;  \br{AB_2B_1B_4}<0 \land \br{AB_3B_1B_4}<0 |_{B_3\in B_2B_1},\ -  \; . 
    \end{split}
\end{equation}
Here $\mathcal{R}_{12}$ is positively oriented (indicated by the $+$), while $\mathcal{R}_{34}$ is negatively oriented.
Notice that both of these regions require $B_1$ to be on the same side of $B_2$ and $B_3$. In other words $B_1$ can not lie between $B_2$ and $B_3$. Further the orientation depends which side of $B_2,B_3$, $B_1$ is on. Then after  we send $B_4$ to the line $P$, we approach another internal boundary, with no further constraints on where $B_4$ can lie.
We thus conclude that the geometry is of the four points $B_i$ on the line $P$ inside the triangle, with $B_1$ not allowed between $B_2$ and $B_3$ and the orientation depending on the relative position of $B_1$ and $B_2,B_3$. 

We can check the geometry more carefully by  exploring  the boundaries of~\eqref{eq:DC4LoopGeom1abbb2} explicitly. To do this we  make the constraint $\br{AB_1B_2B_3}=0$ explicit by expanding $B_1$  as $c_1B_2+c_2B_3$. Then~\eqref{eq:DC4LoopGeom1abbb2} becomes
\begin{equation}
    \label{eq:DC4LoopGeom1abbb1}
    \begin{split}
    \mathcal{R}_{12} &\, =\,  \cA^{(4)}_{dc}   \; \land    \Big( \br{AB_2B_3B_4}>0\land c_1<0\land c_2>0 \Big)\lor \Big( \br{AB_2B_3B_4}<0\land c_1>0\land c_2<0 \Big), \ +   \; ,\\
    \mathcal{R}_{34} &\, =\,  \cA^{(4)}_{dc} \; \land \;  \Big( \br{AB_2B_3B_4}>0\land c_1>0\land c_2<0 \Big)\lor \Big( \br{AB_2B_3B_4}<0\land c_1<0\land c_2>0 \Big),\ -  \; . 
    \end{split}
\end{equation}
and we see  this has two external boundaries, $c_1=0$ and $c_2=0$, and one internal boundary at $\br{AB_2B_3B_4}=0$. The two external boundaries  correspond to sending $B_1\to B_3$  and $B_1\to B_2$ respectively. The limit internal boundary $\br{AB_2B_3B_4}\to0^\pm$ instead corresponds to sending $B_4$ to the line $P$. In this case the geometry is described by the union of a positively oriented region $\mathcal{R}^+_{1234}$ and a negatively oriented region $\mathcal{R}^-_{1234}$ as
\begin{equation}\label{firstAllPP}
\begin{split}
\mathcal{R}^+_{1234}
=\cA^{(4)}_{dc}\land c_1<0\land c_2>0,\ + \; , \\
\mathcal{R}^-_{1234}
=\cA^{(4)}_{dc}\land c_1>0\land c_2<0,\ - \; .
\end{split}
\end{equation}
We clearly see then recalling 
$B_1=c_1B_2+c_2B_3$ that  $B_1$ cannot line in between the point $B_2$ and $B_3$ and the orientation depends on which side of $B2,B_3$ $B_1$ is on, with the position of $B_4$ unconstrained. 

This also reveals that there are still further loop-loop type boundaries we could take even after doing the point-plane-cut. We could  take a residue at $c_2=0$ or $c_3=0$ corresponding to $B_1=B_2$ or $B_3$. The geometry we then obtain corresponds to the 3-loops maximal loop-loop cut $\cA^{(3)}_{\text{mll}}$ but with weight 4 instead of 2 since this time we approached two internal boundaries that is $\br{AB_2B_1B_3}=0$ and $\br{AB_2B_3B_4}=0$. This implies that the residue corresponding to this 4-loop boundary is equal to 2 times the~\eqref{eq:resGeomDC2} all-in-one-point-and-plane 3-loop residue.

Let's now go back to the all-in-one-point cut~\eqref{eq:DC4LoopGeom1}  and explore the only other boundary (modulo permutation of the loop variables), at $\br{AB_2B_1B_4}=0$. 
Notice  that this time it is an external boundary since the 4 regions remain distinct. To be very explicit, we can approach the boundary by parametrizing $B_1$ as $c_1B_2+c_2B_4+c_3Z^*$ and then taking the limit $c_3\to 0^{\pm}$ on~\eqref{eq:DC4LoopGeom1}, which becomes
\begin{equation}
    \label{eq:DC4LoopGeom12}
    \begin{split}
    \mathcal{R}_1 &\ =\  \cA^{(4)}_{dc} \; \land \; c_2\br{AB_2B_3B_4}>0 \;\land -c_1\br{AB_2B_3B_4}>0,\ +   \; ,\\
    \mathcal{R}_2 &\ =\  \cA^{(4)}_{dc} \; \land \; c_2\br{AB_2B_3B_4}<0 \;  \land -c_1\br{AB_2B_3B_4}>0,\ -   \; , \\
    \mathcal{R}_3 &\ =\  \cA^{(4)}_{dc} \; \land \; c_2\br{AB_2B_3B_4}>0 \;  \land -c_1\br{AB_2B_3B_4}<0 ,\ +  \; , \\
    \mathcal{R}_4 &\ =\  \cA^{(4)}_{dc}\; \land \; c_2\br{AB_2B_3B_4}<0 \; \land -c_1\br{AB_2B_3B_4}<0 ,\ -  \; ,
    \end{split}
\end{equation}
where the region 1 and 3 are positively oriented and 2 and 4 are negatively oriented. Here the geometry looks very similar to~\eqref{eq:DC4LoopGeom1abbb1}, but this time $c_1=0$ is unconstrained while $c_2$ is an external boundary. We can see that $c_1$ is free by taking the union of $\mathcal{R}_{1}$ and $\mathcal{R}_{3}$ and expanding the products into the different sign cases. The same goes for the pair $\mathcal{R}_{2},\mathcal{R}_{4}$ .  Because of this we can actually rewrite the~\eqref{eq:DC4LoopGeom12} as
\begin{equation}
    \label{eq:DC4LoopGeom13}
    \begin{split}
    \mathcal{R}_{13} &\ =\  \cA^{(4)}_{dc} \; \land \; c_2 \br{AB_2B_3B_4}>0 ,\ +   \; ,\\
    \mathcal{R}_{24} &\ =\  \cA^{(4)}_{dc} \; \land \; c_2 \br{AB_2B_3B_4}<0  ,\ -
    \end{split}
\end{equation}
This in turn then has  two boundaries, $\br{AB_2B_1B_3}=0$ and $c_2=0$. Setting $\br{AB_2B_1B_3}=0$ obtained by sending $B_3$ to $P$ is described by the union of a positively oriented region and a negatively oriented region as
\begin{equation}
\begin{split}
\mathcal{R}^+_{1234}=\cA^{(4)}_{dc} c_2>0,\ + \; , \\
\mathcal{R}^-_{1234}=\cA^{(4)}_{dc} c_2<0,\ - \; .
\end{split}
\end{equation}
This is the geometry of unconstrained  points $B_i$ with orientation depending on the relative order of $B_1,B_2$.

This point-plane configuration  has a further  (internal) loop-loop   boundary at $c_2=0$ corresponding to  the limit $B_1\to B_2$\,. This boundary then consists of three unconstrained points on the line $B_2,B_3,B_4$.
 This final configuration thus corresponds to the 3-loop maximal cut $\cA^{(3)}_{\text{mll}}$ with weight 2.

\section{An all loop point-plane geometry}
\label{alllpppap}

We here describe in detail the geometry corresponding to the specific all loop point-plane cut described in section~\ref{alllppp}.

We first take the simplest all-in-one-point cut  boundary.
 In this  all loops first intersect the line $A_1B_1$ and then they all slide to the same intersection point in the same order as their labeling. 
At $L$ loops this is given by the inequalities (see~\eqref{eq:DCFinal1Reg2})
\begin{align}
    &\mathcal{R}^{\text{simplest dc}}=\bigcup_{\vec{s}} \mathcal{R}^{\text{simplest dc}}_{\vec{s}}\notag\\
    &\mathcal{R}^{\text{simplest dc}}_{\vec{s}} =
{\mathcal{A}_{\text{ dc}}} \land \; \left( \bigwedge\limits_{a=3}^{L}\bigwedge\limits_{2<i<a}\left\{s_a\br{AB_iB_1B_a} > 0\right\}  \right)\; \qquad  \text{orientation} = \prod_a s_a
    \end{align}
After taking the above all-in-one-point cut we then constrain all loops to lie in the same plane by taking the ordered series of boundaries  $\{\br{AB_2B_1B_L}=0,\br{AB_2B_1B_{L-1}}=0,\cdots,\br{AB_2B_1B_3}=0\}$. To do this, we will parametrize all loops, but $B_1$ and $B_2$, as $B_a= B_1+ b_a B_2 +c_a Z^*$ and we will take the limit $c_a\to 0^\pm$. We start by approaching the boundary $\br{AB_2B_1B_L}$. What we obtain after this first limit is that for all $i$ 
\begin{align}\label{eq:c2Constrains}
    s_L\br{AB_iB_1B_L}>0\to s_L b_L \br{AB_iB_1 B_2}=-s_L 
    s_i b_L<0 \; . 
\end{align}
Since this inequality must hold for all $i$ we have that all $s_i$, apart from $s_L$ must be equal and we can therefore define a single sign $s$ as
\begin{align}
    s:=s_{L-1}=\cdots = s_3 \; .
\end{align}
Moreover, we can see that $b_L$ is actually unconstrained. In fact,  since for $s_L>0$ and $s_L<0$ we have the same orientation,~\eqref{eq:c2Constrains} reduces to $b_L>0\lor b_L<0$. The boundary geometry is then given by
\begin{align}
    \label{eq:DCFinal1Reg2}
    &\mathcal{R}^{\text{simplest dc}}|_{\br{A,B_2B_1B_L}=0}= \mathcal{R}^{\text{simplest dc}}_{s=1}|_{\br{A,B_2B_1B_L}=0}\cup\mathcal{R}^{\text{simplest dc}}_{s=-1}|_{\br{A,B_2B_1B_L}=0}\notag\\
    &\mathcal{R}^{\text{simplest dc}}_{s}|_{\br{A,B_2B_1B_L}=0} =
    {\mathcal{A}_{\text{dc}}} \land \; \left( \bigwedge\limits_{a=3}^{L-1}\bigwedge\limits_{2<i<j}\left\{s\br{AB_iB_1B_a} > 0\right\}  \right)\; \qquad  \text{orientation} = s^{L-4} \; .
    \end{align}
Now let's take the residue $\br{A21B_{L-1}}=0$ by taking the limit $c_L\to 0^\pm$.  What we obtain is that
\begin{align}
   s \br{AB_iB_1B_{L-1}}>0= s \br{AB_iB_1B_2}b_{L-1}>0=b_L<0 \; .
\end{align}
This implies that $b_{L-1}\to0^-$ corresponds to an external boundary of the geometry. Notice also that now the orientation of the two components  of the geometry labeled by $s=1$ and $s-1$ will be given by $s^{L-5}$. The series of boundaries $\br{AB_2B_1B_a}=0$ have a clear recursive structure such that at each step $a$ we get an inequality of the form $b_a<0$ and the orientation of the two components is equal to $s^{L-a-3}$. The recursion ends when we get to the last residue on $\br{AB_1B_2B_3}=0$, for which no brackets of the form $\br{AB_iB_1B_3}$ are present. At that point the two components $\br{AB_1B_2B_3}>0$ and $\br{AB_1B_2B_3}<0$ have the same boundary and opposite orientation and therefore this represents an internal boundary. We can conclude that the geometry of the simplest all-in-one-plane-and-point cut corresponds to
\begin{align}
    \label{eq:PlanePointCut}
    &\mathcal{R}^{\text{simplest dc}}_{s,L}|_{\bigwedge_a\br{A,B_2B_1B_a}=0} =
    {\mathcal{A}_{\text{ dc}}} \land \; \left( \bigwedge\limits_{a=4}^{L-1}b_a < 0 \right) \; ,
    \end{align}
    where the weight of geometry is equal to  2 due to the internal boundary $\br{AB_1B_2B_3}=0$ contribution.

At this point it's straight forward to compute the canonical form of this region for the 4-point MHV amplitude. The inequalities $b_a<0$ for $a=4,\cdots,L-1$ simply tell us that on the line $P$ all $B_a$ must be on the same side of $B_1$ as $B_2$. 
So the full canonical form is as given in~\eqref{eq:predictionSimplest}.

\bibliography{bibliography}{}

\providecommand{\href}[2]{#2}\begingroup\raggedright\begin{thebibliography}{10}

\bibitem{Arkani-Hamed:2013jha}
N.~Arkani-Hamed and J.~Trnka, ``{The Amplituhedron},''
  \href{http://dx.doi.org/10.1007/JHEP10(2014)030}{{\em JHEP} {\bfseries 10}
  (2014) 030}, \href{http://arxiv.org/abs/1312.2007}{{\ttfamily arXiv:1312.2007
  [hep-th]}}.

\bibitem{Arkani-Hamed:2013kca}
N.~Arkani-Hamed and J.~Trnka, ``{Into the Amplituhedron},''
  \href{http://dx.doi.org/10.1007/JHEP12(2014)182}{{\em JHEP} {\bfseries 12}
  (2014) 182}, \href{http://arxiv.org/abs/1312.7878}{{\ttfamily arXiv:1312.7878
  [hep-th]}}.

\bibitem{Herrmann:2022nkh}
E.~Herrmann and J.~Trnka, ``{The SAGEX Review on Scattering Amplitudes, Chapter
  7: Positive Geometry of Scattering Amplitudes},''
  \href{http://arxiv.org/abs/2203.13018}{{\ttfamily arXiv:2203.13018
  [hep-th]}}.

\bibitem{Dennen:2016mdk}
T.~Dennen, I.~Prlina, M.~Spradlin, S.~Stanojevic, and A.~Volovich, ``{Landau
  Singularities from the Amplituhedron},''
  \href{http://dx.doi.org/10.1007/JHEP06(2017)152}{{\em JHEP} {\bfseries 06}
  (2017) 152}, \href{http://arxiv.org/abs/1612.02708}{{\ttfamily
  arXiv:1612.02708 [hep-th]}}.

\bibitem{Prlina:2017azl}
I.~Prlina, M.~Spradlin, J.~Stankowicz, S.~Stanojevic, and A.~Volovich,
  ``{All-Helicity Symbol Alphabets from Unwound Amplituhedra},''
  \href{http://dx.doi.org/10.1007/JHEP05(2018)159}{{\em JHEP} {\bfseries 05}
  (2018) 159}, \href{http://arxiv.org/abs/1711.11507}{{\ttfamily
  arXiv:1711.11507 [hep-th]}}.

\bibitem{Arkani-Hamed:2016byb}
N.~Arkani-Hamed, J.~L. Bourjaily, F.~Cachazo, A.~B. Goncharov, A.~Postnikov,
  and J.~Trnka, \href{http://dx.doi.org/10.1017/CBO9781316091548}{{\em
  {Grassmannian Geometry of Scattering Amplitudes}}}.
\newblock Cambridge University Press, 4, 2016.
\newblock \href{http://arxiv.org/abs/1212.5605}{{\ttfamily arXiv:1212.5605
  [hep-th]}}.

\bibitem{Britto:2005fq}
R.~Britto, F.~Cachazo, B.~Feng, and E.~Witten, ``{Direct proof of tree-level
  recursion relation in Yang-Mills theory},''
  \href{http://dx.doi.org/10.1103/PhysRevLett.94.181602}{{\em Phys. Rev. Lett.}
  {\bfseries 94} (2005) 181602},
  \href{http://arxiv.org/abs/hep-th/0501052}{{\ttfamily arXiv:hep-th/0501052}}.

\bibitem{Even-Zohar:2021sec}
C.~Even-Zohar, T.~Lakrec, and R.~J. Tessler, ``{The Amplituhedron BCFW
  Triangulation},'' \href{http://arxiv.org/abs/2112.02703}{{\ttfamily
  arXiv:2112.02703 [math-ph]}}.

\bibitem{Arkani-Hamed:2009ljj}
N.~Arkani-Hamed, F.~Cachazo, C.~Cheung, and J.~Kaplan, ``{A Duality For The S
  Matrix},'' \href{http://dx.doi.org/10.1007/JHEP03(2010)020}{{\em JHEP}
  {\bfseries 03} (2010) 020}, \href{http://arxiv.org/abs/0907.5418}{{\ttfamily
  arXiv:0907.5418 [hep-th]}}.

\bibitem{Arkani-Hamed:2017tmz}
N.~Arkani-Hamed, Y.~Bai, and T.~Lam, ``{Positive Geometries and Canonical
  Forms},'' \href{http://dx.doi.org/10.1007/JHEP11(2017)039}{{\em JHEP}
  {\bfseries 11} (2017) 039}, \href{http://arxiv.org/abs/1703.04541}{{\ttfamily
  arXiv:1703.04541 [hep-th]}}.

\bibitem{Dian:2021idl}
G.~Dian and P.~Heslop, ``{Amplituhedron-like geometries},''
  \href{http://dx.doi.org/10.1007/JHEP11(2021)074}{{\em JHEP} {\bfseries 11}
  (2021) 074}, \href{http://arxiv.org/abs/2106.09372}{{\ttfamily
  arXiv:2106.09372 [hep-th]}}.

\bibitem{Galashin:2018fri}
P.~Galashin and T.~Lam, ``{Parity duality for the amplituhedron},''
  \href{http://dx.doi.org/10.1112/S0010437X20007411}{{\em Compos. Math.}
  {\bfseries 156} no.~11, (2020) 2207--2262},
  \href{http://arxiv.org/abs/1805.00600}{{\ttfamily arXiv:1805.00600
  [math.CO]}}.

\bibitem{Mohammadi:2020plf}
F.~Mohammadi, L.~Monin, and M.~Parisi, ``{Triangulations and Canonical Forms of
  Amplituhedra: A Fiber-Based Approach Beyond Polytopes},''
  \href{http://dx.doi.org/10.1007/s00220-021-04160-5}{{\em Commun. Math. Phys.}
  {\bfseries 387} no.~2, (2021) 927--972},
  \href{http://arxiv.org/abs/2010.07254}{{\ttfamily arXiv:2010.07254
  [math.CO]}}.

\bibitem{Kojima:2020tjf}
R.~Kojima and C.~Langer, ``{Sign Flip Triangulations of the Amplituhedron},''
  \href{http://dx.doi.org/10.1007/JHEP05(2020)121}{{\em JHEP} {\bfseries 05}
  (2020) 121}, \href{http://arxiv.org/abs/2001.06473}{{\ttfamily
  arXiv:2001.06473 [hep-th]}}.

\bibitem{Blot:2022geq}
X.~Blot and J.-R. Li, ``{The amplituhedron crossing and winding numbers},''
  \href{http://arxiv.org/abs/2206.03435}{{\ttfamily arXiv:2206.03435
  [math.CO]}}.

\bibitem{Franco:2014csa}
S.~Franco, D.~Galloni, A.~Mariotti, and J.~Trnka, ``{Anatomy of the
  Amplituhedron},'' \href{http://dx.doi.org/10.1007/JHEP03(2015)128}{{\em JHEP}
  {\bfseries 03} (2015) 128}, \href{http://arxiv.org/abs/1408.3410}{{\ttfamily
  arXiv:1408.3410 [hep-th]}}.

\bibitem{Galloni:2016iuj}
D.~Galloni, ``{Positivity Sectors and the Amplituhedron},''
  \href{http://arxiv.org/abs/1601.02639}{{\ttfamily arXiv:1601.02639
  [hep-th]}}.

\bibitem{Arkani-Hamed:2018rsk}
N.~Arkani-Hamed, C.~Langer, A.~Yelleshpur~Srikant, and J.~Trnka, ``{Deep Into
  the Amplituhedron: Amplitude Singularities at All Loops and Legs},''
  \href{http://dx.doi.org/10.1103/PhysRevLett.122.051601}{{\em Phys. Rev.
  Lett.} {\bfseries 122} no.~5, (2019) 051601},
  \href{http://arxiv.org/abs/1810.08208}{{\ttfamily arXiv:1810.08208
  [hep-th]}}.

\bibitem{mathresidues}
I.~A. Aizenberg and A.~P. Yuzhakov, ``{ Integral representations and residues
  in multidimensional complex analysis},'' {\em Translations of Mathematical
  Monographs, AMS Volume} {\bfseries 58} (1983) .

\bibitem{Eden:2017fow}
B.~Eden, P.~Heslop, and L.~Mason, ``{The Correlahedron},''
  \href{http://dx.doi.org/10.1007/JHEP09(2017)156}{{\em JHEP} {\bfseries 09}
  (2017) 156}, \href{http://arxiv.org/abs/1701.00453}{{\ttfamily
  arXiv:1701.00453 [hep-th]}}.

\bibitem{Langer:2019iuo}
C.~Langer and A.~Yelleshpur~Srikant, ``{All-loop cuts from the
  Amplituhedron},'' \href{http://dx.doi.org/10.1007/JHEP04(2019)105}{{\em JHEP}
  {\bfseries 04} (2019) 105}, \href{http://arxiv.org/abs/1902.05951}{{\ttfamily
  arXiv:1902.05951 [hep-th]}}.

\bibitem{Bourjaily:2016evz}
J.~L. Bourjaily, P.~Heslop, and V.-V. Tran, ``{Amplitudes and Correlators to
  Ten Loops Using Simple, Graphical Bootstraps},''
  \href{http://dx.doi.org/10.1007/JHEP11(2016)125}{{\em JHEP} {\bfseries 11}
  (2016) 125}, \href{http://arxiv.org/abs/1609.00007}{{\ttfamily
  arXiv:1609.00007 [hep-th]}}.

\bibitem{Mond1}
J.~Rao, ``{4-particle Amplituhedron at 3-loop and its Mondrian Diagrammatic
  Implication},'' \href{http://dx.doi.org/10.1007/JHEP06(2018)038}{{\em JHEP}
  {\bfseries 06} (2018) 038}, \href{http://arxiv.org/abs/1712.09990}{{\ttfamily
  arXiv:1712.09990 [hep-th]}}.

\bibitem{Mond2}
Y.~An, Y.~Li, Z.~Li, and J.~Rao, ``{All-loop Mondrian Diagrammatics and
  4-particle Amplituhedron},''
  \href{http://dx.doi.org/10.1007/JHEP06(2018)023}{{\em JHEP} {\bfseries 06}
  (2018) 023}, \href{http://arxiv.org/abs/1712.09994}{{\ttfamily
  arXiv:1712.09994 [hep-th]}}.

\bibitem{Mond3}
J.~Rao, ``{All-loop Mondrian Reduction of 4-particle Amplituhedron at Positive
  Infinity},'' \href{http://dx.doi.org/10.1016/j.nuclphysb.2020.115086}{{\em
  Nucl. Phys. B} {\bfseries 957} (2020) 115086},
  \href{http://arxiv.org/abs/1910.14612}{{\ttfamily arXiv:1910.14612
  [hep-th]}}.

\bibitem{Kojima:2018qzz}
R.~Kojima, ``{Triangulation of 2-loop MHV Amplituhedron from Sign Flips},''
  \href{http://dx.doi.org/10.1007/JHEP04(2019)085}{{\em JHEP} {\bfseries 04}
  (2019) 085}, \href{http://arxiv.org/abs/1812.01822}{{\ttfamily
  arXiv:1812.01822 [hep-th]}}.

\bibitem{Kojima:2020gxs}
R.~Kojima and J.~Rao, ``{Triangulation-free Trivialization of 2-loop MHV
  Amplituhedron},'' \href{http://dx.doi.org/10.1007/JHEP10(2020)140}{{\em JHEP}
  {\bfseries 10} (2020) 140}, \href{http://arxiv.org/abs/2007.15650}{{\ttfamily
  arXiv:2007.15650 [hep-th]}}.

\bibitem{Herrmann:2020qlt}
E.~Herrmann, C.~Langer, J.~Trnka, and M.~Zheng, ``{Positive geometry, local
  triangulations, and the dual of the Amplituhedron},''
  \href{http://dx.doi.org/10.1007/JHEP01(2021)035}{{\em JHEP} {\bfseries 01}
  (2021) 035}, \href{http://arxiv.org/abs/2009.05607}{{\ttfamily
  arXiv:2009.05607 [hep-th]}}.

\bibitem{Mason:2009qx}
L.~Mason and D.~Skinner, ``{Dual Superconformal Invariance, Momentum Twistors
  and Grassmannians},''
  \href{http://dx.doi.org/10.1088/1126-6708/2009/11/045}{{\em JHEP} {\bfseries
  11} (2009) 045}, \href{http://arxiv.org/abs/0909.0250}{{\ttfamily
  arXiv:0909.0250 [hep-th]}}.

\bibitem{Arkani-Hamed:2021iya}
N.~Arkani-Hamed, J.~Henn, and J.~Trnka, ``{Nonperturbative negative geometries:
  amplitudes at strong coupling and the amplituhedron},''
  \href{http://dx.doi.org/10.1007/JHEP03(2022)108}{{\em JHEP} (2022) 108},
  \href{http://arxiv.org/abs/2112.06956}{{\ttfamily arXiv:2112.06956
  [hep-th]}}.

\bibitem{Damgaard:2019ztj}
D.~Damgaard, L.~Ferro, T.~Lukowski, and M.~Parisi, ``{The Momentum
  Amplituhedron},'' \href{http://dx.doi.org/10.1007/JHEP08(2019)042}{{\em JHEP}
  {\bfseries 08} (2019) 042}, \href{http://arxiv.org/abs/1905.04216}{{\ttfamily
  arXiv:1905.04216 [hep-th]}}.

\bibitem{Lukowski:2020bya}
T.~\L{}ukowski and R.~Moerman, ``{Boundaries of the amplituhedron with
  amplituhedronBoundaries},''
  \href{http://dx.doi.org/10.1016/j.cpc.2020.107653}{{\em Comput. Phys.
  Commun.} {\bfseries 259} (2021) 107653},
  \href{http://arxiv.org/abs/2002.07146}{{\ttfamily arXiv:2002.07146
  [hep-th]}}.

\bibitem{Ferro:2020lgp}
L.~Ferro, T.~\L{}ukowski, and R.~Moerman, ``{From momentum amplituhedron
  boundaries toamplitude singularities and back},''
  \href{http://dx.doi.org/10.1007/JHEP07(2020)201}{{\em JHEP} {\bfseries 07}
  no.~07, (2020) 201}, \href{http://arxiv.org/abs/2003.13704}{{\ttfamily
  arXiv:2003.13704 [hep-th]}}.

\bibitem{Moerman:2021cjg}
R.~Moerman and L.~K. Williams, ``{Grass trees and forests: Enumeration of
  Grassmannian trees and forests, with applications to the momentum
  amplituhedron},'' \href{http://arxiv.org/abs/2112.02061}{{\ttfamily
  arXiv:2112.02061 [math.CO]}}.

\bibitem{Arkani-Hamed:2014bca}
N.~Arkani-Hamed, J.~L. Bourjaily, F.~Cachazo, A.~Postnikov, and J.~Trnka,
  ``{On-Shell Structures of MHV Amplitudes Beyond the Planar Limit},''
  \href{http://dx.doi.org/10.1007/JHEP06(2015)179}{{\em JHEP} {\bfseries 06}
  (2015) 179}, \href{http://arxiv.org/abs/1412.8475}{{\ttfamily arXiv:1412.8475
  [hep-th]}}.

\bibitem{Damgaard:2021qbi}
D.~Damgaard, L.~Ferro, T.~Lukowski, and R.~Moerman, ``{Kleiss-Kuijf relations
  from momentum amplituhedron geometry},''
  \href{http://dx.doi.org/10.1007/JHEP07(2021)111}{{\em JHEP} {\bfseries 07}
  (2021) 111}, \href{http://arxiv.org/abs/2103.13908}{{\ttfamily
  arXiv:2103.13908 [hep-th]}}.

\bibitem{He:2021llb}
S.~He, C.-K. Kuo, and Y.-Q. Zhang, ``{The momentum amplituhedron of SYM and
  ABJM from twistor-string maps},''
  \href{http://dx.doi.org/10.1007/JHEP02(2022)148}{{\em JHEP} {\bfseries 02}
  (2022) 148}, \href{http://arxiv.org/abs/2111.02576}{{\ttfamily
  arXiv:2111.02576 [hep-th]}}.

\bibitem{Huang:2021jlh}
Y.-t. Huang, R.~Kojima, C.~Wen, and S.-Q. Zhang, ``{The orthogonal momentum
  amplituhedron and ABJM amplitudes},''
  \href{http://dx.doi.org/10.1007/JHEP01(2022)141}{{\em JHEP} {\bfseries 01}
  (2022) 141}, \href{http://arxiv.org/abs/2111.03037}{{\ttfamily
  arXiv:2111.03037 [hep-th]}}.

\bibitem{Ferro:2021dub}
L.~Ferro and R.~Moerman, ``{The Grassmannian for celestial superamplitudes},''
  \href{http://dx.doi.org/10.1007/JHEP11(2021)187}{{\em JHEP} {\bfseries 11}
  (2021) 187}, \href{http://arxiv.org/abs/2107.07496}{{\ttfamily
  arXiv:2107.07496 [hep-th]}}.

\bibitem{Arkani-Hamed:2019vag}
N.~Arkani-Hamed, S.~He, G.~Salvatori, and H.~Thomas, ``{Causal Diamonds,
  Cluster Polytopes and Scattering Amplitudes},''
  \href{http://arxiv.org/abs/1912.12948}{{\ttfamily arXiv:1912.12948
  [hep-th]}}.

\bibitem{Arkani-Hamed:2022cqe}
N.~Arkani-Hamed, A.~Hillman, and S.~Mizera, ``{Feynman polytopes and the
  tropical geometry of UV and IR divergences},''
  \href{http://dx.doi.org/10.1103/PhysRevD.105.125013}{{\em Phys. Rev. D}
  {\bfseries 105} no.~12, (2022) 125013},
  \href{http://arxiv.org/abs/2202.12296}{{\ttfamily arXiv:2202.12296
  [hep-th]}}.

\bibitem{John:2020jww}
R.~R. John, R.~Kojima, and S.~Mahato, ``{Weights, Recursion relations and
  Projective triangulations for Positive Geometry of scalar theories},''
  \href{http://dx.doi.org/10.1007/JHEP10(2020)037}{{\em JHEP} {\bfseries 10}
  (2020) 037}, \href{http://arxiv.org/abs/2007.10974}{{\ttfamily
  arXiv:2007.10974 [hep-th]}}.

\bibitem{Jagadale:2022rbl}
M.~Jagadale and A.~Laddha, ``{Towards Positive Geometries of Massive Scalar
  field theories},'' \href{http://arxiv.org/abs/2206.07979}{{\ttfamily
  arXiv:2206.07979 [hep-th]}}.

\bibitem{Lukowski:2022fwz}
T.~Lukowski, R.~Moerman, and J.~Stalknecht, ``{Pushforwards via Scattering
  Equations with Applications to Positive Geometries},''
  \href{http://arxiv.org/abs/2206.14196}{{\ttfamily arXiv:2206.14196
  [hep-th]}}.

\bibitem{Arkani-Hamed:2017fdk}
N.~Arkani-Hamed, P.~Benincasa, and A.~Postnikov, ``{Cosmological Polytopes and
  the Wavefunction of the Universe},''
  \href{http://arxiv.org/abs/1709.02813}{{\ttfamily arXiv:1709.02813
  [hep-th]}}.

\bibitem{Benincasa:2022gtd}
P.~Benincasa, ``{Amplitudes meet Cosmology: A (Scalar) Primer},''
  \href{http://arxiv.org/abs/2203.15330}{{\ttfamily arXiv:2203.15330
  [hep-th]}}.

\end{thebibliography}\endgroup
\bibliographystyle{utphys.bst}

\end{document}